\documentclass{jfm}
\usepackage{graphicx}
\usepackage{epstopdf, epsfig}
\usepackage{amsmath}
\usepackage{eurosym}
\usepackage{graphicx}
\usepackage{caption,subcaption}
\usepackage{float}

\usepackage{color}
\usepackage{soul}

\newcommand{\beq}{\begin{equation}}
\newcommand{\eeq}{\end{equation}}
\newcommand{\bu}{{\bf u}} 
 
\newcommand{\bF}{{\bf f}}
\newcommand{\bk}{{\bf k}}
\newcommand{\bq}{{\bf q}}
\newcommand{\bp}{{\bf p}}
\newcommand{\bx}{{\bf x}}

\newcommand{\cN}{{\mathcal{N}}}
\newcommand{\cE}{{\mathcal{E}}}

\newcommand{\cT}{{\mathcal{T}}}

\newcommand{\bul}{{\bf u}_{_L}}
\newcommand{\buf}{{\bf u}_{_F}}
\newcommand{\but}{{\bf u}_{_T}}
\newcommand{\kf}{{k_{f}}}

\shorttitle{Thermal equilibrium of large scales}
\shortauthor{A. Alexakis and M-E. Brachet}

\title{On the thermal equilibrium state of large scale flows}

\author{Alexandros Alexakis \aff{1}
  \corresp{\email{alexakis@lps.ens.fr}},
Marc-Etienne Brachet \aff{1}}

\affiliation{ \aff{1} Laboratoire de Physique Statistique, D\'epartement de Physique de l’ Ecole Normale Sup\'erieure,
PSL Research University, Universit\'e Paris Diderot, Sorbonne Paris Cit\'e, Sorbonne Universit\'es,
UPMC Univ. Paris 06, CNRS, 75005 Paris, France}

\begin{document}

\maketitle
\begin{abstract}
In a forced  three-dimensional turbulent flow the scales larger than the forcing scale have been conjectured to reach a thermal equilibrium state forming a $k^2$ energy spectrum.  
In this work we examine the properties of these large scales in turbulent flows with the use of numerical simulations. 
We show that the choice of forcing can strongly affect the behavior of the large scales.
A spectrally-dense forcing (a forcing that acts on {\it all modes} inside a finite-width spherical shell) with long correlation times may lead to strong deviations from the $k^2$ energy spectrum, while a spectrally-sparse forcing (a forcing that acts only on a few modes) with short correlated time-scale can reproduce the thermal spectrum. 
The origin of these deviations is analysed and the involved mechanisms is unraveled by examining: 
(i) the number of triadic interactions taking place, 
(ii) the spectrum of the non-linear term, 
(iii) the amplitude of interactions and the fluxes due to different scales, and 
(iv) the transfer function between different shells of wavenumbers.
It is shown that the spectrally-dense forcing allows for numerous triadic interactions that couple one large scale mode with two forced modes and this leads to an excess of energy input in the large scales. 
This excess of energy is then moved back to the small-scales by self-interactions of the large-scale modes and by interactions with the turbulent small-scales.
The overall picture that arises from the present analysis is that the large scales in a turbulent flow resemble a reservoir that is in (non-local) contact with a second out-of equilibrium reservoir consisting of the smaller (forced, turbulent and dissipative) scales. 
If the injection of energy at the large scales from the forced modes is relative weak (as is the case for the spectrally sparse forcing) 
then the large-scale spectrum remains close to a thermal equilibrium and the role of long-range interactions is to set the global energy (temperature) of the equilibrium state. 
If, on the other hand, the long-range interactions are dominant (as is the case for the spectrally dense forcing), the large-scale self-interactions cannot respond fast enough to bring the system into equilibrium. Then the large scales deviate from the equilibrium state with
energy spectrum that may display exponents different from the $k^2$ spectrum.
\end{abstract}


\section{Introduction}        
\label{sec:intro}             

In a typical three-dimensional high-Reynolds turbulent flow, energy that is injected by an external force at a particular scale (from now on the {\it forcing scale}) is transferred by nonlinear interactions to smaller and smaller scales. This process continues until small enough scales are reached, (the {\it dissipation scales}), such that energy is dissipated by viscous forces. At late times the flow reaches a statistical steady state at which there is a continuous flux of energy from the forcing scale to the small dissipation scales. The statistical properties at these intermediate scales, between the forcing scale and the dissipation scale (that we will refer as the {\it turbulent scales}), have been extensively studied by both numerical simulations and experiments in the past decades (\cite{frisch95}). The flow properties at the turbulent scales are determined by the energy flux and lead to a power-law energy spectrum that, to a close approximation, it is given by the Kolmogorov energy spectrum $E(k) \propto \epsilon^{2/3} k^{-5/3}$ where $k$ is the wavenumber and $\epsilon$ is the per unit-mass energy injection rate. 
The statistical properties however of the flow at scales larger than the forcing scale (which we will refer as the {\it large scales}) have received very little investigation. In the absence of any anisotropy caused by the domain geometry, rotation or other mechanisms there is no net flux of energy to the large scales \citep{alexakis2018cascades}. Since there is zero average flux of energy at these scales it has been conjectured that these scales can be described by a {\it thermal equilibrium} state. 
 
Thermal equilibrium states are realized in isolated system conserving a number of invariants that 
determine the system's statistical properties at late times.
In fluid dynamics, equilibrium spectra are realized for the truncated Euler equations where only a finite number of Fourier modes are kept:
\beq
\partial_t \bu + \mathbb{P}_K[\bu \cdot \nabla \bu +\nabla P] =0. \label{TEE}
\eeq
Here $\bf u$ is an incompressible velocity field, $P$ is the pressure and $\mathbb{P}_K$ is a projection operator that sets to zero all Fourier modes except those that belong to a particular set $K$ (here chosen to be a sphere centered at the origin with radius $k_{max}$). 
The truncated Euler equations conserve exactly the two quadratic invariants of the Euler Equations, 
$$ \mathrm{Energy} \quad \mathcal{E}=\frac{1}{2}\int |{\bf u}|^2 dx^3 \quad \mathrm{and\,\, Helicity} \quad \mathcal{H}=\frac{1}{2}\int {\bf u\cdot\nabla \times u} dx^3.$$
The distribution of these invariants among the different degrees of freedom are quantified by the energy and helicity spherically averaged spectra $E(k),H(k)$ respectively defined as
\[ E(k)=\frac{1}{2}\sum_{k\le |{\bf k}| <k+1} |\tilde{\bf u}_{\bf k}|^2 \quad \mathrm{and} \quad  
   H(k)=\frac{1}{2}\sum_{k\le |{\bf k}| <k+1}  \tilde{\bf u}_{\bf k}\cdot(i{\bf k} \times \tilde{\bf u}_{\bf -k}) \]
where $\tilde{\bf u}$ is the Fourier transform of $\bf u$ and a triple periodic cubic domain has been assumed.
One can then consider the statistical equilibrium state based on Liouville's theorem \cite{LEE:1952p4100} and Gaussian equipartition ensemble \cite{OrszagHouches} of this system that leads to the \cite{kraichnan1973helical} predictions for $E(k),H(k)$:
\beq
E(k) = \frac{4\pi \alpha k^2}{\alpha^2 - \beta^2 k^2 },\qquad  H(k) = \frac{4 \pi \beta k^4}{\alpha^2 - \beta^2 k^2 }.
\label{eq:EkTh}
\eeq 
In fluid dynamics this state is referred to as an {\it absolute equilibrium} and it is equivalent to a thermal equilibrium state in statistical physics.
The coefficients $ \alpha$ and $\beta$ are determined by imposing the conditions
$$\mathcal{E}=\sum_{\bf k} E(k) \quad \mathrm{and}\quad  \mathcal{H}=\sum_{\bf k} H(k)$$ 
where $\mathcal{E}$ and $\mathcal{H}$ are the initial energy and helicity respectively.
For zero helicity $\mathcal{H} = \beta=0$ the energy spectrum reduces to equipartition of energy among all Fourier modes.  
For $\beta\ne 0$ a $k^2$ spectrum is also followed for small $k$ but when the flow is strongly helical $\beta \sim \alpha/k_{max}$ 
a near singular behavior is observed at the wavenumber $k_c=\alpha/\beta> k_{max}$. 
Similarly, the helicity spectrum follows $H(k) \propto  \beta/\alpha^2 k^4$ for small $k$ and is also singular at $k_c > k_{max}$.
\begin{figure}                 
  \centering
  \includegraphics[width=0.48\textwidth]{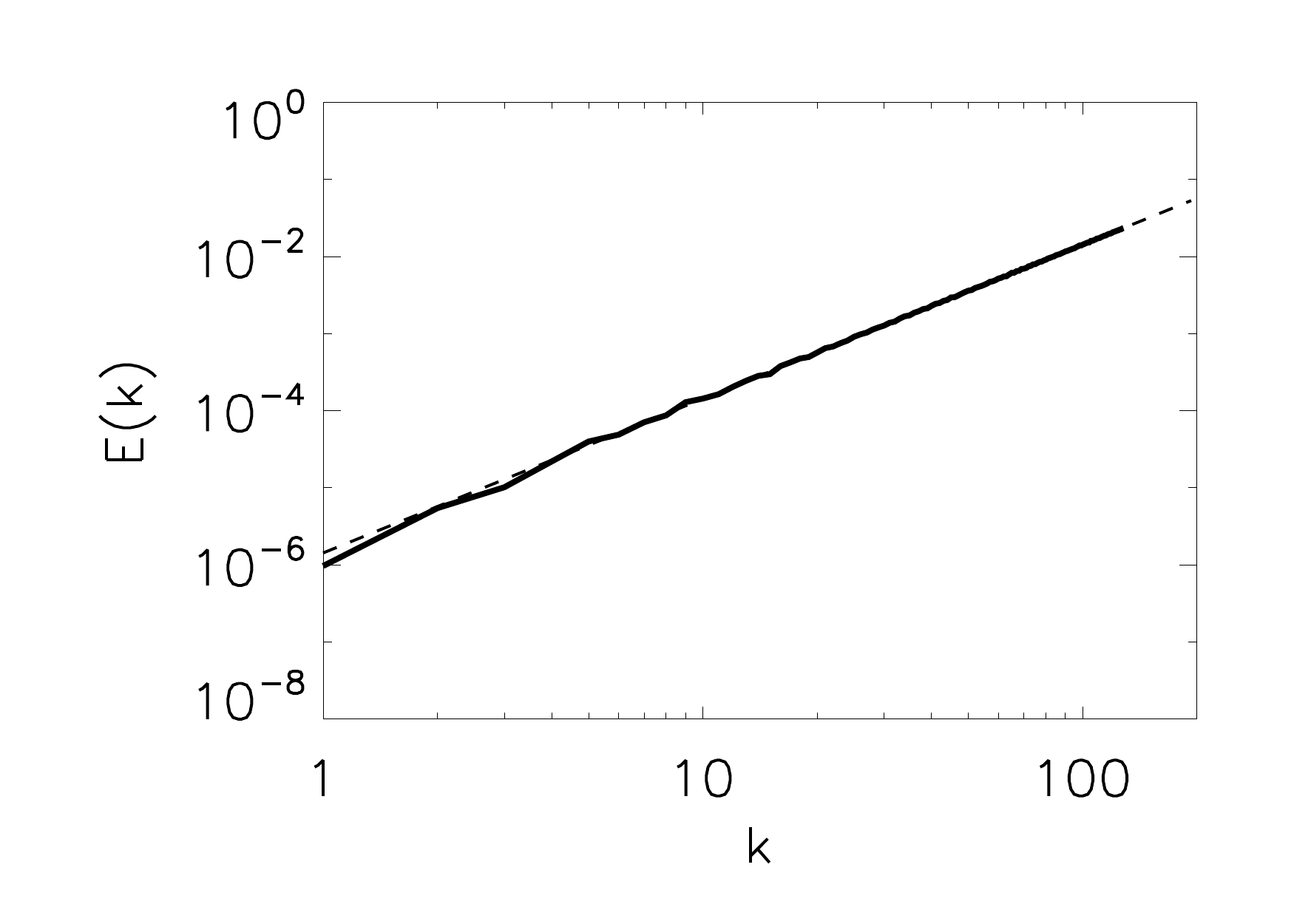}
  \includegraphics[width=0.48\textwidth]{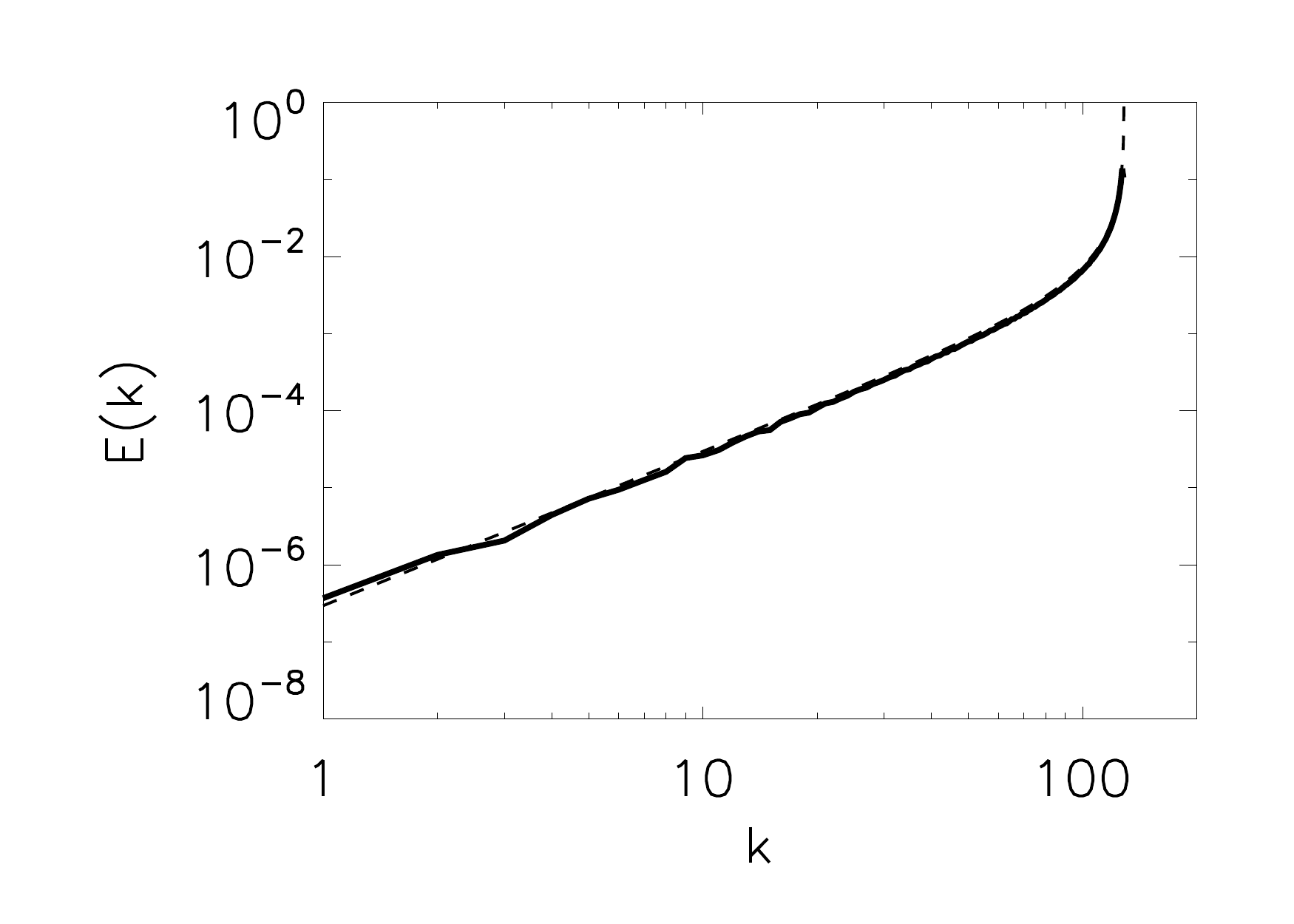}
  \caption{ The energy spectra from two simulations of the truncated Euler equations with $k_{max}=128$ and zero helicity (left) and $\mathcal{H}/\mathcal{E}k_{max}=0.82$ (right). The dashed lines show the theoretical predictions  given in eq. \ref{eq:EkTh}. }
  \label{fig:specEuler}
\end{figure}                   
A realization of the two spectra at late times obtained from two different numerical simulations of the truncated Euler equations with 
$k_{max}=128$ and zero helicity (left) and $\mathcal{H}/\mathcal{E}k_{max}=0.82$ (right) is shown in figure \ref{fig:specEuler}. 
The dashed lines show the theoretical predictions  given in eq. \ref{eq:EkTh}.
Furthermore, besides the energy spectum,
the correlation time $\tau_{\bf k}$ of the different Fourier modes can be calculated for the thermal equilibrium state and scales like $\tau_{\bf k} \propto k^{-1} \mathcal{E}^{-1/2}$ for non-helical flows while $\tau_{\bf k} \propto k^{-1/2} \mathcal{H}^{-1/2}$  for strongly helical flows \citep{cichowlas2005effective,cameron2017effect}. These predictions have been verified with numerical simulations of the truncated Euler equations in numerous investigations 
\citep{cichowlas2005effective, krstulovic2009cascades, cameron2017effect}.

Recent numerical simulations \citep{dallas2015statistical,cameron2017effect} have revealed that the properties of the large scales in a forced turbulent flow, are close to those predicted 
by the equilibrium statistical mechanics in \cite{kraichnan1973helical}. (We need to note that the steady state problem considered here differs from the one of the large
scale structure in decaying turbulence, although it leads to the prediction of a similar spectrum discussed in \cite{saffman1967large,ishida2006decay,krogstad2010grid}.)
At steady state the energy spectra at large scales were shown to be close to a $k^2$ power-law while the 
correlation time were also compatible with the $\tau_{\bf k} \propto k^{-1} \mathcal{E}^{-1/2}$ prediction 
at least for some range of scales. The agreement with the spectral and temporal predictions 
would indicate that the large scales in a turbulent flow are in equilibrium and can be described by such dynamics.
However some notable deviations both in the spectra and in the correlation time scales were also observed in \cite{dallas2015statistical,cameron2017effect} and some care needs to be taken.
This is what we will try to investigate in this work. 

Strictly speaking absolute equilibrium statistical mechanics can be applied to isolated systems such that no energy injection
or dissipation takes place. The large scales in dissipative Navier-Stokes equations are different in many respects from the absolute equilibrium 
of the truncated Euler equations. First of all in the truncated Euler equations the energy and helicity are conserved
and determined solely by the initial conditions. For the large scales of the Navier Stokes equations however, 
there is a constant exchange of energy with the forcing and turbulent scales by the nonlinearity that couples all modes. 
The large scales reach an equilibrium state such that only on average there is zero energy exchange with the forced and the turbulent scales. 
The energy contained in these scales is thus not determined by initial conditions but by the equilibration processes with the forced and turbulent scales.
In other words the large scales in a turbulent flow resemble a reservoir that is in a (non-local) contact with a second out-of equilibrium reservoir consisting 
of the smaller (forced, turbulent and dissipative) scales.

This point of view leads to two possibilities. If the energy exchange fluctuations between the large and the small scales
are relative weak compared to the large-scale self-interactions then one expects that the large scale spectrum 
 will be indeed close to a thermal equilibrium state and will be universal. The role of the long-range interactions (between the large and the small scales) 
will only be to set the global energy (temperature) and helicity of the equilibrium state without altering the functional form of the spectrum 
that is determined by the local large-scale interactions. If on the other hand the long-range interactions
are dominant, so that the  large-scale self-interactions cannot respond fast enough to bring the system in equilibrium 
then the large scales can deviate from the equilibrium state and their statistical properties will be determined by the forcing 
and turbulent scales. Furthermore, if it is the interactions with the turbulent scales that determine the large scale spectrum we expect again the large scale spectrum to be universal. 
If however it is the interactions with the forcing scales that dominate, the large scale spectrum will not be universal and will depend on the details of the forcing.

In this work we try to answer these questions with a set of numerical simulations. 
Our findings show that at least for the examined values of the scale-separations
both situations are feasible. The remaining presentation of this work is as follows. In section \ref{sec:simulations} we describe the exact setup we are going to investigate and 
present the numerical simulations used. In section \ref{sec:spectra} we present the resulting large-scale energy spectra, while in section \ref{sec:analysis} we present an analysis of the results 
by looking at the number of interacting triads, the amplitude of the nonlinearity spectrum, the flux due to interactions of different scales and different helicity and the energy transfer among the different scales.
We draw our conclusions in the last section.

\section{Setup and Numerical simulations} 
\label{sec:simulations}         

\begin{table}                                                                                                                         
\centering                                                                                                                            
\begin{tabular}{l|ccccccc}
RUN \qquad & \quad Helicity \quad & correlation time &  forced modes     & $k_f$ &$\delta k_f$ &  \quad $\nu$ \quad          \\ \hline 
NS0        & Non-helical  &  $\delta t=0$          &  $(\pm k_f,0,0), \,\, (0,\pm k_f,0), \,\, (0,0,\pm k_f)$  & 40 & 0 & $7\cdot 10^{-19}$   \\
NS1        & Non-helical  &  $\delta t=1$          &  $(\pm k_f,0,0), \,\, (0,\pm k_f,0), \,\, (0,0,\pm k_f)$  & 40 & 0 & $7\cdot 10^{-19}$   \\
NS8        & Non-helical  &  $\delta t=\infty$     &  $(\pm k_f,0,0), \,\, (0,\pm k_f,0), \,\, (0,0,\pm k_f)$  & 40 & 0 & $5\cdot 10^{-19}$    \\
NM0        & Non-helical  &  $\delta t=0$          &   All modes $\bf k$ in $k_f-\delta k_f \le|{\bf k}| \le k_f$    & 40 & 4 & $7\cdot 10^{-19}$   \\
NM1        & Non-helical  &  $\delta t=1$          &   All modes $\bf k$ in $k_f-\delta k_f \le|{\bf k}| \le k_f$    & 40 & 4 & $7\cdot 10^{-19}$  \\
NM8        & Non-helical  &  $\delta t=\infty$     &   All modes $\bf k$ in $k_f-\delta k_f \le|{\bf k}| \le k_f$    & 40 & 4 & $6\cdot 10^{-19}$    \\
HS0        & Helical      &  $\delta t=0$          &   $(\pm k_f,0,0), \,\, (0,\pm k_f,0), \,\, (0,0,\pm k_f)$ & 40 & 0 & $5\cdot 10^{-19}$   \\
HS1        & Helical      &  $\delta t=1$          &   $(\pm k_f,0,0), \,\, (0,\pm k_f,0), \,\, (0,0,\pm k_f)$ & 40 & 0 & $5\cdot 10^{-19}$   \\
HS8        & Helical      &  $\delta t=\infty$     &   $(\pm k_f,0,0), \,\, (0,\pm k_f,0), \,\, (0,0,\pm k_f)$ & 40 & 0 & $5\cdot 10^{-19}$   \\
HM0        & Helical      &  $\delta t=0$          &   All modes $\bf k$ in $k_f-\delta k_f \le|{\bf k}| \le k_f$    & 40 & 4 & $5\cdot 10^{-19}$   \\
HM1        & Helical      &  $\delta t=1$          &   All modes $\bf k$ in $k_f-\delta k_f \le|{\bf k}| \le k_f$    & 40 & 4 & $5\cdot 10^{-19}$  \\
HM8        & Helical      &  $\delta t=\infty$     &   All modes $\bf k$ in $k_f-\delta k_f \le|{\bf k}| \le k_f$    & 40 & 4 & $5\cdot 10^{-19}$ 
\end{tabular}
\caption{Table of runs. For all runs the resolution in each direction was $N=1024$. The $\delta t=\infty$ implies that the forcing was constant in time, 
$\delta t=1$ implies that the forcing was changed randomly approximately every turnover time and $\delta t=0$ impies that
the forcing changed randomly every time step.}
\label{tableofruns}                                                                                                                   
\end{table}                                                                                                                           

To investigate the dynamics of the large scales we performed a number of numerical simulations.
The simulations follow the flow of an incompressible and unit density fluid in a triple periodic cube of size $2\pi$.
The flow satisfies the hyper-viscous Navier-Stokes equation:
\beq
\partial_t \bu + \bu \cdot \nabla \bu = -\nabla P - \nu \Delta^4 \bu + \bF
\eeq  
where $\bu$ is the incompressible ($\nabla \cdot \bu=0$) velocity field, $P$ is the pressure, $\nu$ is the hyper-viscosity
and $\bF$ is the an externally imposed forcing. Since we are mostly interested in the behavior of the large scales we have chosen a forcing that is concentrated around large Fourier wavenumbers with $|{\bf k}| \sim k_f=40$. The use of hyper-viscosity was found to be necessary to avoid any molecular viscosity effects in the large scales.
Carrying out the present simulations with regular viscosity $\nu$ at high enough Reynolds number $Re=u_{rms}/(\nu k_f)$ so that the flow is turbulent while maintaining 
 a large scale separation between the forcing scale and the domain size is not feasible with the available computational resources. 
The simulations were performed using the pseudospectral {\sc Ghost}-code \citep{mininni2011hybrid} with a second order Runge-Kutta method for the time advancement and 2/3 rule for dealiasing. 
For all runs we used a computational grid of size $N=1024$ in each of the three directions, and we tuned the value of the hyper-viscous coefficient so that the flow is well resolved.

To test the effect of forcing on the large scale modes $12$ 
different forcing functions were used that varied in helicity, their correlation time and the number of forced modes. 
In particular regarding the helicity two options were considered.
Either the forcing was chosen such that every realization was fully helical $\nabla \times \bF = k_f \bF$ 
or it had exactly zero helicity $\langle \bF \cdot \nabla \times \bF \rangle=0$ (where brackets stand for spacial average). 
We refer to these two types of forcing as the helical and the non-helical forcing.

The second parameter we varied, was the number of Fourier modes that were forced. 
Two choices were examined. In the first choice the forced modes ${\bf k}=(k_x,k_y,k_z)$ were the six modes on the faces of the $k_f$-cube 
$(\pm k_f,0,0), \,\, (0,\pm k_f,0), \,\, (0,0,\pm k_f)$ with $k_f=40$. This case corresponds to an ABC forcing if helical, 
or to its non-helical version sometimes referred as the CBA forcing \citep{cameron2017effect}.
We refer to the flows with this forcing as the {\it six-mode} forced flows.
The other choice was to force all Fourier modes inside a spherical shell of external radius $k_f=40$ and internal radius $k_f-\delta k_f=36$. 
This forcing corresponds to a random (almost) isotropic forcing and we will refer to this forcing as {\it multi-mode} forcing.

Finally, the last parameter we varied was the correlation time of the forcing.
The phases of the forcing modes $\tilde{\bF}_{\bf k}$ were changed randomly every time interval $\delta t$. 
Three choices for the correlation time $\delta t$ were made: (a) it was either infinite, $\delta t=\infty$, (so that $\bF$ was independent of time), 
(b) it was finite and close to the turnover time $\delta t \simeq 1/(k_f\mathcal{E}^{1/2})$, or (c) the phases were changed every numerical time step. In the last case the forcing is approximately delta-correlated in time fixing in this way the energy injection rate $\epsilon$. 
The parameters of all the runs are given in Table \ref{tableofruns}.

\section{Large scale energy spectra} 
\label{sec:spectra}                  

We begin by examining the energy spectra of the different flows. The energy spectra are outputted frequently throughout the numerical
simulation and are time-averaged in the steady state regime. This averaging is particularly important for the time dependent six-mode forced runs that displayed large fluctuations in the large-scale energy spectra, and a spectrum calculated from a single time realization of the flow field can considerably deviate from the time-averaged value. 

\begin{figure}                 
  \centering
  \includegraphics[width=0.48\textwidth]{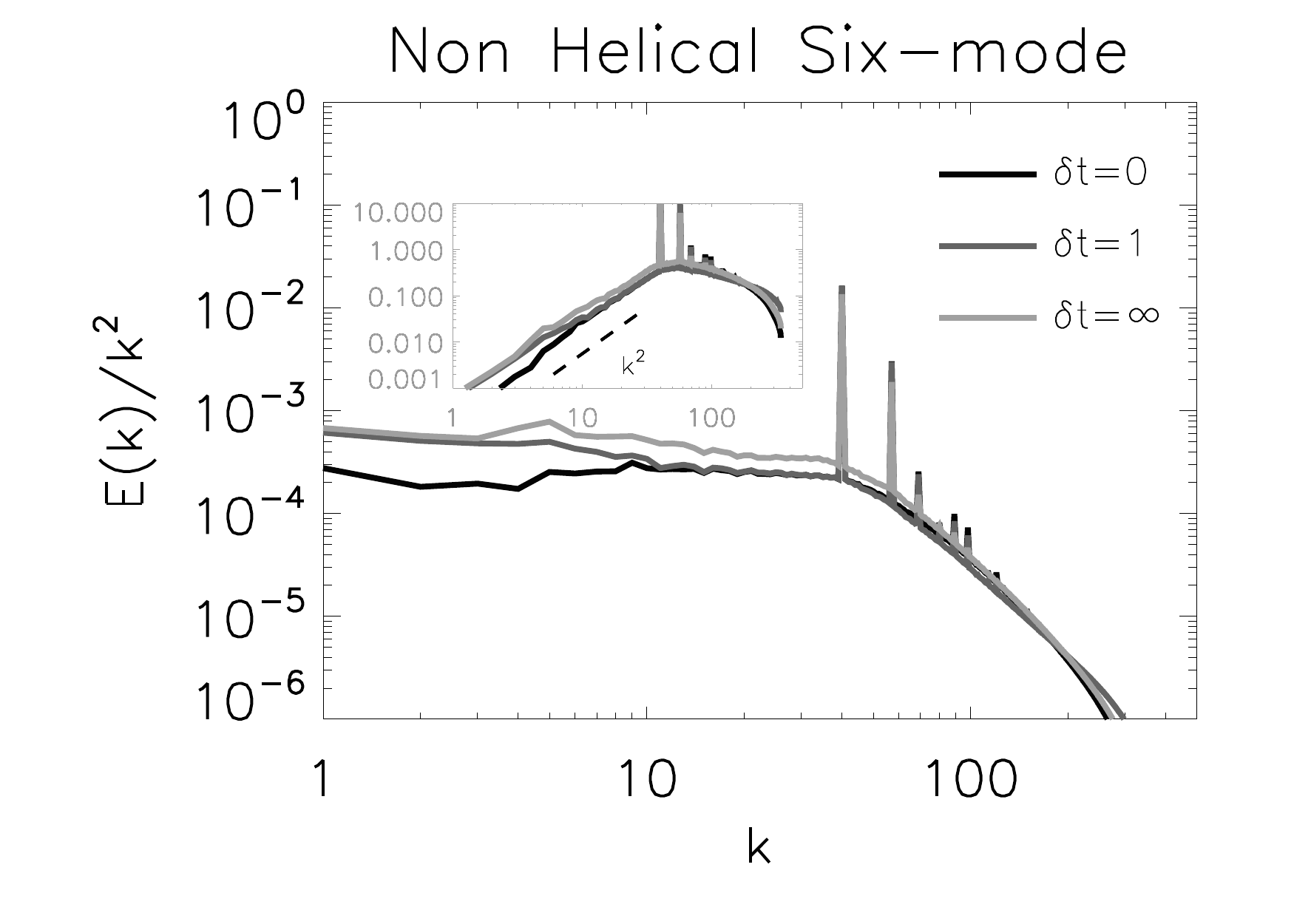}
  \includegraphics[width=0.48\textwidth]{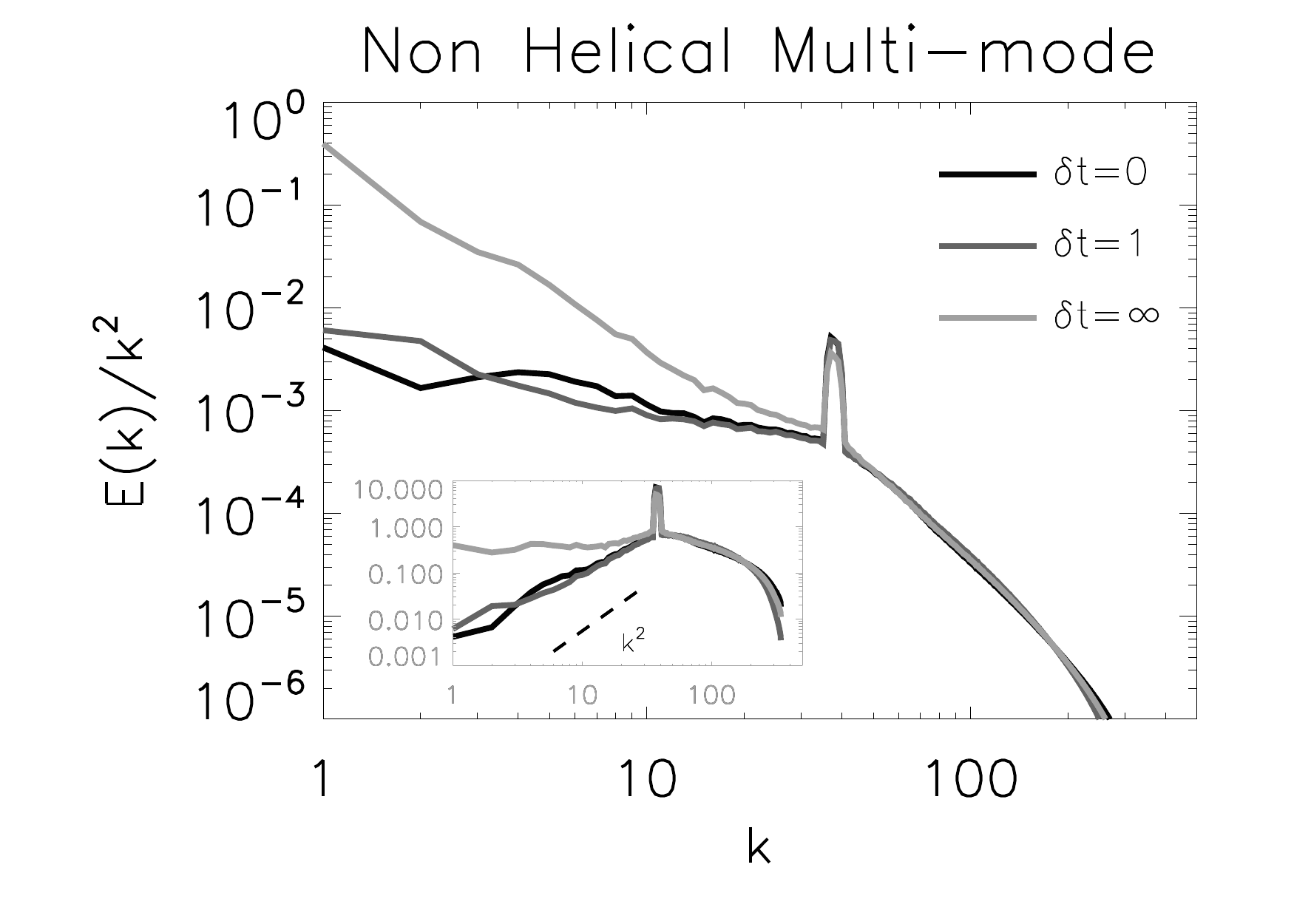}
  \includegraphics[width=0.48\textwidth]{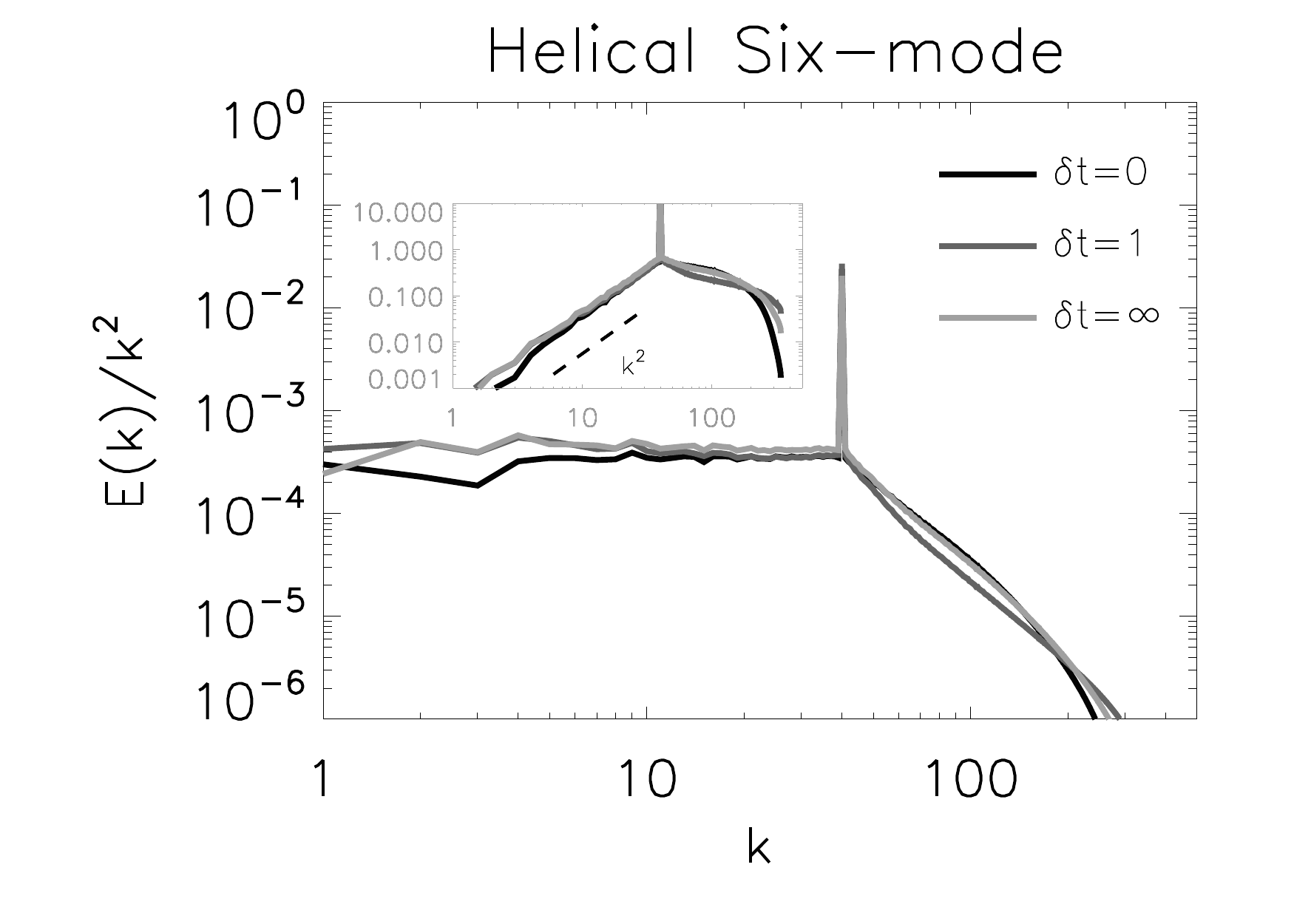}
  \includegraphics[width=0.48\textwidth]{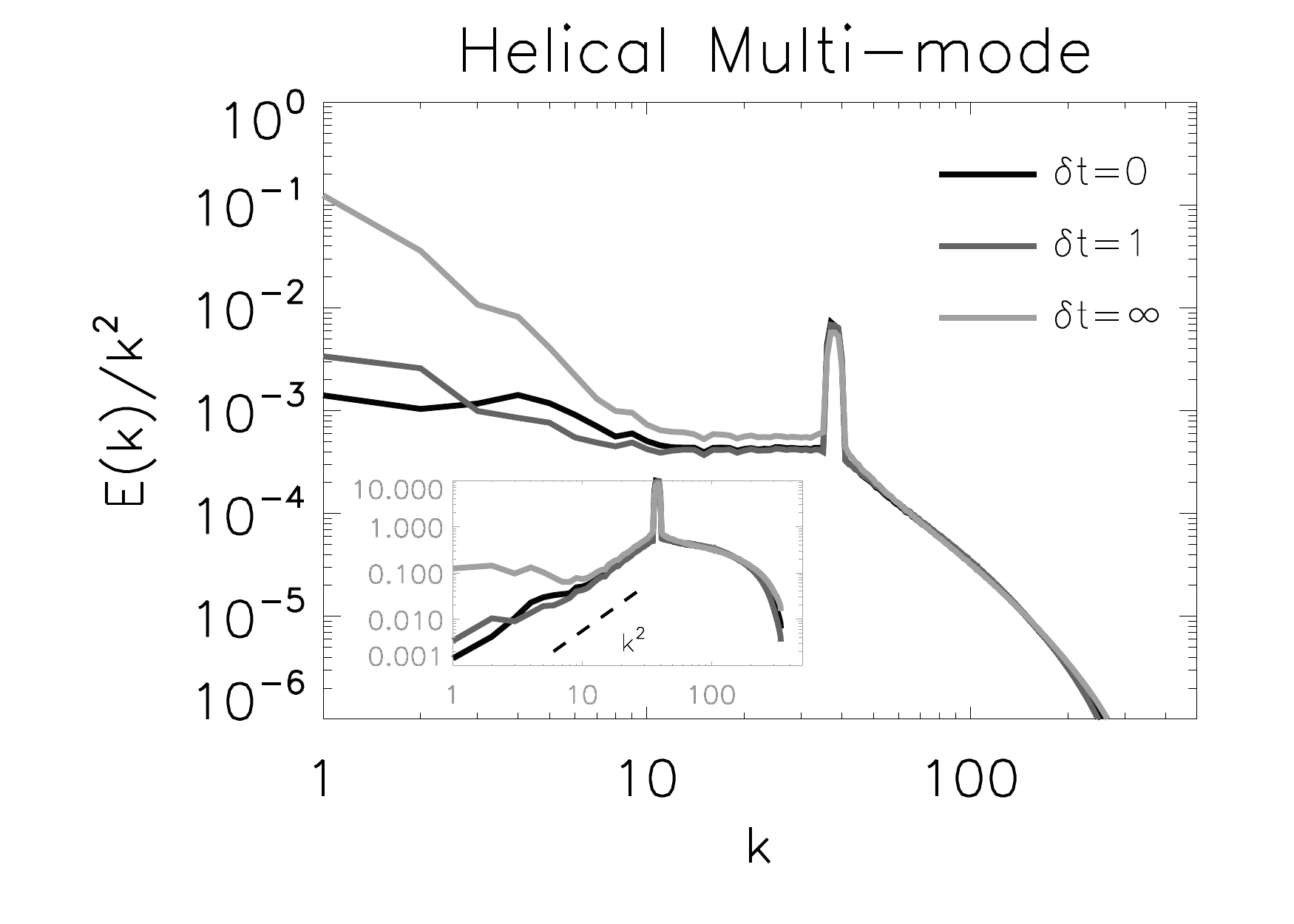}
  \caption{  Energy spectra $E(k)$ compensated by $k^{-2}$ for the 12 different runs given in table \ref{tableofruns}.
  The top/bottom panels show the spectra for the non-helical/helical flows 
  and the left/right panels show the spectra for the six-mode/multi-mode forced flows.
  The insets show the same spectra uncompensated. }
  \label{fig:spec1}
\end{figure}

The spectra are shown in the four panels of fig. \ref{fig:spec1} for the 12 different runs examined in this work. 
They are compensated by $k^{-2}$ so that a thermal spectrum $k^2$ will appear as flat.
Non-compensated spectra are plotted in the insets.
Non-helical runs are displayed on the top panels while helical runs are displayed in the bottom panels. 
Runs with six-mode forcing are displayed in the panels on the left while runs with multi-mode forcing are displayed in the panels on the right. The three different lines in each panel correspond to the three different correlation times used. The darkest line corresponds to the delta-correlated in time forcing, while the lightest gray line corresponds to the time-independent forcing. 

The differences of the energy spectra in the large scales among the different runs are striking.
Flows with six-mode forcing are very close to the thermal equilibrium spectrum $E(k)\propto k^2$.
This is most clear for the helical flows (bottom left panel of figure \ref{fig:spec1}) for which
all three cases show a clear $k^2$ scaling. We need to note here that although the forcing was fully helical
in these flows the amount of helicity that was transferred in the large scales remained minimal. For this reason
these flows also equilibrate to a thermal state with $\beta\simeq0$ (see eq. \ref{eq:EkTh}). 

Non-helical six-mode forcing flows also saturate close to the thermal equilibrium spectrum (top left panel of figure \ref{fig:spec1}).
The spectrum of the flow with the delta correlated forcing is particularly close to a $k^2$ spectrum while the flows
with finite correlation time and infinite correlation time showed a slightly smaller exponent than $2$. This was also observed
for the infinite-correlation-time non-helical forcing in \cite{cameron2017effect}. The series of peaks in the spectrum that appear for these 
flows at wavenumbers larger than the forcing are due to self-interactions between forcing modes that excite
first velocity modes with wavevectors of module $\sqrt{2} \kf$. 

The flows with a multi-mode forcing deviate from the thermal prediction (right panels of figure \ref{fig:spec1}). 
This effect is relatively weak for the flows with short-time correlated forcing but very strong for the time independent forcing where a strong peak (for the compensated spectra) appears at the largest scales of the spectrum at $k=1$. In the later case the energy concentrated in the large scales is comparable to the energy at the forced scales.  This is true both for the helical and the non-helical forcing however the deviation is stronger for the non-helical forcing. The non-compensated spectra appear almost flat in this case.
The helical runs appear to satisfy the $k^2$ law for a short range close to $k_f$ but have a strong deviation at the largest scales $k<10$.

\begin{figure}                 
  \centering
  \includegraphics[width=0.48\textwidth]{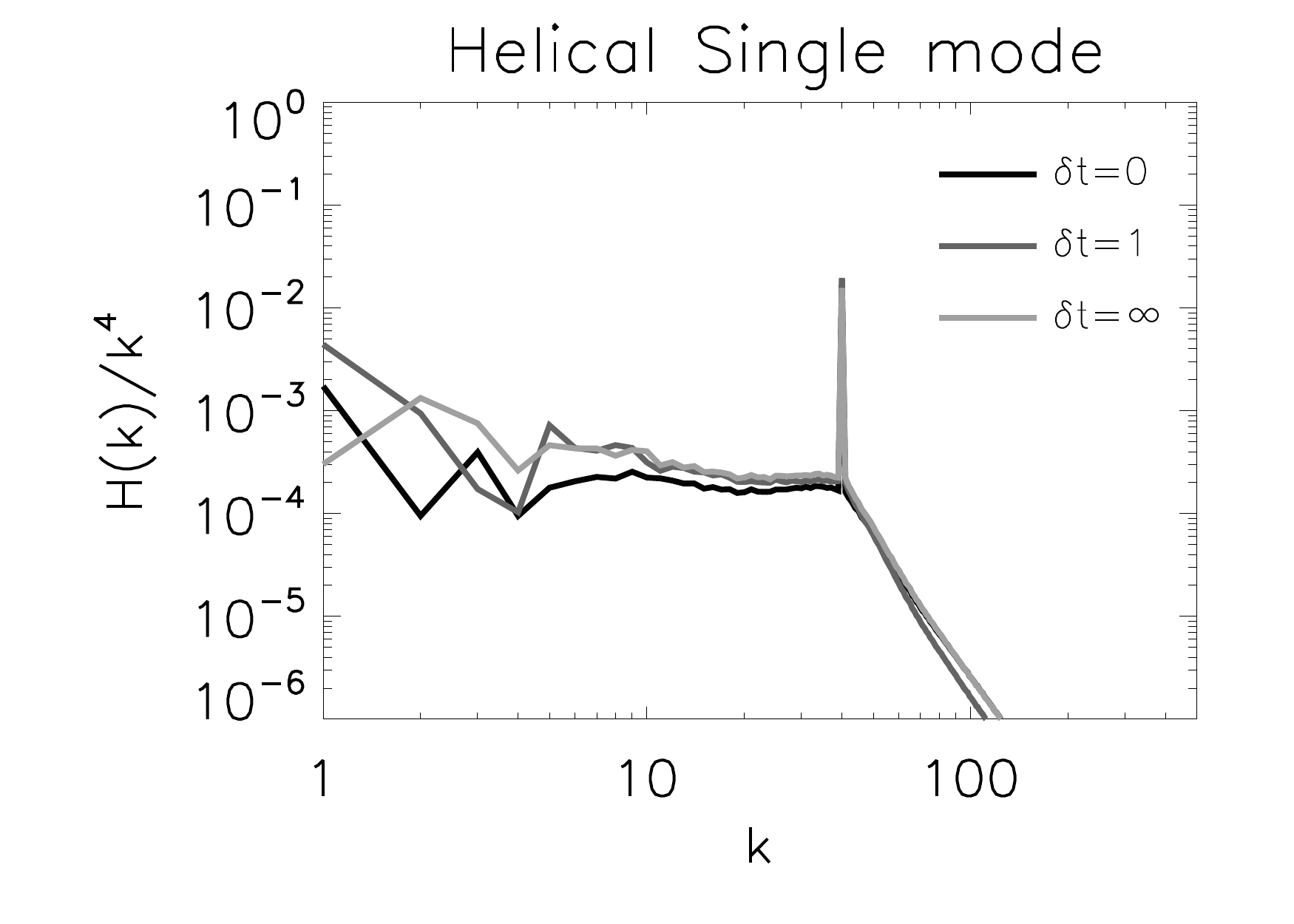}
  \includegraphics[width=0.48\textwidth]{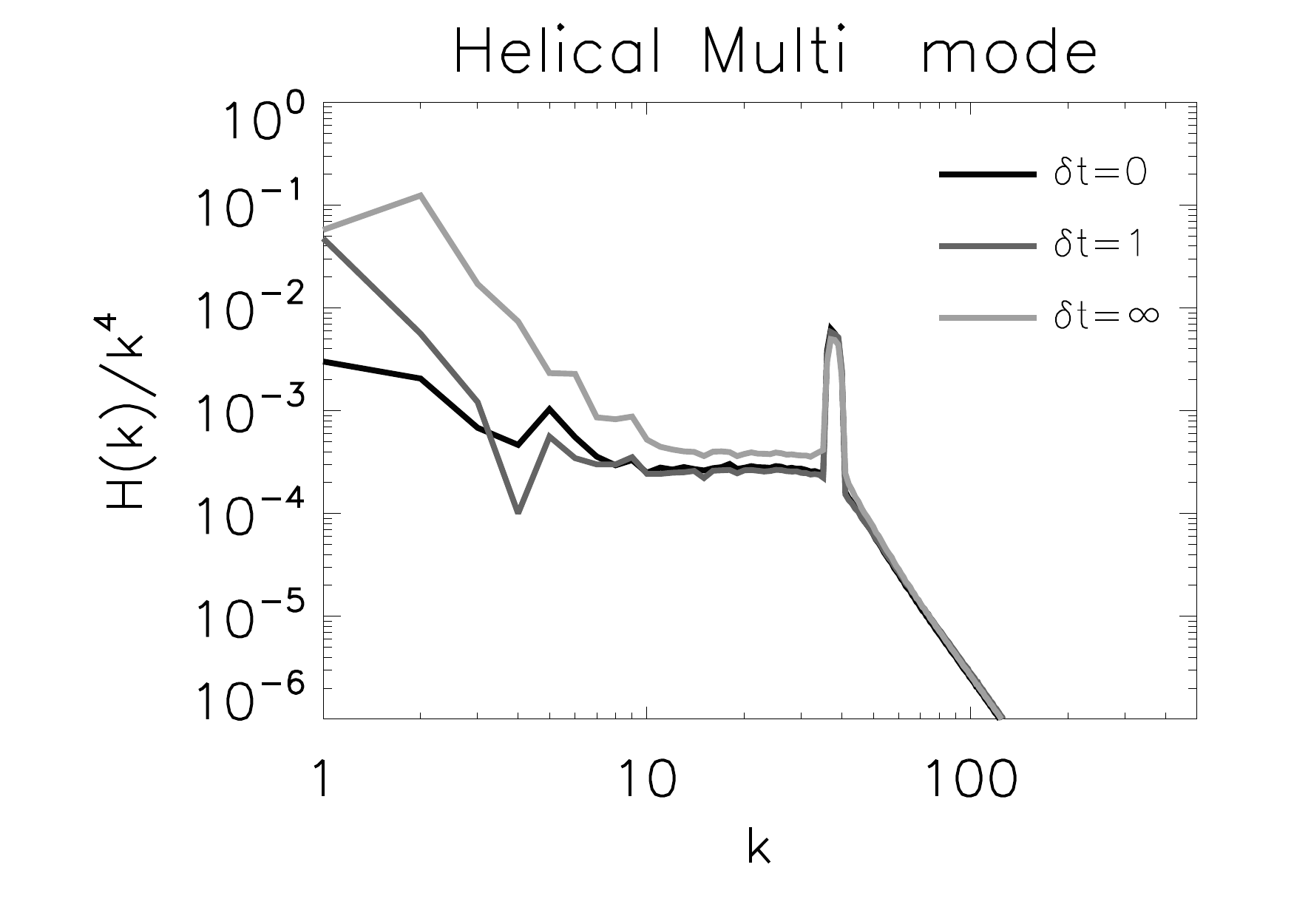}
  \caption{  Helicity spectra $H(k)$ compensated by $k^{-4}$ for the 6 different helical runs given in table \ref{tableofruns}.
  the left pale is for the six-mode forced flows and the right panel is for the multi-mode forced flows. }
  \label{fig:spec2}
\end{figure}                   

For completeness we also show in figure \ref{fig:spec2} the helicity spectra $H(k)$ compensated by $k^{-4}$.
Similar to the energy spectra the helical six-mode forced runs have a helicity spectrum close to the absolute equilibrium solution $k^4$ 
for all correlation times. However, because $H(k)$ is a sign-indefinite quantity the fluctuations are larger.
For the multi-scale forcing we see again significant deviations from the $k^4$ power-law.
This effect is again strongest for the infinite correlated forcing for which a large peak at $k=1$ and $k=2$ 
appears for the compensated spectra.

\section{Analysis}   
\label{sec:analysis} 

The presented spectra suggest that the large scale spectrum does not have a universal character and can be effected by the details of the forcing that excites the flow. In what follows we try to analyze the origin of these deviations by looking at the number of interacting triads, the amplitude of the nonlinearities at the large scales and the energy transfer properties of the flow at large scales.

\subsection{Interacting triads} 
\label{sec:analysis1} 

Perhaps the strong deviations for the multi-mode forcing could have been anticipated.  
In the case that only six modes are forced, the forced modes 
 do not directly interact with large scale modes, 
while the multi-mode forcing allows forcing mode interactions that directly couple with the large scales.

In more detail, for the six-mode forcing the wavenumbers that are forced are given by 
${\bf k}_1,{\bf k}_2\in [(\pm k_f,0,0), \,\, (0,\pm k_f,0), \,\, (0,0,\pm k_f)]$.
If we consider two velocity modes with wavenumbers ${\bf k}_1,{\bf k}_2$ that belong to these forcing modes
then they can interact with a third wavenumber $\bf q$ that forms the triad ${\bf q}+{\bf k}_1+{\bf k}_2=0$. 
The allowed $\bf q$'s for which there is non-zero energy transfer 
are ${\bf q}\in [(\pm k_f,\pm k_f,0), \,\, (\pm k_f,0,\pm k_f), \,\, (0,\pm k_f,\pm k_f)]$. The $\bf q=0$ 
as well as the ${\bf q}= [(\pm 2k_f,0,0), \,\, (0,\pm 2k_f,0), \,\, (0,0,\pm 2k_f)]$ cases although allowed by the 
triad condition $ {\bf q}+{\bf k}_1+{\bf k}_2=0$ they lead to zero nonlinearity and do not transfer any energy. 
Thus the forced velocity modes only excite directly modes with $|q| = \sqrt{2} k_f$ and therefore smaller scales than the forcing scale. 
This does not mean that large scales cannot be excited. They can be excited by interactions of the form 
${\bf k}+{\bf q}_1+{\bf q}_2=0$ where only $k\in [(\pm k_f,0,0), \,\, (0,\pm k_f,0), \,\, (0,0,\pm k_f)]$
or by the subsequent interactions between the large and turbulent scales. However since the forced velocity modes are in general stronger, 
at large scales, interactions which involve only one or no forced mode $\bf k$  tend to be weaker than those with two forced modes.

A multi-mode forcing, on the other hand, allows forcing mode interactions that directly effect the large scales. This occurs because, among the many modes that reside inside the spherical shell of external radius $k_f$ and width $\delta k_f$ (that we denote as $K_F$), 
one can find many combinations of forced velocity modes ${\bf k}_1,{\bf k}_2$ that can form a triad ${\bf q}+{\bf k}_1+{\bf k}_2=0$  provided that $|{\bf q}|\le 2k_f$.
Thus energy can be transferred  directly to a large scale modes $ q<k_f$. 
More precisely it is shown in appendix \ref{app:nmodes} that the number of triads $N_Q$ that are allowed between the modes inside a
spherical shell $Q$ of radius $q$ and width $1$ with the forcing modes at $K_F$ are given by 
\beq N_Q \simeq 16 \pi^2 k_f^2 \delta k_f   \,\, q^2  \quad \mathrm{for} \quad  q \ll \delta k_f \ll k_f \eeq and
\beq N_Q \simeq 8  \pi^2 k_f^2 \delta k_f^2 \,\, q    \quad \mathrm{for} \quad  \delta k_f \ll q \ll k_f .\eeq
The multi-mode forcing thus leads to interactions with the forcing modes that have a power-law distribution with 
the modulus of the large scale wavenumbers $q$. The index of this power-law depends on the relative 
magnitude of $q$ with $\delta k_f$.

Therefore, the density of forced modes can alter significantly the number of allowed triads that couple 
forced modes with large scale modes. The `{\it sparse}' six-mode forcing leads to no direct interactions
while the `{\it dense}' multi-scale forcing lead to a power law distribution of such triads.
It is thus not surprising that the forced mode density can effect the large scale spectrum.

\subsection{Spectrum of the nonlinearity} 
\label{sec:analysis2} 

Besides the density of these interactions their amplitude should also be examined to draw conclusions.
Energy injected at the forcing scales is redistributed among all Fourier modes of the flow 
by the nonlinear term  of the Navier-Stokes equation.
To understand how the large scales come to thermal equilibrium and the origins of the deviations from it 
we analyze the nonlinear term of the Navier-Stokes equation by looking at its spectrum and its different components.

The nonlinearity of the Navier-Stokes equation ${\cN(x)} $ is given by 
\beq
{\bf \cN(x)}=  \bu \cdot \nabla \bu +\nabla P =  \bu \cdot \nabla \bu - \nabla \Delta^{-1} \nabla \cdot ( \bu \cdot \nabla \bu) 
\eeq
where in the second equality we have written an explicit expression for the pressure 
\beq P= - \Delta^{-1} \nabla \cdot ( \bu \cdot \nabla \bu) \label{eq:Press}\eeq 
with  $\Delta^{-1}$ standing for the inverse Laplacian.   
The nonlinearity $\cN(\bx)$ is a divergence-free vector field that depends on space. We can therefore define its Fourier transform 
\beq
\tilde{\cN}(\bk) =\frac{1}{(2\pi)^3}\int  {\cN}(\bx) e^{i\bk \bx} dx^3
\eeq
and define its spectrum as 
\beq
E_{\cN} (k) = \sum_{k\le|\bk| < k+1} | \tilde{\cN}(\bk) |^2. 
\eeq
The spectrum $E_{\cN}(k)$ gives a measure of the amplitude of the non-linear term 
at the given shell of wavenumbers $k$.  

\begin{figure}                 
  \centering
  \includegraphics[width=0.48\textwidth]{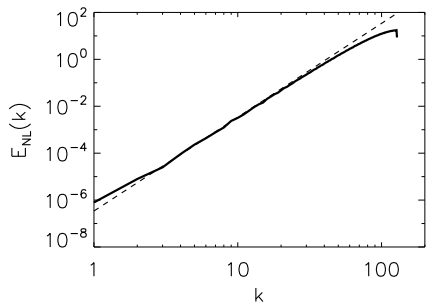}
  \caption{ The spectrum $E_{\cN}(k)$ of the nonlinearity obtained from the truncated Euler simulations
            (solid line) and compared with the theoretical prediction \ref{NLth1}.   }
  \label{fig:Nspecth1}
\end{figure}

For the truncated Euler equations where the flow reaches a thermal equilibrium state the spectrum of the nonlinearity can be calculated exactly. 
This is done in appendix \ref{app:absSpec} and leads to the prediction 
\beq
E_{\cN} (q) = \frac{ 14\, \cE^2}{5\, k_{max}^3}\,\, q^4.  \label{NLth1}
\eeq
A comparison of this result with $E_\cN(q)$ obtained from numerical simulations of flows obeying the truncated Euler eqs. \ref{TEE} is shown in figure \ref{fig:Nspecth1}. 
Similar estimates (although no-longer rigorous) can be made for the forced flows if some extra assumptions are made.
These calculations are presented in appendix \ref{app:nonlin} and lead to the predictions  
\beq                                      
E_{\cN} (q) \propto   q^4                 
\label{eq:SpecNonlin1}                    
\eeq                                      
if the energy spectrum $E(k)$ varies smoothly over distances of order $q$ or 
\beq                                      
E_{\cN} (q) \propto   q^3                 
\label{eq:SpecNonlin2}                    
\eeq                                      
if the interactions are dominated by interactions with modes in a thin spherical shell
(as for example with the forced modes in the multi-mode forcing flows) with $\delta k_f \ll q \ll k_f$.
\begin{figure}                 
  \centering
  \includegraphics[width=0.48\textwidth]{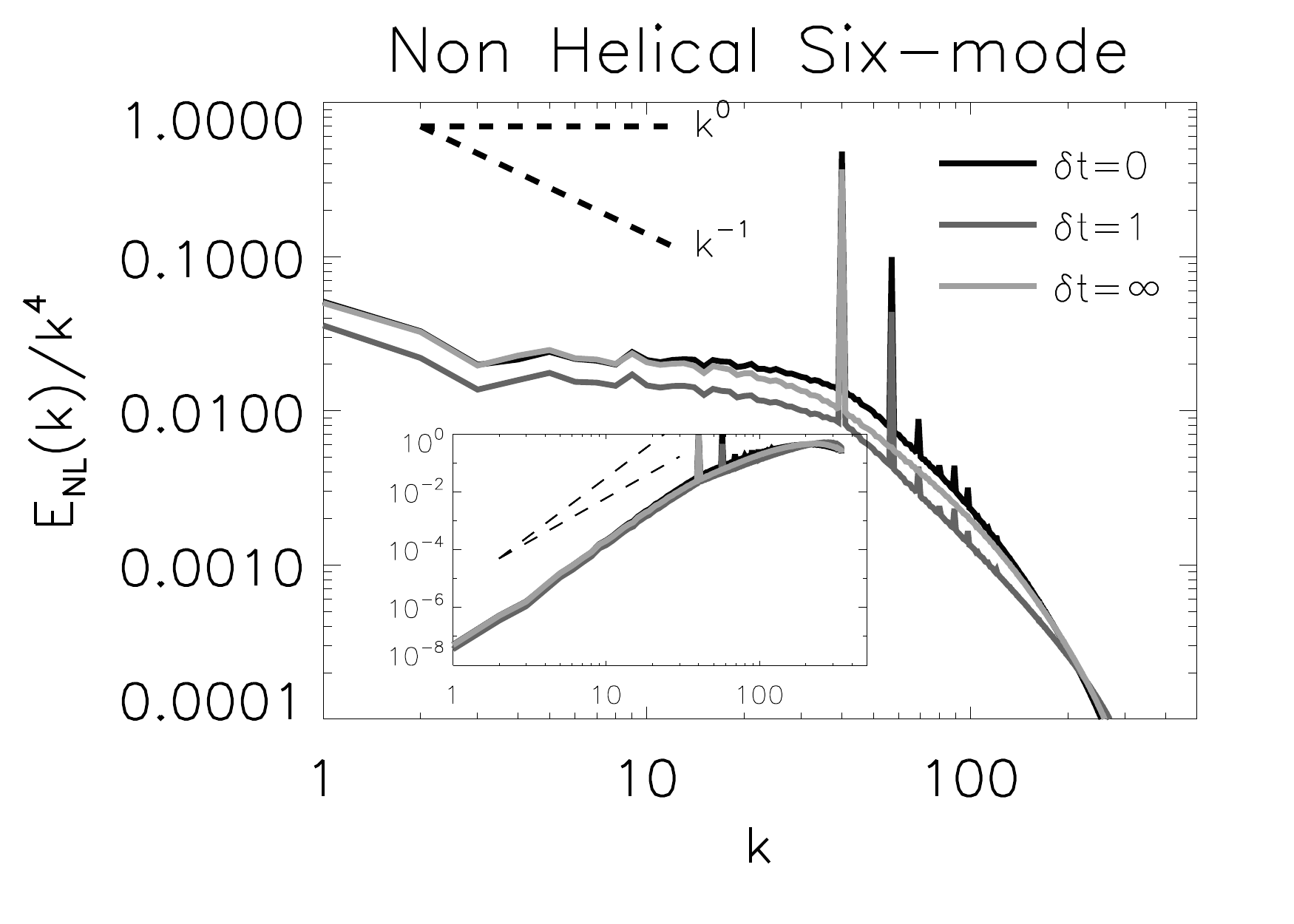}
  \includegraphics[width=0.48\textwidth]{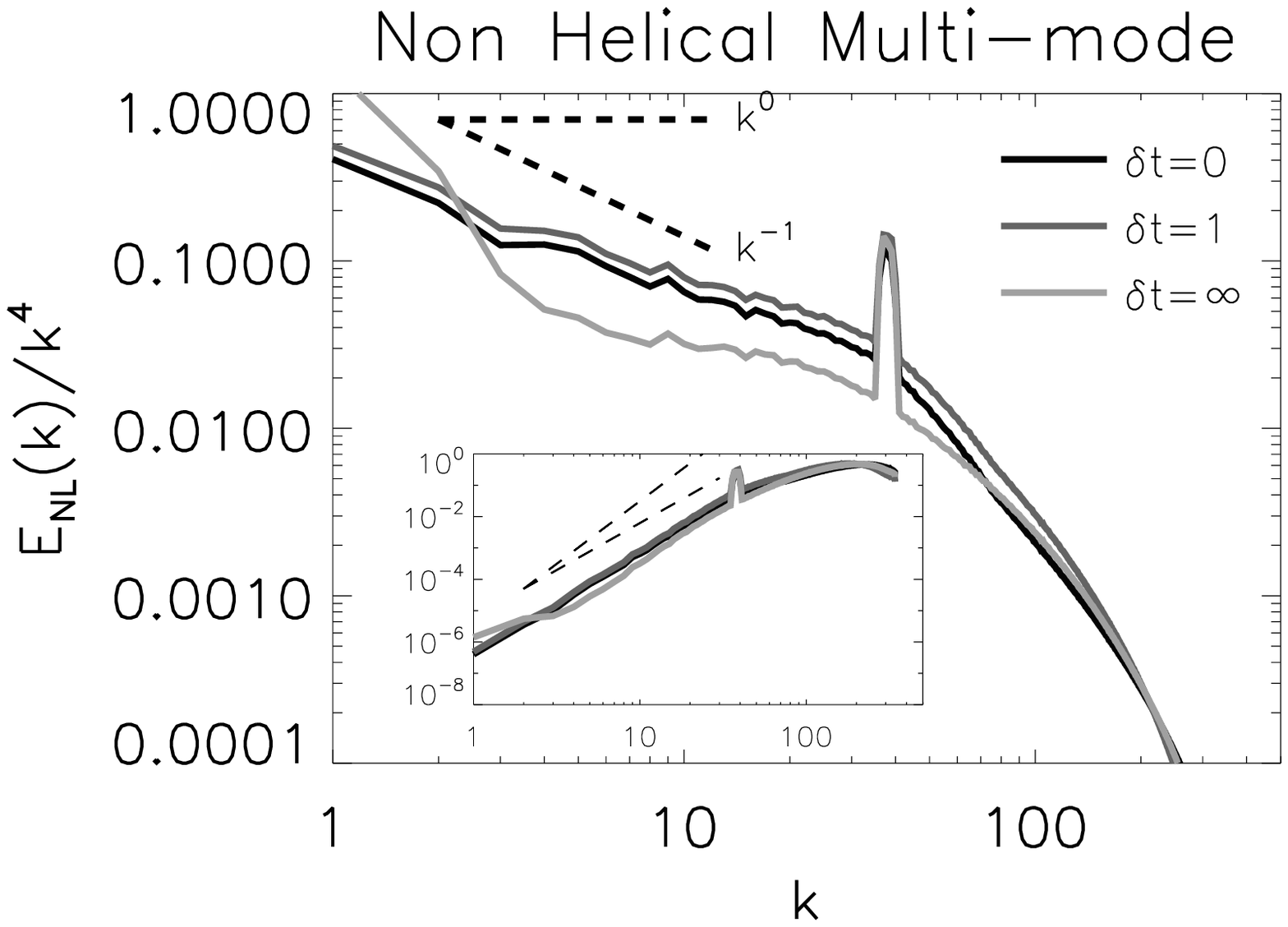}
  \includegraphics[width=0.48\textwidth]{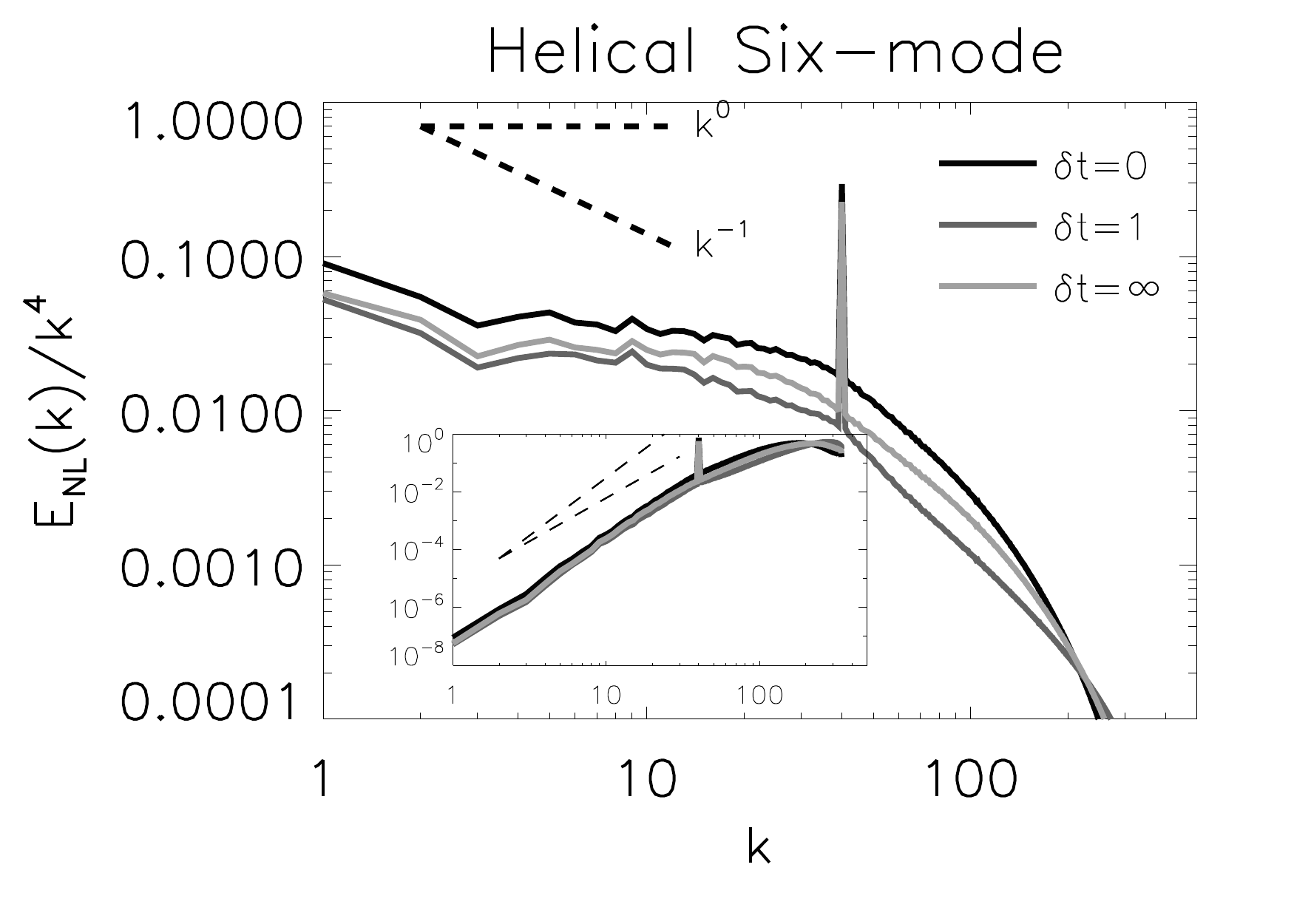}
  \includegraphics[width=0.48\textwidth]{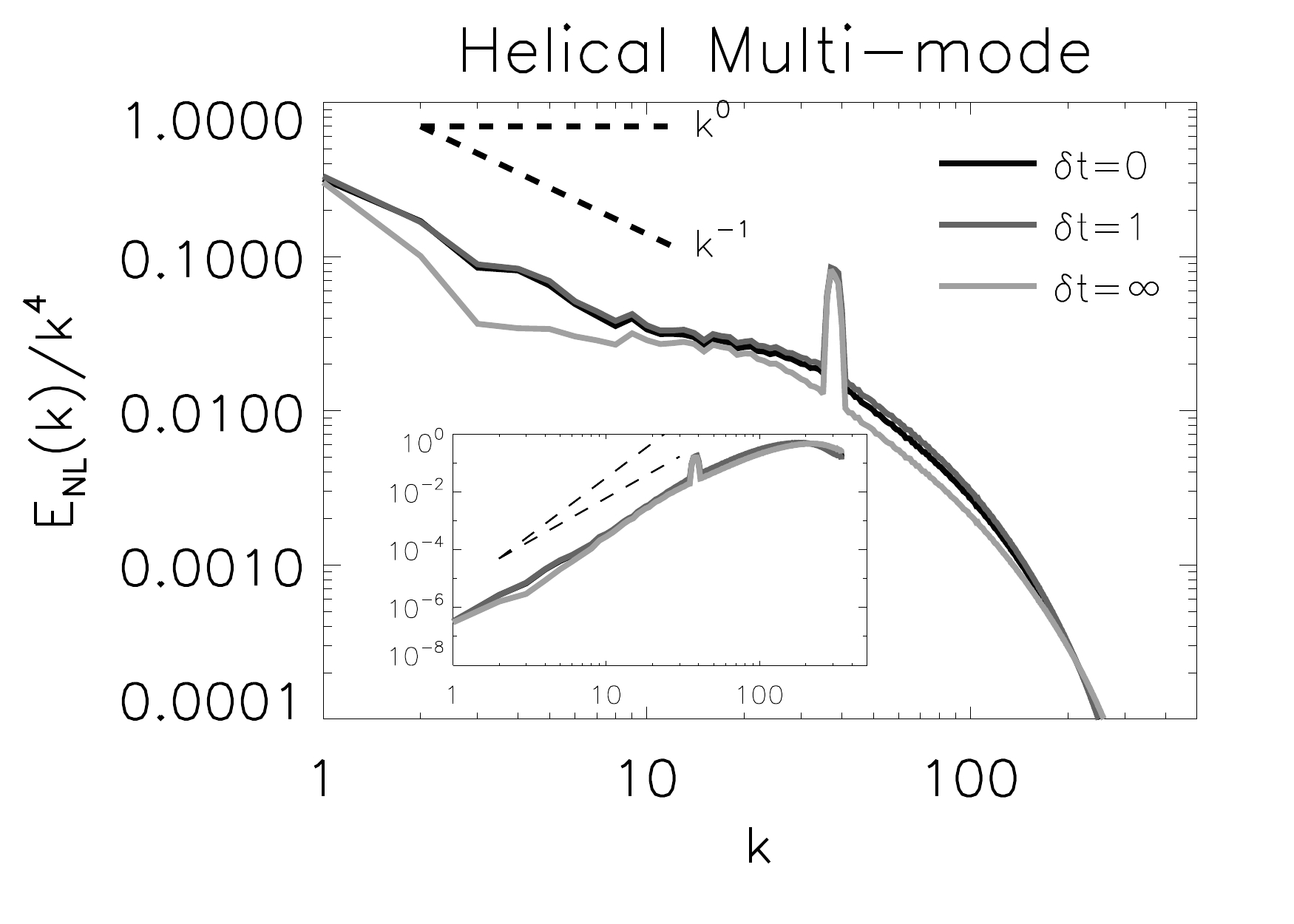}
  \caption{ The spectrum $E_{\cN}(k)$ of the nonlinearity, normalized by $k^4$. 
  The top/bottom panels are for the non-helical/helical flows 
  and the left/right panels show the spectra for the six-mode/multi-mode forced flows.  }
  \label{fig:Nspec2}
\end{figure}                   

The compensated spectra $E_{\cN}(k)/k^4$ for the twelve runs examined are shown in figure \ref{fig:Nspec2}.
The six-mode forcing simulations (for which there are no direct interactions with the forced modes at large scales)
result in a spectrum for the non-linearity close to $E_{\cN}(k) \propto k^4$. 
These are the flows that were also shown to develop energy spectra close to the ones predicted by the thermal equilibrium.
The scaling appears to be valid both for the helical and the non-helical runs although perhaps more clear for the non-helical runs.
The multi-mode forced flows (that allow direct interactions with the forced modes) resulted in spectra that are closer to $E_{N}(k) \propto k^3$. 
This suggests that the forcing modes that are restricted in a this spherical shell are important for the evolution of the large scale 
modes as they can deform the spectrum of the nonlinearity in the large scales.

An other quantity that is of interest is the spectrum of the pressure. 
\beq
E_{P} (k) = \sum_{k\le|\bk| < k+1} | \tilde{P}(\bk) |^2
\label{eq:SpecPress}
\eeq
where $\tilde{P}$ stands for the Fourier transform of the pressure field obtained by \ref{eq:Press}.
The great advantage of the pressure spectum as opposed to the spectrum of the nonlinearity 
is that pressure can be  measured in the laboratory and thus this prediction can also be tested in experiments.

\begin{figure}                 
  \centering
  \includegraphics[width=0.48\textwidth]{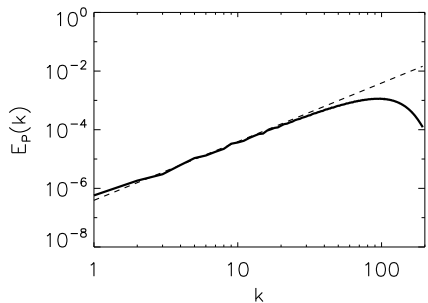}
  \caption{ The presure spectrum $E_{P}(k)$ obtained from the truncated Euler simulations
            (solid line) and compared with the theoretical prediction \ref{NLth2}.   }
  \label{fig:Nspecth2}
\end{figure}

The same arguments that were made in appendix \ref{app:nonlin} for the nonlinearity can be made for the gradient of the pressure $\nabla P$.
We can therefore also make a prediction for the pressure spectrum.  The pressure spectrum then should scale like $E_P(k)\propto E_{N}(k)k^{-2}$ and therefore 
it is expected to scale like $k^{2}$ for the six-mode forced flows and like $k^{1}$ for the multi-mode forced flows. For the truncated Euler flows at thermal equilibrium, 
the pressure can be evaluated exactly (see appendix \ref{app:absSpec}) and it is given by
\beq
E_{P} (q) = \frac{ 16 \cE^2}{5 k_{max}^3} k^2.  \label{NLth2}
\eeq
A comparison of this result with $E_P(q)$ obtained from numerical simulations is shown in figure \ref{fig:Nspecth2} showing excellent agreement. 
\begin{figure}                 
  \centering
  \includegraphics[width=0.48\textwidth]{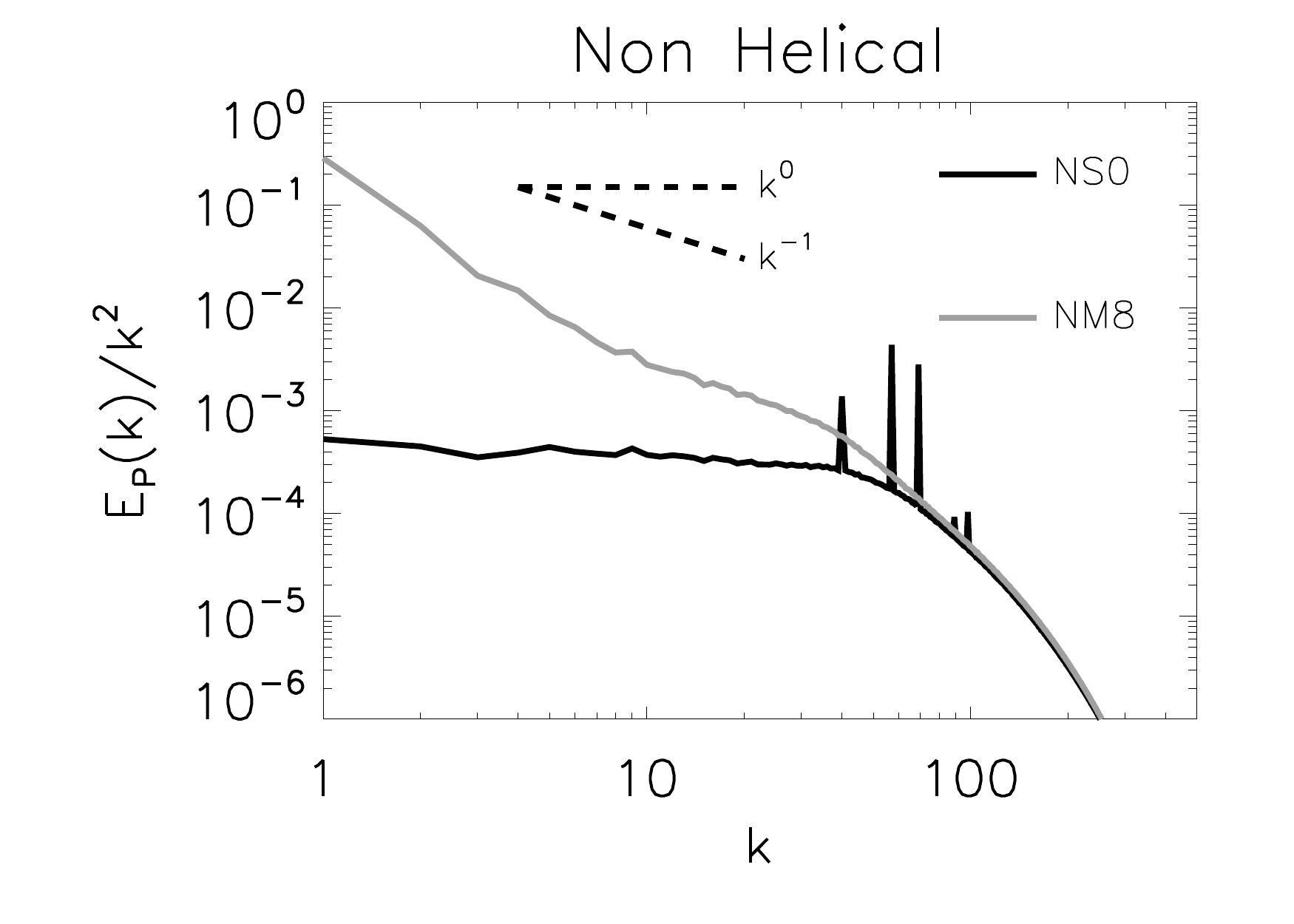}
  \includegraphics[width=0.48\textwidth]{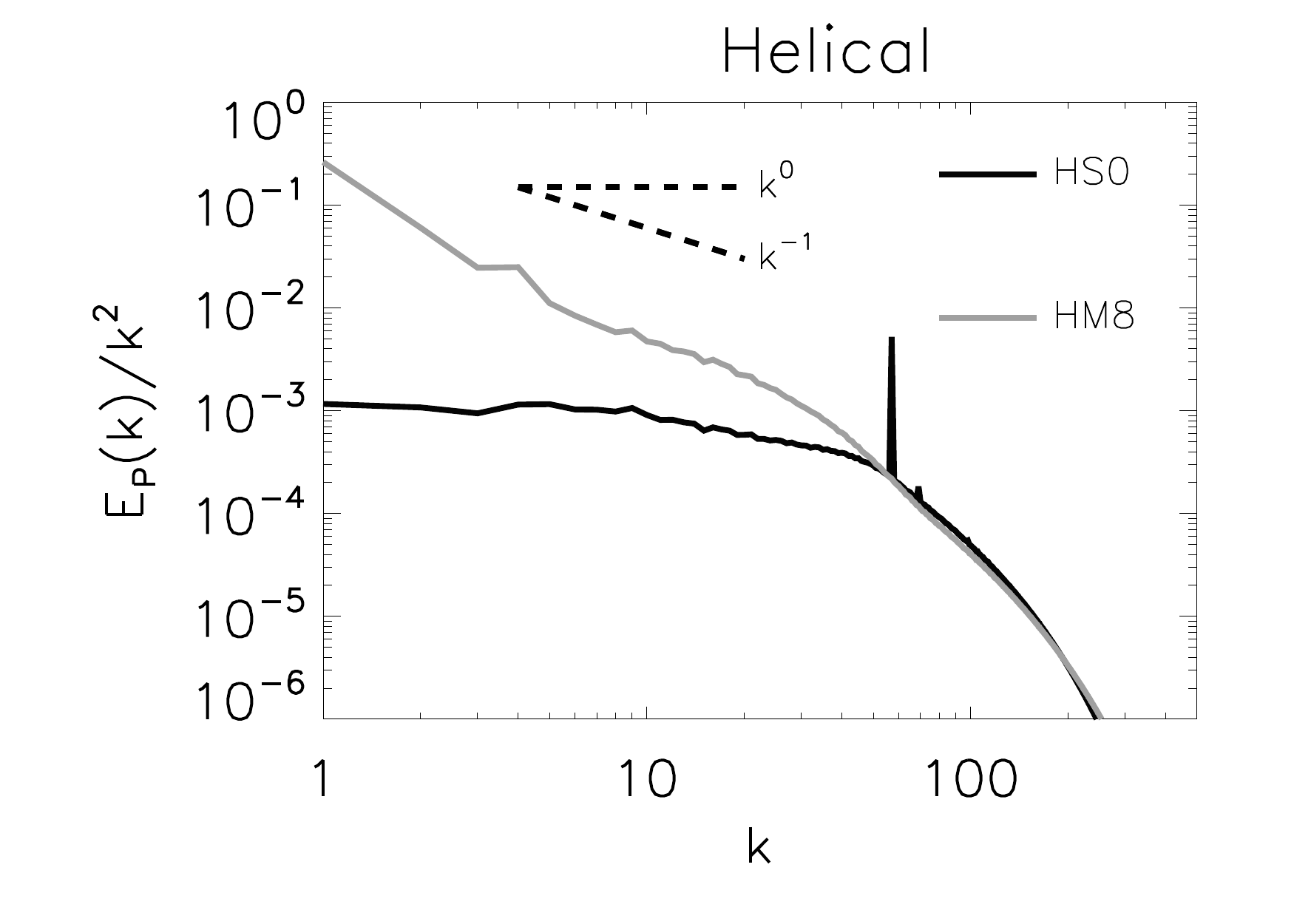}
  \caption{ Spectrum of the pressure field $E_{P}(k)$ for the non-helical runs NS0,NM8 (left panel)
  and for the helical runs HS0,HN8 (right panel). The dark lines correspond to the six-mode forced runs with $\delta t=0$ (NS0,HS0) 
  and the light gray lines correspond to the multi-mode forced runs with $\delta t=\infty$.  }
  \label{fig:Pspec}
\end{figure}                   
For the forced runs, the pressure spectrum is shown in figure \ref{fig:Pspec} the for cases $NS0,HS0,NM8,HM8$
normalized by $k^2$. The spectra are compatible with the afore mentioned predictions with the six-mode forced runs
being close to a $k^2$ spectrum and the multi-mode forced runs closer to a $k^1$ spectrum. 


\subsection{ The effect of different scales}
\label{sec:analysis3} 

To further illuminate the role of interactions among different scales  and understand 
which ones lead the large scales to reach an equilibrium we decompose the velocity field in to three components 
\beq
\bu = \bul + \buf + \but
\eeq
the large scale flow $\bul$, the forcing scale flow $\buf$, and the turbulent scales flow $\but$. 
The three flows are defined as:
\beq
\bul = \sum_{             |\bk| <   k_f-\delta k_f} \tilde{\bu}_\bk e^{i\bk\bx}, \qquad
\buf = \sum_{k_f-\delta k_f \le |\bk| \le k_f     } \tilde{\bu}_\bk e^{i\bk\bx}, \qquad
\but = \sum_{k_f      <   |\bk|             } \tilde{\bu}_\bk e^{i\bk\bx} \label{uscl}
\eeq
Given this decomposition the nonlinearity can be written as the sum of 9 terms explicitly given by 
\beq
{\bf \cN(x)}=  \sum_{I,J} \mathbb{P}[\bu_{_I} \cdot \nabla \bu_{_J} ]
\eeq
where $\mathbb{P}$ stands for the projector operator to incompressible flows,
$I,J$ stand for the indexes $L,F,T$ and the sum is over all possible permutations.
If we symmetrize over the change of two indexes we obtain the following 6 nonlinear terms: 
\begin{eqnarray}
\cN_{LL} = \mathbb{P} [\bu_{_L} \cdot \nabla \bu_{_L} ],\quad 
\cN_{FF} = \mathbb{P} [\bu_{_F} \cdot \nabla \bu_{_F} ],\quad
\cN_{TT} = \mathbb{P} [\bu_{_T} \cdot \nabla \bu_{_T} ]   \nonumber   \\
\cN_{LF} = \mathbb{P} [\bu_{_L} \cdot \nabla \bu_{_F} + \bu_{_F} \cdot \nabla \bu_{_L} ],\quad 
\cN_{LT} = \mathbb{P} [\bu_{_L} \cdot \nabla \bu_{_T} + \bu_{_T} \cdot \nabla \bu_{_L} ], \label{eq:nl6} \\  \mathrm{and} \quad 
\cN_{TF} = \mathbb{P} [\bu_{_T} \cdot \nabla \bu_{_F} + \bu_{_F} \cdot \nabla \bu_{_T} ]. \qquad  \qquad \qquad \qquad\nonumber
\end{eqnarray}
The first three represent the nonlinearity due to self interactions of the large scales, forcing scales and turbulent scales
while the remaining three represent cross-interactions. The sum of all six terms recovers the nonlinearity $\cN(x)$.

\begin{figure}
  \centering
  \includegraphics[width=0.48\textwidth]{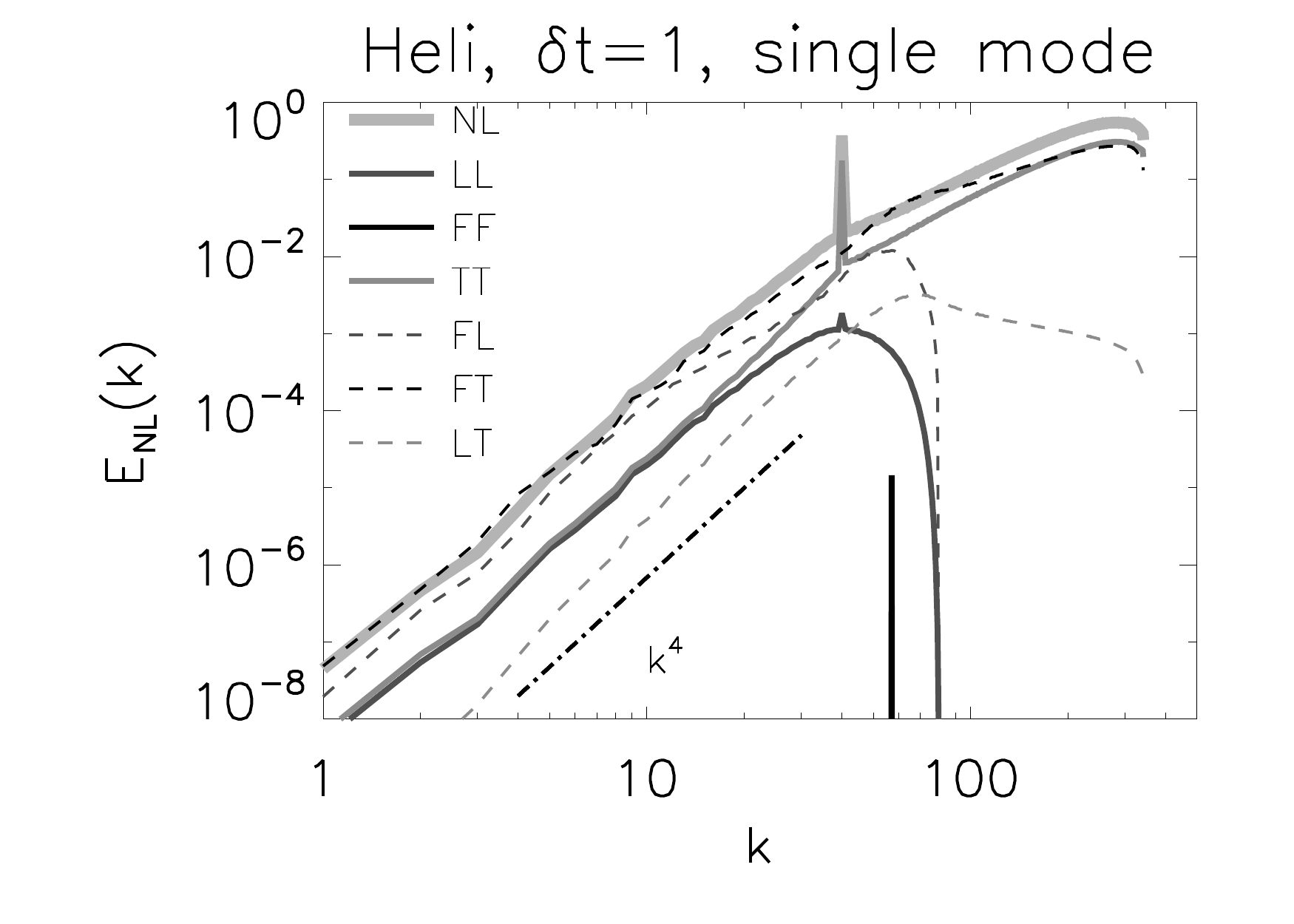}
  \includegraphics[width=0.48\textwidth]{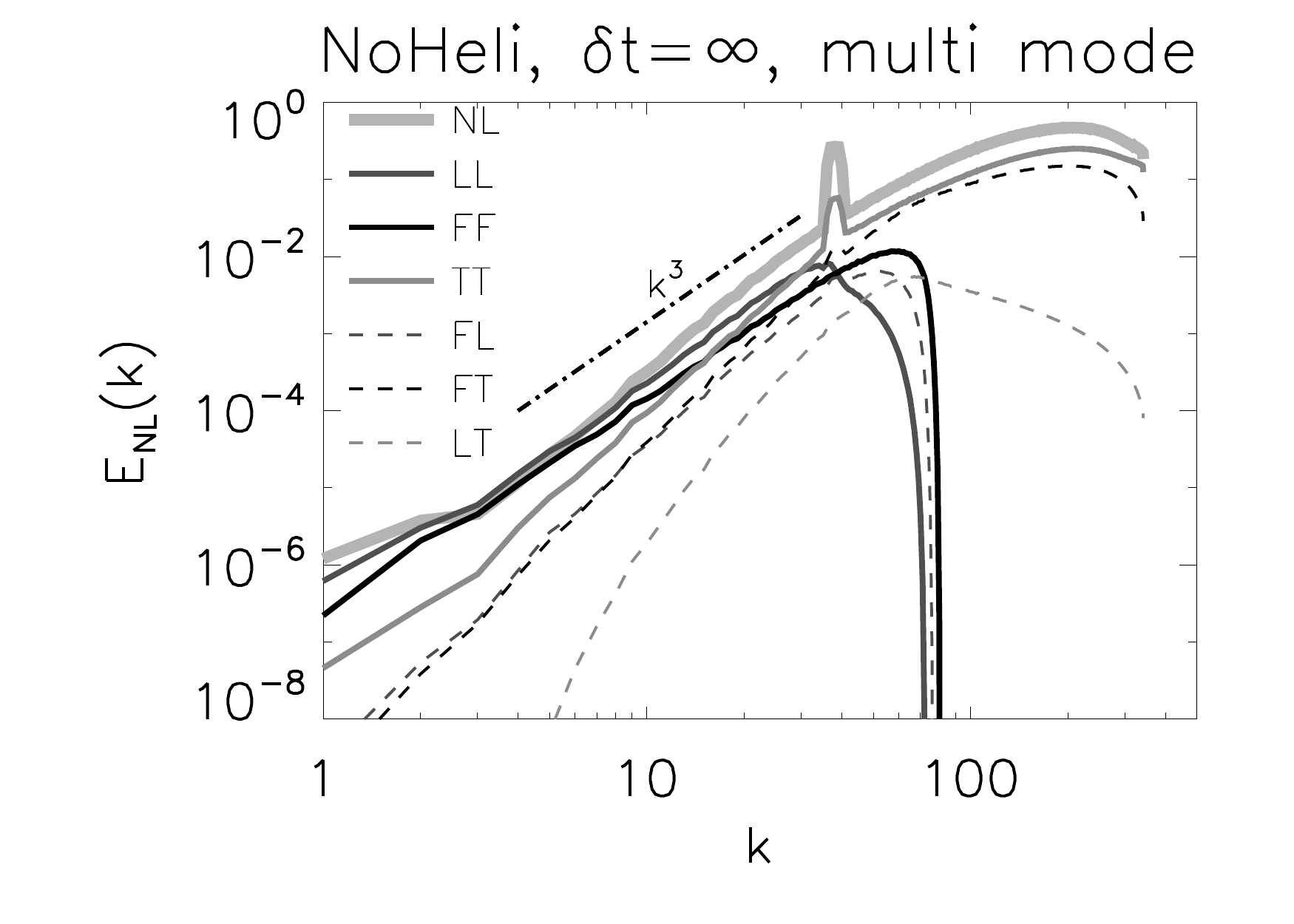}
  \caption{ The spectra for the six different nonlinear terms given in \ref{eq:nl6} for the flows  
            HS1 (helical, six-mode forced, $\delta t=1$ left panel) and NM8 (non-helical, multi-mode forced, $\delta t=\infty$, right panel). 
            The index $NL$ indicates the spectrum of the full nonlinearity, while the remaining indexes indicate the spectra of the nonlinearities as given in eq. \ref{eq:nl6}. }         
  \label{fig:nl6}
\end{figure}

As before we can calculate the spectra for each of the six nonlinear terms.
We have done this for the six-mode forced run HS1 (helical, six-mode, $\delta t=1$) that is a characteristic example that displays a thermal equilibrium spectrum and
and NM8 (non-helical, multi-mode, $\delta t=\infty$) that is a characteristic example of a flow that deviates from this spectrum.
The spectra of the six nonlinear terms for these flows are shown in figure \ref{fig:nl6}. 
In both cases the small scales are dominated by TT interactions as expected.
At large scales however differences can be seen.
For the flow HS1 the FF interactions are absent. They only appear as a single peak at $k=\sqrt{2}\,k_f$.
The most dominant interactions in the large scales appear to be the FL and FT interactions followed by LL and TT interactions.
All terms appear to display a $k^4$ power-law at large scales. In this case therefore it appears that all scales play a role for 
the formation of the large scale spectrum.

On the other hand for the flow NM8 the most dominant interactions are with the forcing modes FF and the self-interactions of the large scales LL. 
The FF and LL interactions appear to follow a clear $k^3$ power-law at large scales while the other terms appear to have a slightly steeper behavior.
This implies that in this case the large scale spectrum is determined by a balance between the forcing scales and the large scale self-interactions.
Furthermore since these interactions follow a less steep scaling from the rest they become more dominant as smaller wavenumbers are reached.

This analysis therefore demonstrates that for the  multi-mode forcing it is the interactions with forcing modes that dominate and 
the deviations from the thermal equilibrium spectrum can be attributed to them. The self-interactions of the large scales,  
which are of similar amplitude, try to restore the equilibrium as we will demonstrate in the next subsections.

\subsection{Energy Fluxes}             
\label{sec:analysis4} 

The spectra examined in the previous section give some information regarding 
the amplitude of different interactions in the large scales. However they do not 
provide direct information on how much energy is added or extracted from these scales 
due to those particular interactions. The rate of exchange of energy can be extracted by 
looking at the flux of energy.

The flux of energy through a spherical shell in Fourier space of radius $k$ is defined as
\beq
\Pi(k) = \langle \bu^{<k} \cdot \bf \cN \rangle 
\label{eq:Pi}
\eeq
where $\bu^{<k}$ is the velocity field filtered so that only Fourier modes of wavenumbers smaller than $k$ are retained.
For 3D  high $Re$ turbulent flows the time averaged flux  $\Pi(k)$ is zero for wavenumbers smaller than the forcing,
it is equal to the energy injection rate in the inertial scales and drops back to zero at the dissipative
scales. The fact that $\Pi(k)$ is zero at large scales expresses that there is no mean transfer of energy
to the large scales. This is true for all the simulated flows, and thus $\Pi(k)$ alone can not provide 
information for the exchange of between different scales. 

Some insight however can be gained if we look separately the role played by different scales in cascading the energy. 
To that end we define  the partial fluxes 
\[
\Pi_{L}(k)= \langle {\bf u}^{<k} \cdot (\bul \cdot \nabla {\bf u} ) \rangle, \quad
\Pi_{F}(k)= \langle {\bf u}^{<k} \cdot (\buf \cdot \nabla {\bf u} ) \rangle\quad  \mathrm{and} \quad  \] \beq
\Pi_{T}(k)= \langle {\bf u}^{<k} \cdot (\but \cdot \nabla {\bf u} ) \rangle, \label{eq:pi3}
\eeq
where $\bul,\buf$ and $\but$ are given by \ref{uscl}. 
Then $\Pi_{L}(k)$ can be interpreted as the flux of energy due to interactions with the large scales,
     $\Pi_{F}(k)$ can be interpreted as the flux of energy due to interactions with the forced scales and
     $\Pi_{T}(k)$ can be interpreted as the flux of energy due to interactions with the turbulent scales.
Adding the three recovers the total flux $ \Pi(k) = \Pi_{L}(k) + \Pi_{F}(k) + \Pi_{T}(k)$. 

\begin{figure}
  \centering
  \includegraphics[width=0.48\textwidth]{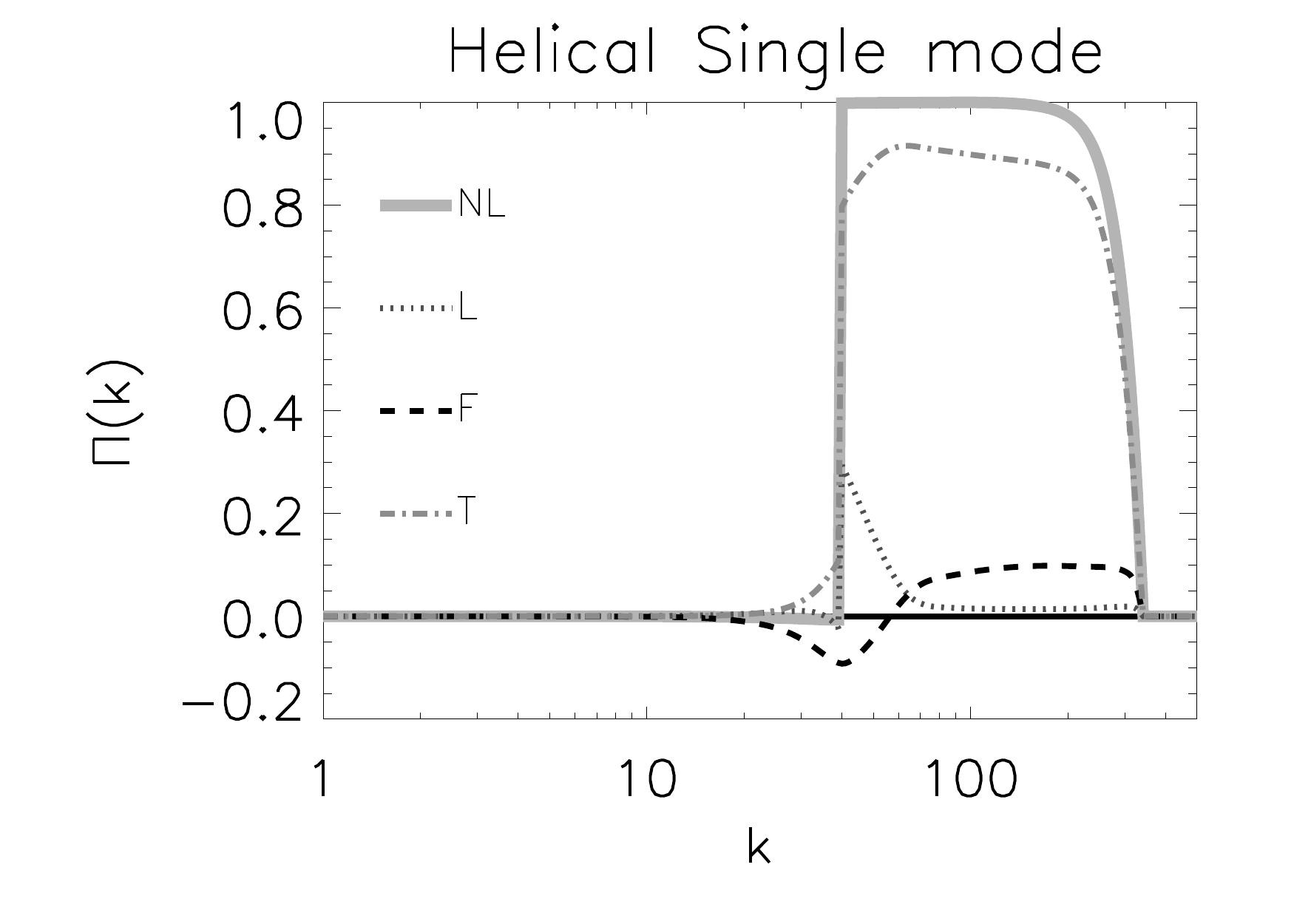}
  \includegraphics[width=0.48\textwidth]{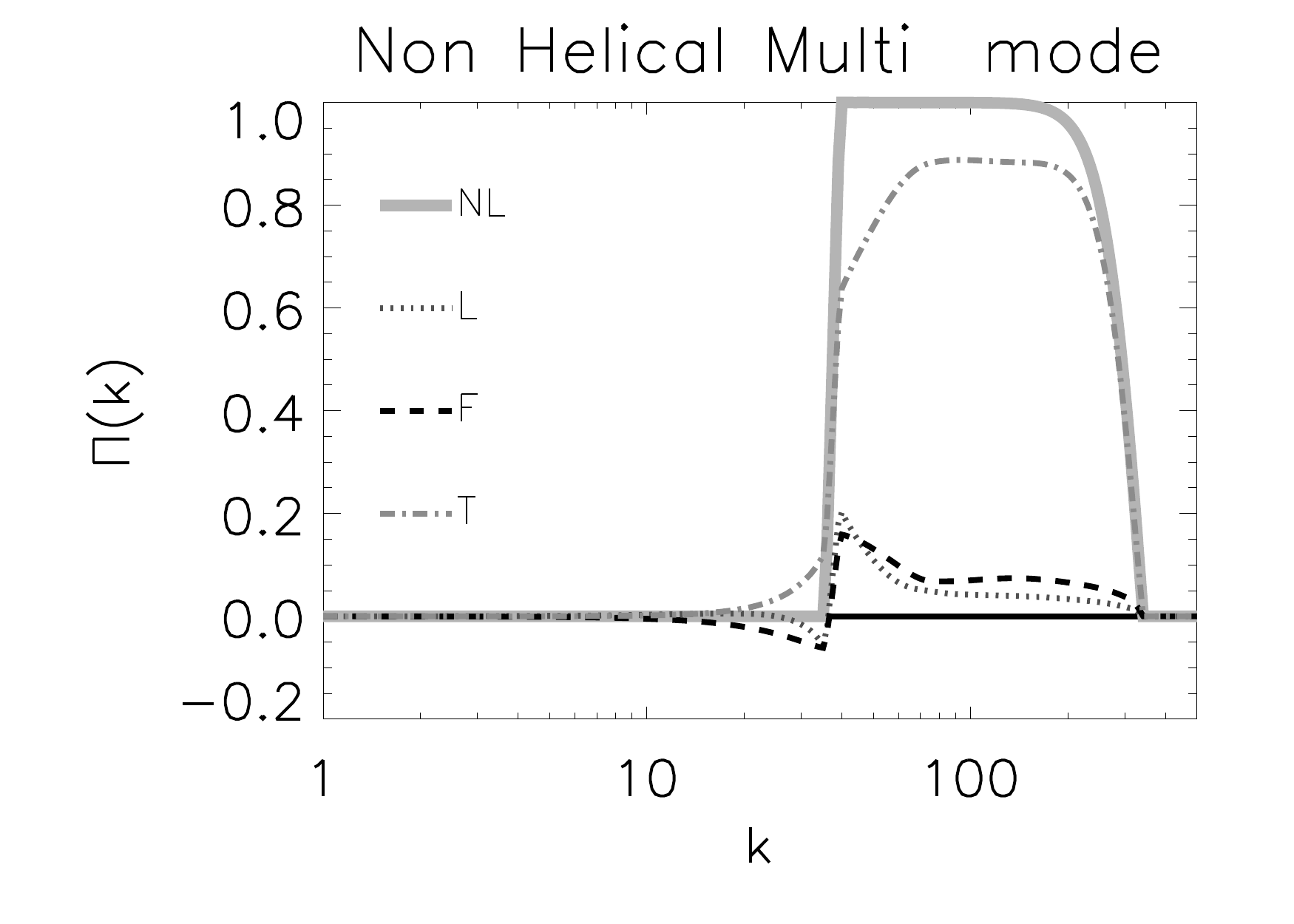}
  \includegraphics[width=0.48\textwidth]{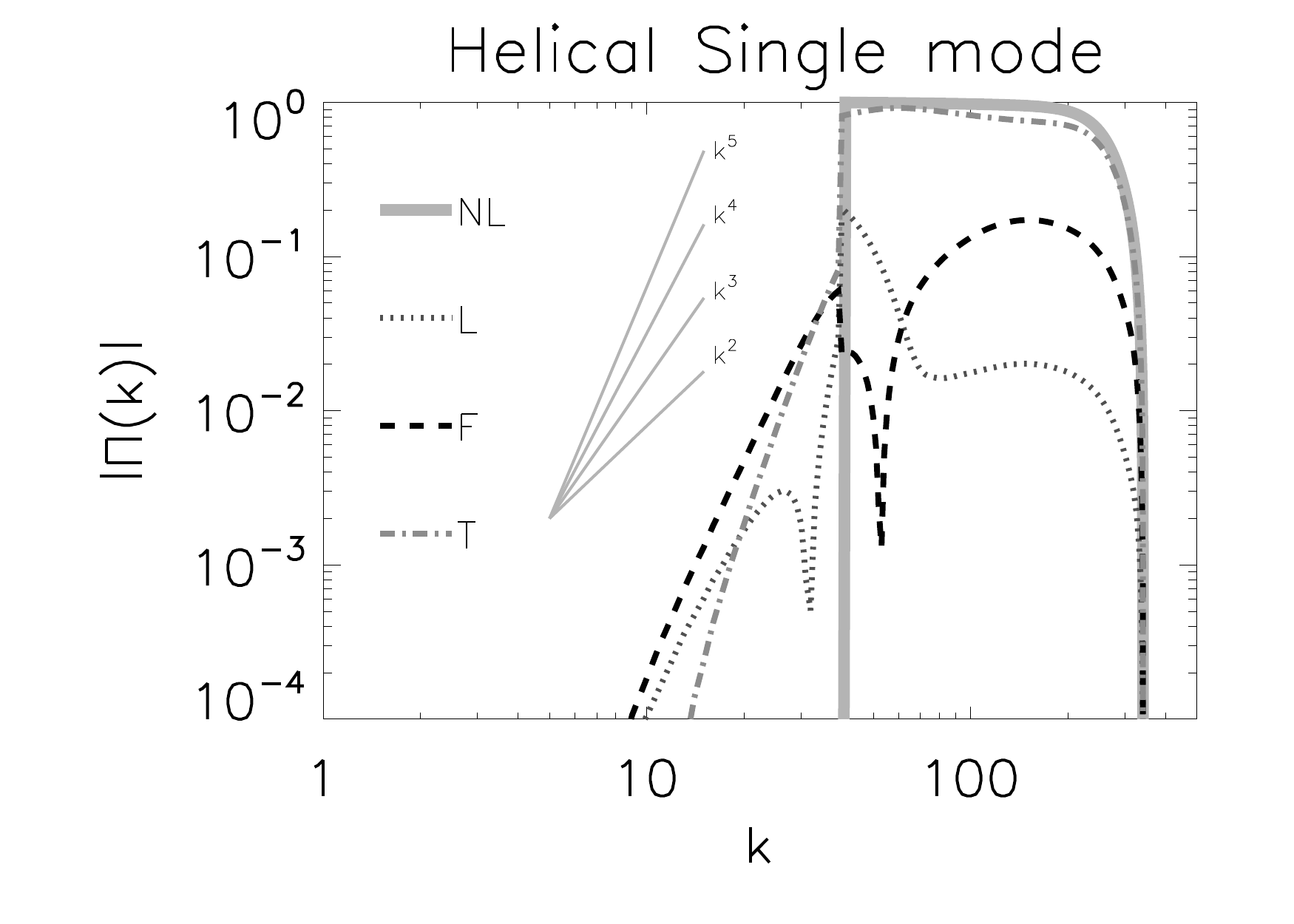}
  \includegraphics[width=0.48\textwidth]{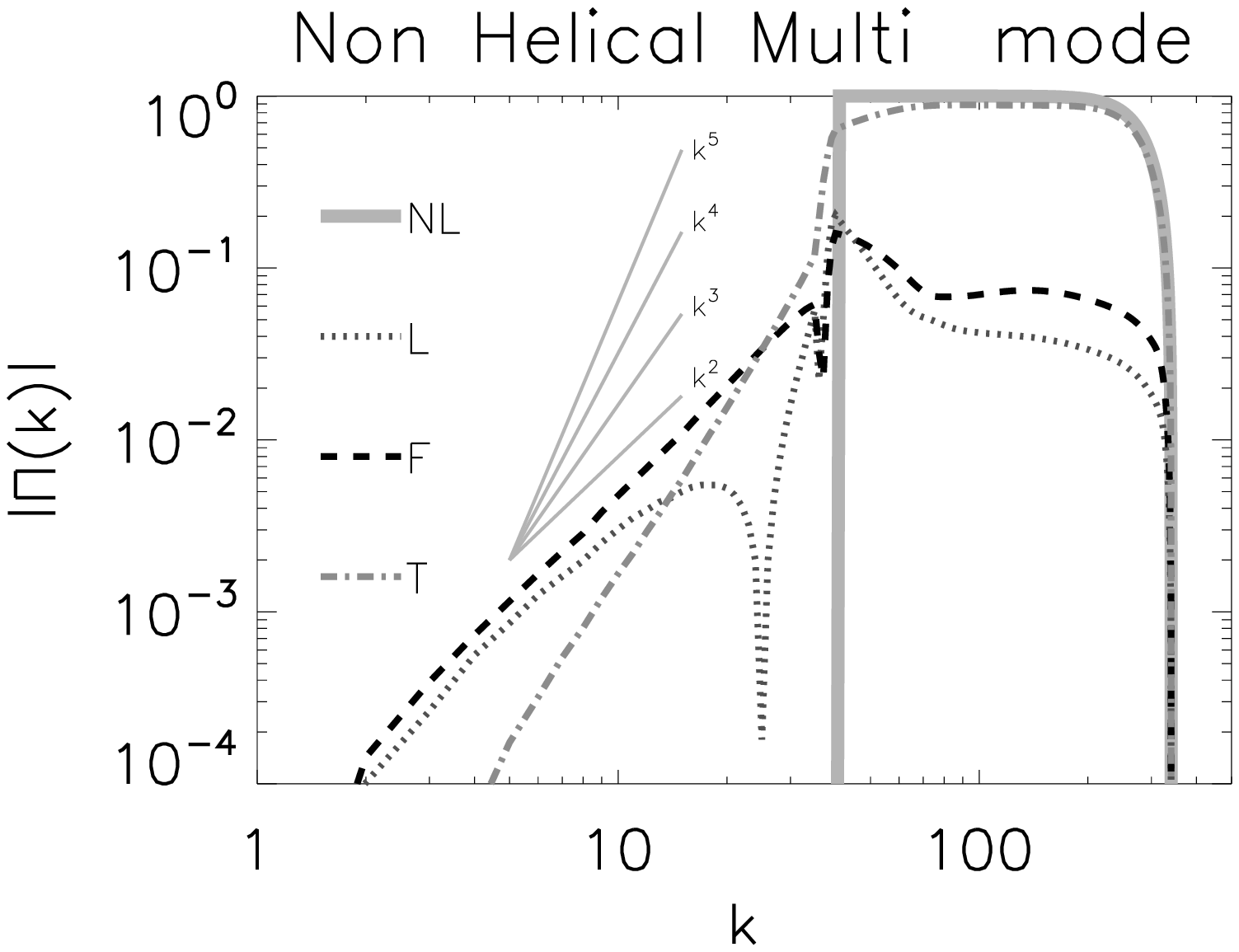}
  \caption{ The decomposed energy fluxes given $\Pi_{L}(k),\Pi_{F}(k),\Pi_{T}(k)$ in \ref{eq:pi3} along with the 
  total energy flux $\Pi(k)$ (marked by NL) for the flow HS0 left panel and the fow NM8 in the right panel. 
  Top panels are in lin-log scale while bottom panes show the absolute value in log-log scale. }
  \label{fig:flux1}
\end{figure}

The three fluxes along with the total flux for the flows HS1 and NM8  are plotted in figure \ref{fig:flux1} in linear-logarithmic scale (top panel)
and their absolute values in logarithmic scale (bottom panels). 
In the small scales all fluxes are positive indicating a forward cascade of energy to the small scales, with the interactions of the turbulent scales dominating. 
This is a well known result that has been investigated in detail for high Reynolds number flows in \cite{mininni2008nonlocal,eyink2009localness1,aluie2009localness}.
In the large scales however the three fluxes play different roles. In particular,  $\Pi_{T}(k)$ is positive indicating that interactions with the turbulent scales transfer energy to smaller scales, while $\Pi_{F}(k)$ is negative indicating that interactions with the forcing scales transfer energy to large scales. The flux due to interactions with the large scales $\Pi_{L}(k)$ is positive for small enough $k$ while it changes sign at a wave number close to $k_f$. This implies that the equilibrium at large scales is achieved by the forcing scales transferring energy to the large scales, while interactions with the turbulent scales and the large scales try to remove the excess of energy by transferring it back to the small scales.

This equilibration is best seen in the logarithmic plot of the fluxes where the two processes can be compared.
For very small $k$ the inverse transfer of energy due to interactions with the forced scales is balanced by the 
flux due to large scale interactions while as the forcing scale is approached the transfer due to the turbulent
scales becomes more dominant. The partial fluxes appear to display a power-law behavior that depends on the type of forcing.
For the six-mode forcing a steep power-law is observed that is close to $k^5$ or $k^4$.  For the multi-mode forcing
where deviations from the thermal equilibrium spectrum are observed a much less steep power-law closer to $k^2$ is observed.

The picture that arises from these results is that the forcing scales disrupt the thermal equilibrium solution 
by transferring energy to the large scales and that local large scale self interactions self-adjust to bring this
energy back to the small scales. If the effect of the forced scales is weak the adjustment of the large scales does not
disrupt the thermal equilibrium solutions while for multi-mode forcing (that is more effective at injecting energy to the large scales) 
the dynamics of the large scales need to change significantly to re-compensate for this excess input of energy.

\subsection{Helical Decomposition}             
\label{sec:analysis5} 

\begin{figure}
  \centering
  \includegraphics[width=0.48\textwidth]{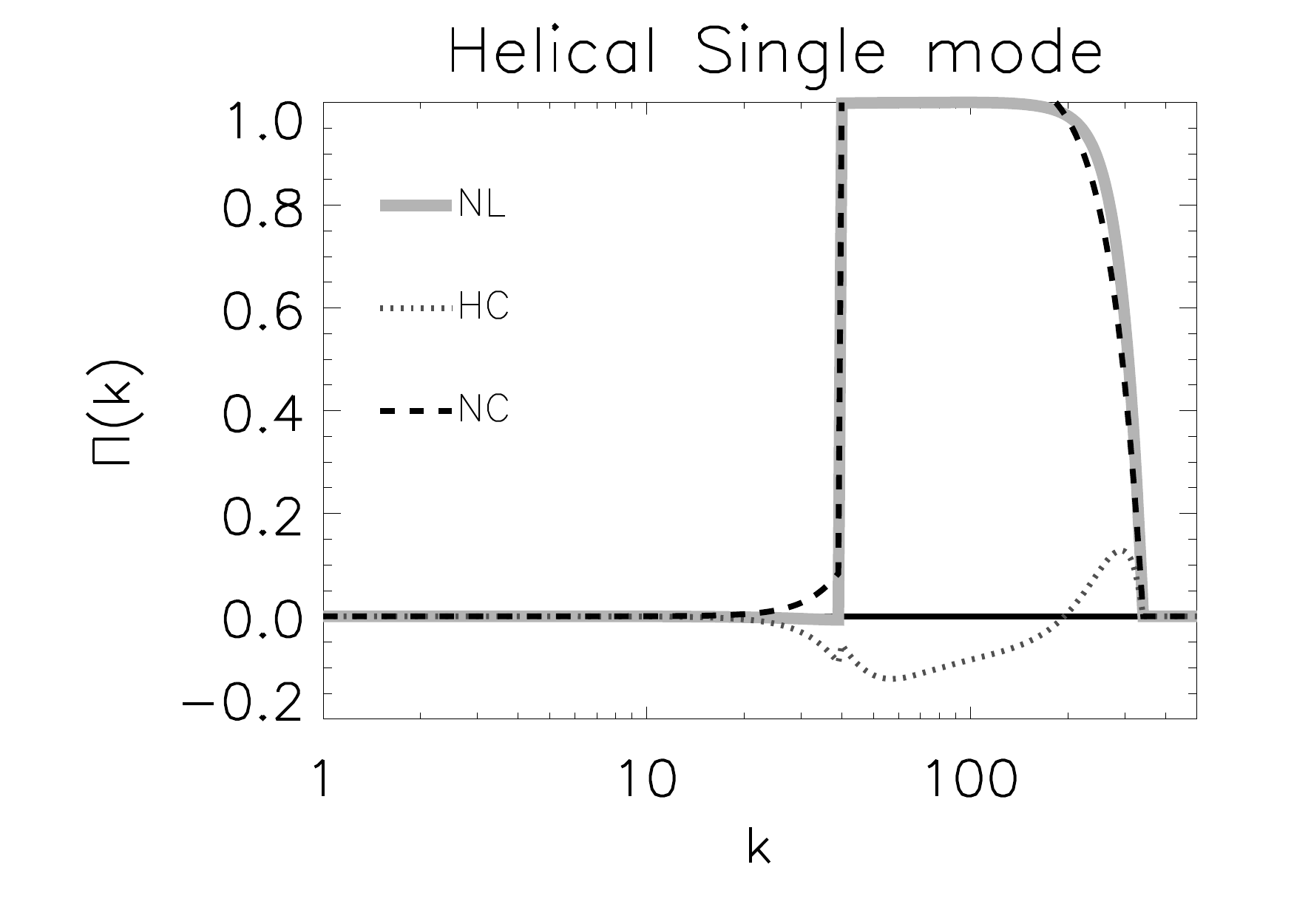}
  \includegraphics[width=0.48\textwidth]{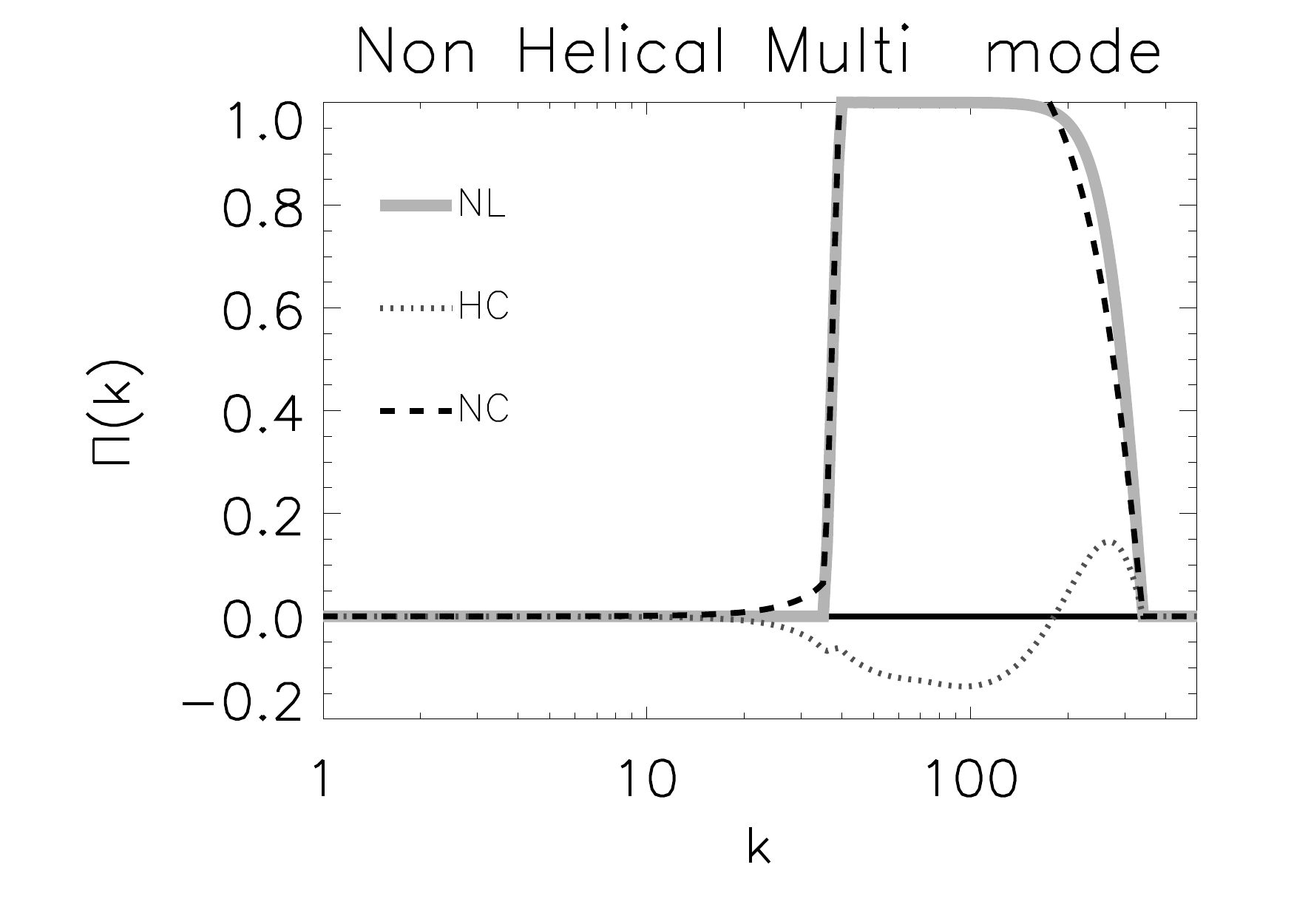}
  \includegraphics[width=0.48\textwidth]{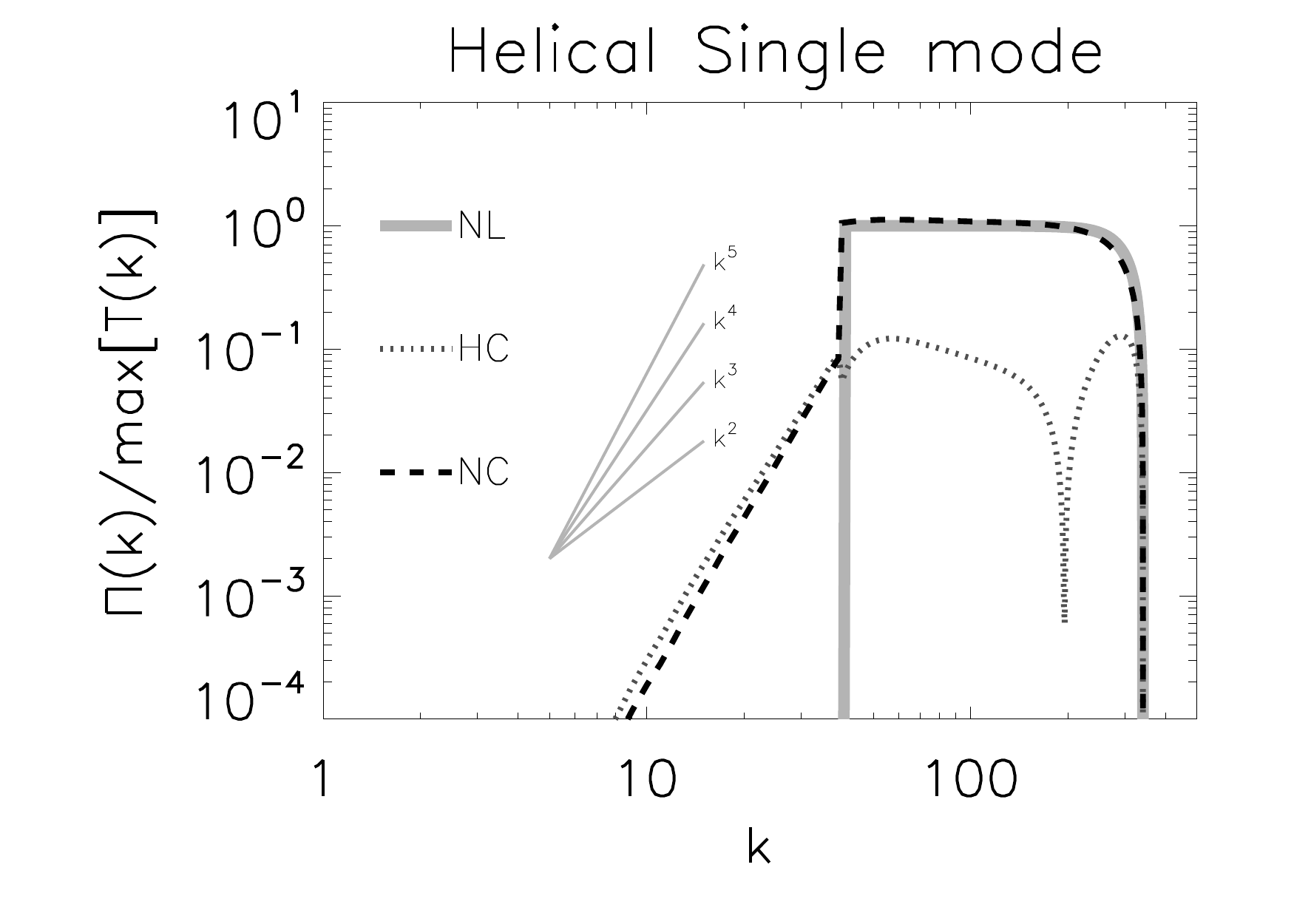}
  \includegraphics[width=0.48\textwidth]{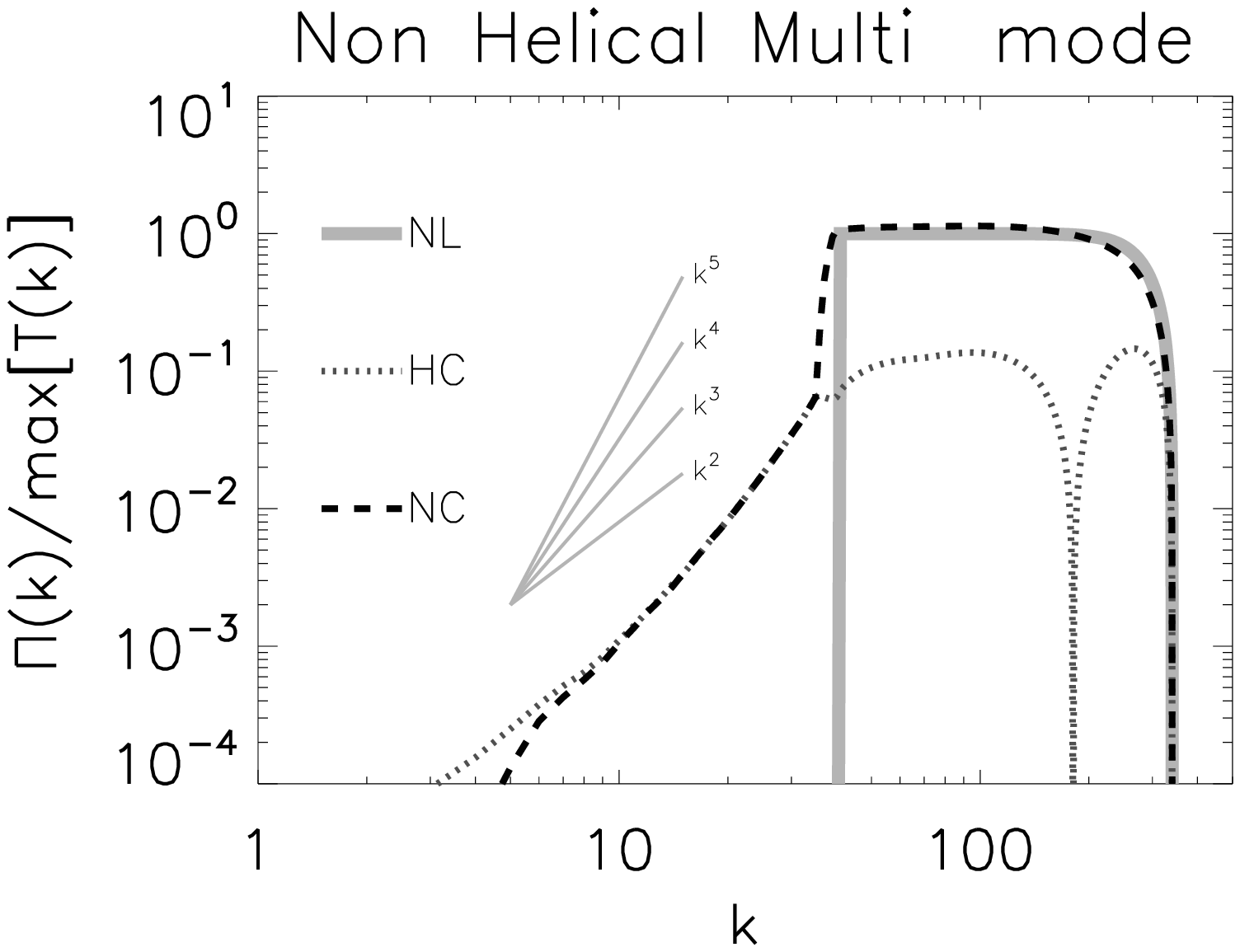}
  \caption{ Decomposed fluxes using the helical decomposition for the flow HS0 left panel and the fow NM8 in the right panel.
            The index $HC$ stands for homochiral and $NC$ stands for non-homochiral (ie hetrochiral).
            The total flux using the full nonlinear term is marked by NL. Top panels are in lin-log scale 
            while bottom panes show the absolute value in log-log scale.  }
  \label{fig:homo}
\end{figure}

An other direction for analyzing the energy fluxes has been discussed recently and comes from decomposing the velocity field in helical modes 
\cite{craya1958contributiona, Lesieur72, herring1974approach}. In this way every Fourier mode is written as the sum of two modes
one with positive helicity and one with negative helicity $$\tilde{\bf u}_\bk = \tilde{u}_\bk^+ {\bf h}^+_\bk +   \tilde{u}_\bk^- {\bf h}_\bk^-$$
where $\bf h_k^\pm$ are eigenfunctions of the curl operator $i \bk \times {\bf h}^\pm_\bk = \pm k {\bf h}^\pm_\bk$ (see appendix \ref{app:nonlin}). 
This decomposition splits the interactions among different modes to interactions that are 
homochiral (involve only modes with the same sign of helicity) or heterochiral (involve modes of both signs of helicity). 
Homochiral interactions tend to transfer on average energy to the large scales while 
heterochiral interactions tend to transfer energy on average to the small scales. This was first conjectured by 
\cite{waleffe1992nature} based on the stability properties of isolated triads, and discussed in may works 
\citep{waleffe1993inertial, chen2003joint, rathmann2017pseudo, moffatt2014note}. 
The homochiral interactions, when isolated so that the flow is driven only by them, they to lead to an inverse cascade 
\citep{biferale2012inverse, biferale2013split, sahoo2017discontinuous, sahoo2018energy}. 
In \cite{alexakis2017helically} it was also shown that even in non-helical turbulence the homochiral interactions, 
although sub-dominant, transfer energy inversely in the inertial range.

It is thus worth looking also the role played by homochiral and heterochiral in the large scale equilibrium situation.
Following  \citep{alexakis2017helically} we define the homochiral flux as 
\beq 
\Pi_{HC} (k) =\left\langle {\bf (u^+)}^{<k} \cdot ({\bf u^+} \cdot \nabla {\bf u^+} ) \right\rangle 
             +\left\langle {\bf (u^-)}^{<k} \cdot ({\bf u^-} \cdot \nabla {\bf u^-} ) \right\rangle 
\eeq
and the heterochiral flux as
\beq 
\Pi_{NC} (k) = \Pi(k) - \Pi_{HC}(k)
\eeq
where the vector fields $\bf u^+$ and $\bf u^-$ are defined as
\beq {\bf u^\pm } = \sum_\bk  {\bf h}^\pm_\bk u^\pm_k e^{i\bk \bf x} . \eeq 
and the ${<k}$ upper index stands for the filtering such that only Fourier modes of wavenumbers smaller
than k are retained as in \ref{eq:Pi}.

In figure \ref{fig:homo} the total flux $\Pi(k)$  and the homochiral flux $\Pi_{HC}(k)$ and heterochiral flux $\Pi_{NC}(k)$
are shown for the same runs as in figure \ref{fig:flux1}. The top panels are in linear-logarithmic scale 
while the bottom panels are in log-log scale and display the absolute value. As shown in  \cite{alexakis2017helically}
the flux due to homochiral interactions is negative at almost all scales while the flux due to 
heterochiral interactions is positive. In the small turbulent scales the homochiral inverse flux is sub-dominant,
while in the large scales the two counter-directed fluxes come in balance. In the large scales 
the two fluxes appear to follow a power law that is close to $k^5$ or $k^4$ for the six-mode 
forcing while closer to $k^3$ for the multi-mode forcing. 

\begin{figure}
  \centering
  \includegraphics[width=0.48\textwidth,height=0.3\textwidth]{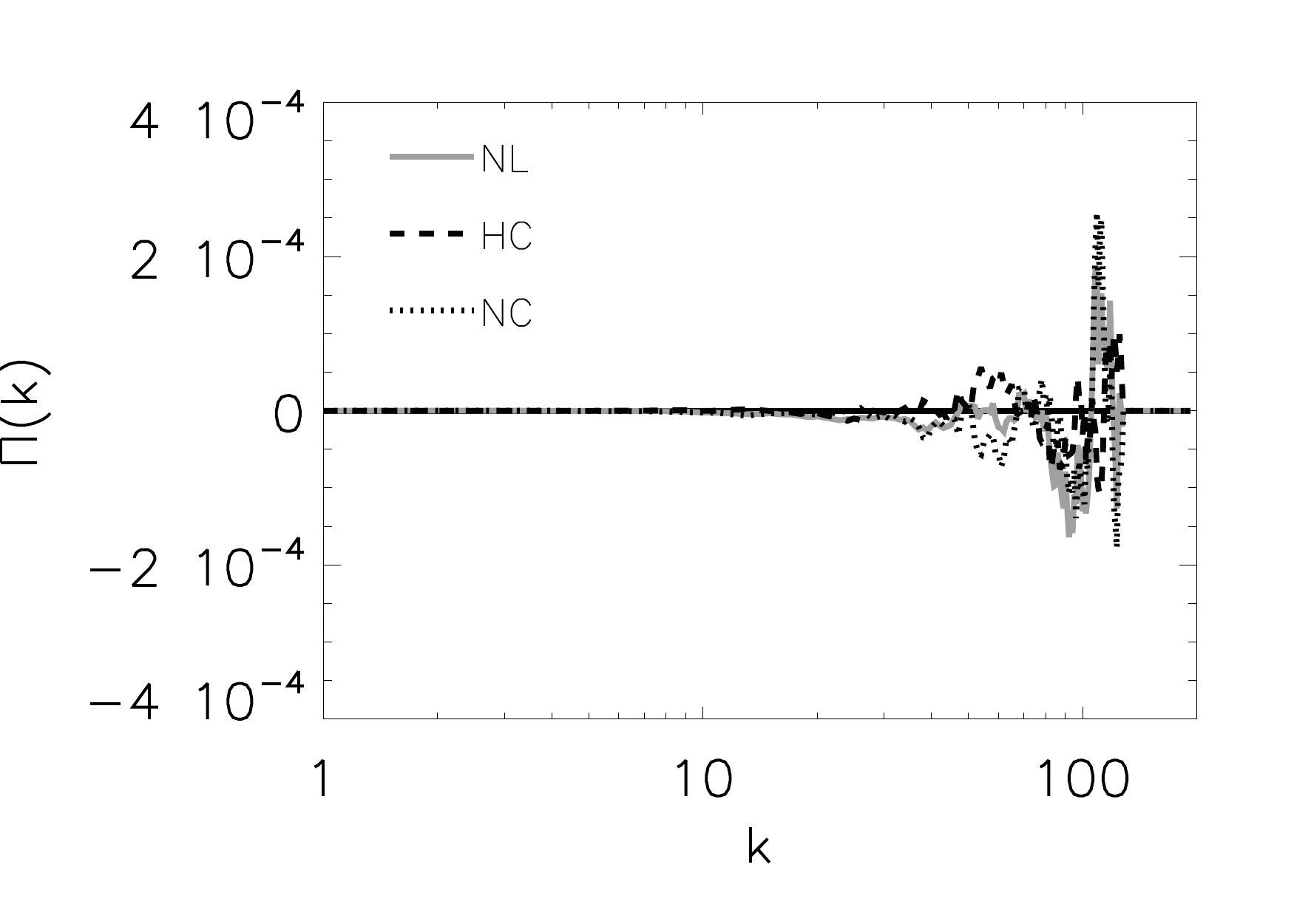}
  \caption{ Homochiral and heterochiral fluxes for truncated Euler equation system at thermal equilibrium. }
  \label{fig:homoEuler}
\end{figure}

We note that this organized inverse flux from the homochiral interactions does not appear in simulations of the truncated Euler equations
in thermal equilibrium that result in zero net flux. This is demonstrated in figure \ref{fig:homoEuler} that shows the two fluxes $\Pi_{HC}(k)$ and $\Pi_{NC}(k)$ from a simulation of the truncated Euler equations. The two fluxes although averaged over many outputs they 
appear noisy with no preferential direction of cascade. Thus the non-zero and sign definite value of the fluxes 
$\Pi_{HC}(k)$ and $\Pi_{NC}(k)$ that was observed in the forced runs indicate a deviation from the thermal equilibrium. 


\subsection{Energy shell to shell transfers}          
\label{sec:analysis6} 

We end this section by examining the shell to shell transfer functions $\cT(K,Q)$
that express the rate energy is transferred from one shell of wavenumbers $K< |{\bf k }| < K+1$ 
to an other shell of wavenumbers $Q< |{\bf k }| < Q+1$. We define these transfer functions as 
\beq
\cT(K,Q) = -\langle \bu_K (\bu \cdot \nabla ) \bu_Q \rangle 
\eeq
where $\bu_K$ and $\bu_Q$ are the velocity field filtered so that only the wavenumbers at shell $K$ and $Q$ are kept respectively. 
These transfer functions have been studied extensively in the literature \cite{domaradzki1990local, alexakis2005imprint, verma2005local, mininni2006large, verma2007energy, domaradzki2007analysis, mininni2008nonlocal, eyink2009localness1,  aluie2009localness}. If $\cT(K,Q)<0$ it means that the shell $K$ is giving energy to the shell $Q$ while if  $\cT(K,Q)>0$ the shell $K$ is receives energy from the shell $Q$.
The transfer $\cT(K,Q)$ satisfies the relation $\cT(K,Q)=- \cT(Q,K) $ that reflects the conservation of energy by the nonlinear term. 

\begin{figure}
  \centering
  \includegraphics[width=0.68\textwidth]{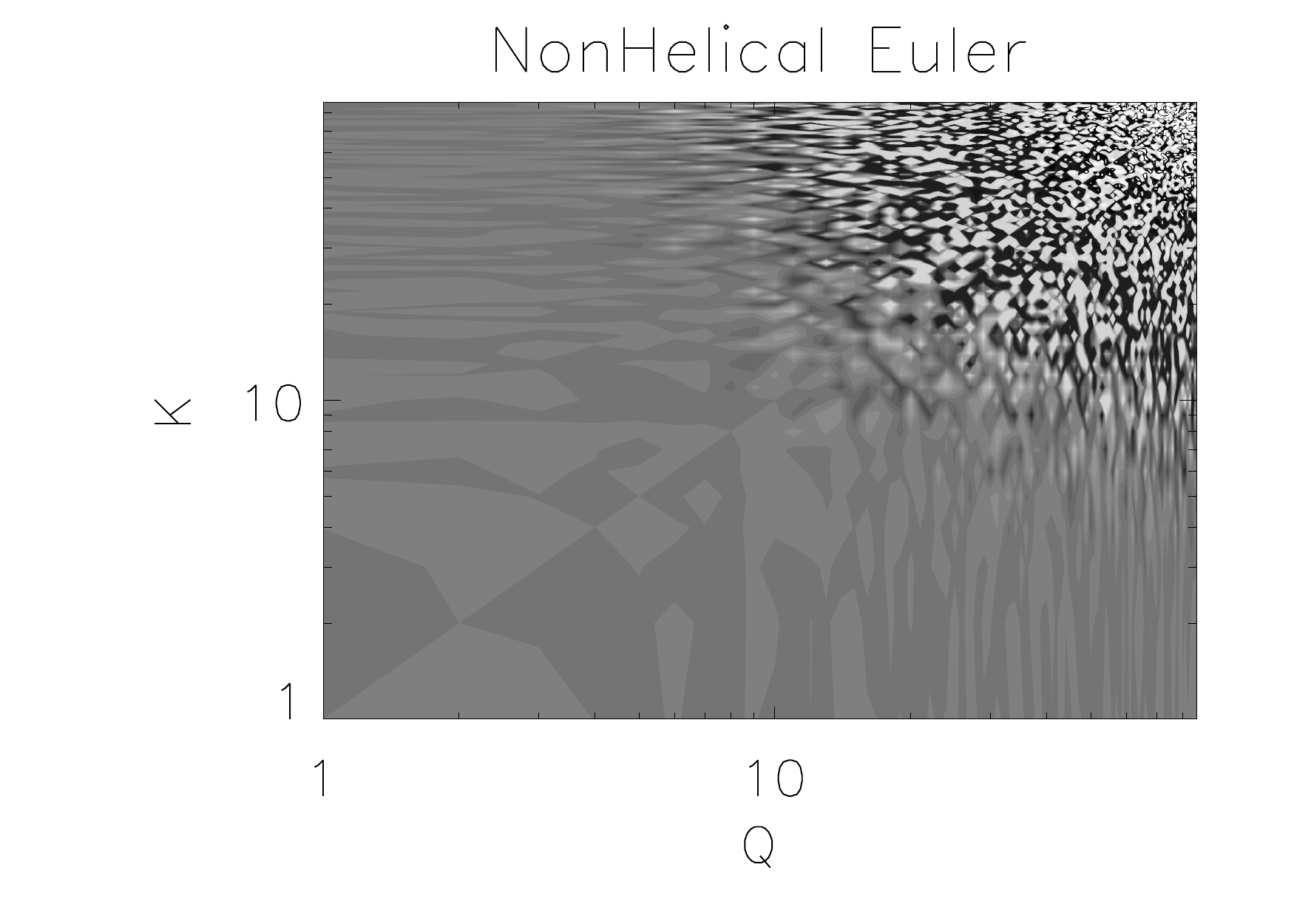}
  \caption{ A gray scale image shell to shell transfer function $\cT(K,Q)$ for a truncated Euler flow.   }
  \label{fig:tranEULER}
\end{figure}

We calculate $\cT(K,Q)$ for the truncated Euler flow and it is displayed in figure \ref{fig:tranEULER} where dark colors imply negative values of  $\cT(K,Q)$, while light colors imply positive values. 
As expected for the thermal flows the transfer function appears as noise. This is because at the absolute equilibrium state there
is no preferential direction of transfer of energy from any set of wavenumbers to any other. If averaged over many outputs $\cT(K,Q)$ will become zero.

\begin{figure}
  \centering
  \includegraphics[width=0.68\textwidth]{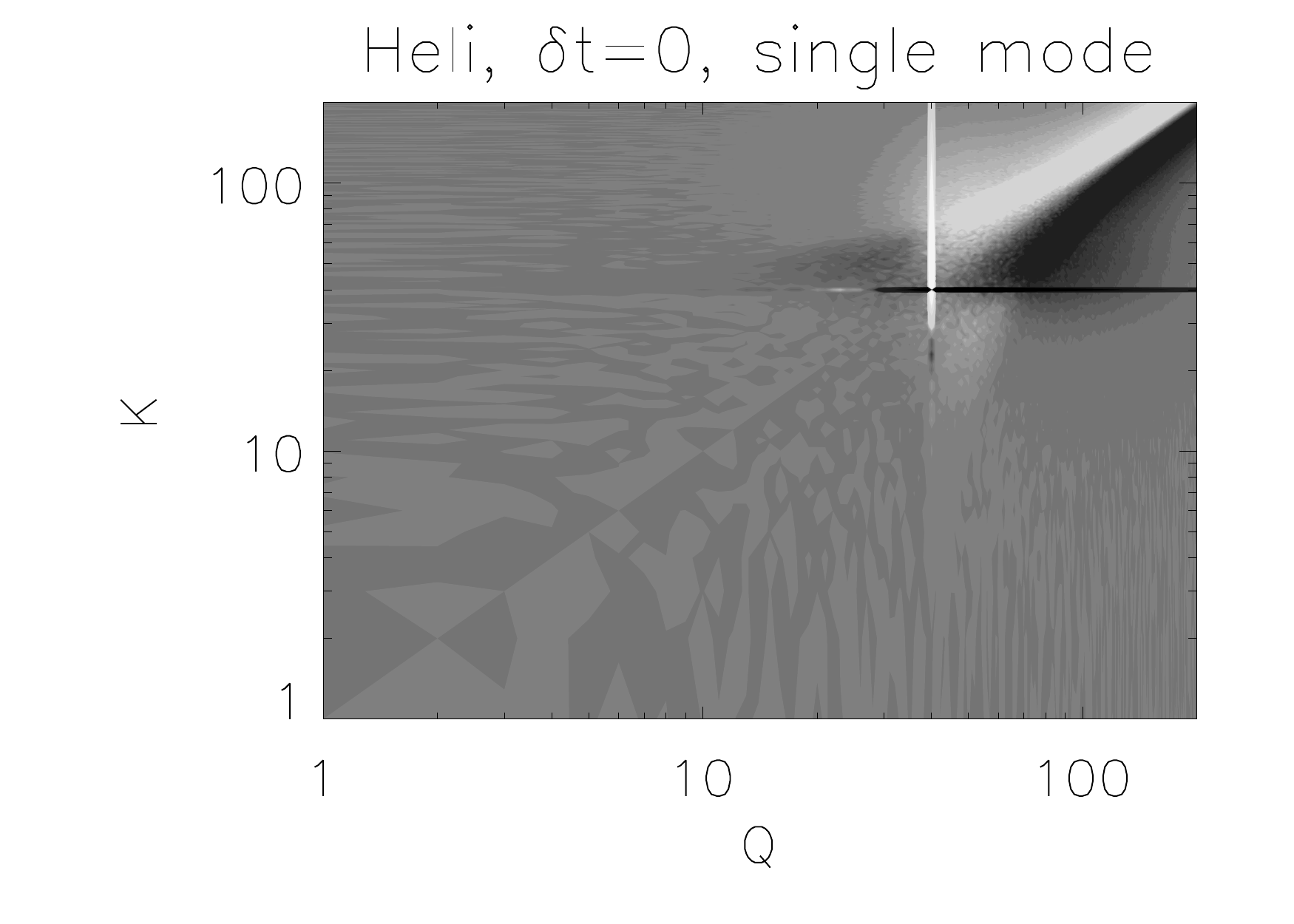}
  \caption{ A gray scale image of the shell to shell transfer function $\cT(K,Q)$ for the run HS0.   }
  \label{fig:tranH0}
\end{figure}

The situation is different when we investigate the flows obeying the forced Navier-Stokes equation.
In figure \ref{fig:tranH0} we plot the transfer function $\cT(K,Q)$ for the flow HS0 (helical, six-mode, $\delta t=0$) that displayed 
a thermal energy spectrum in the large scales. 
The dark horizontal line at $K=k_f=40$ and the bright vertical line at $Q=k_f$ represent the transfer of energy from the forced modes 
that interact and transfer energy to almost all wavenumbers.
For values of both $Q$ and $K$ larger than the forcing wavenumber, $\cT(K,Q)$ displays the standard behavior for the forward cascade with negative (dark) values below the diagonal $Q=K$ and positive 
values (light) above the diagonal indicating that energy is transferred from the small wavenumbers to the large.   
For values of both $Q$ and $K$ smaller than the forcing wavenumber, $\cT(K,Q)$ is almost zero. 
An exchange of energy with the large scales is observed only with the forcing scale (bright and dark line at $Q=40$ and $K=40$) that inject energy to the large scales,
and some exchange (both positive and negative) with the turbulent scales indicated by the light and bright patches in the top left quadrant and bottom right quadrant. 

\begin{figure}
  \centering
  \includegraphics[width=0.48\textwidth]{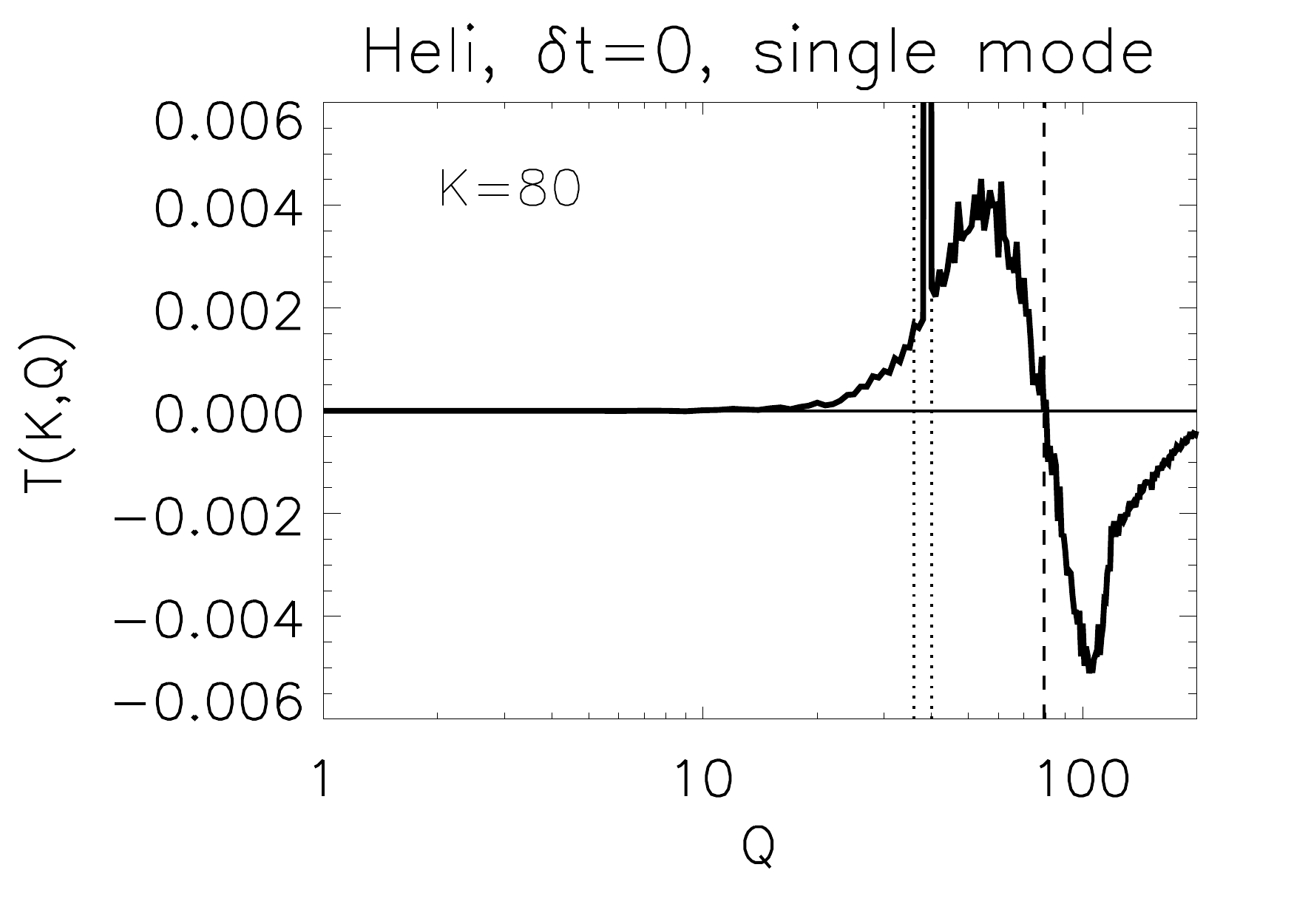}
  \includegraphics[width=0.48\textwidth]{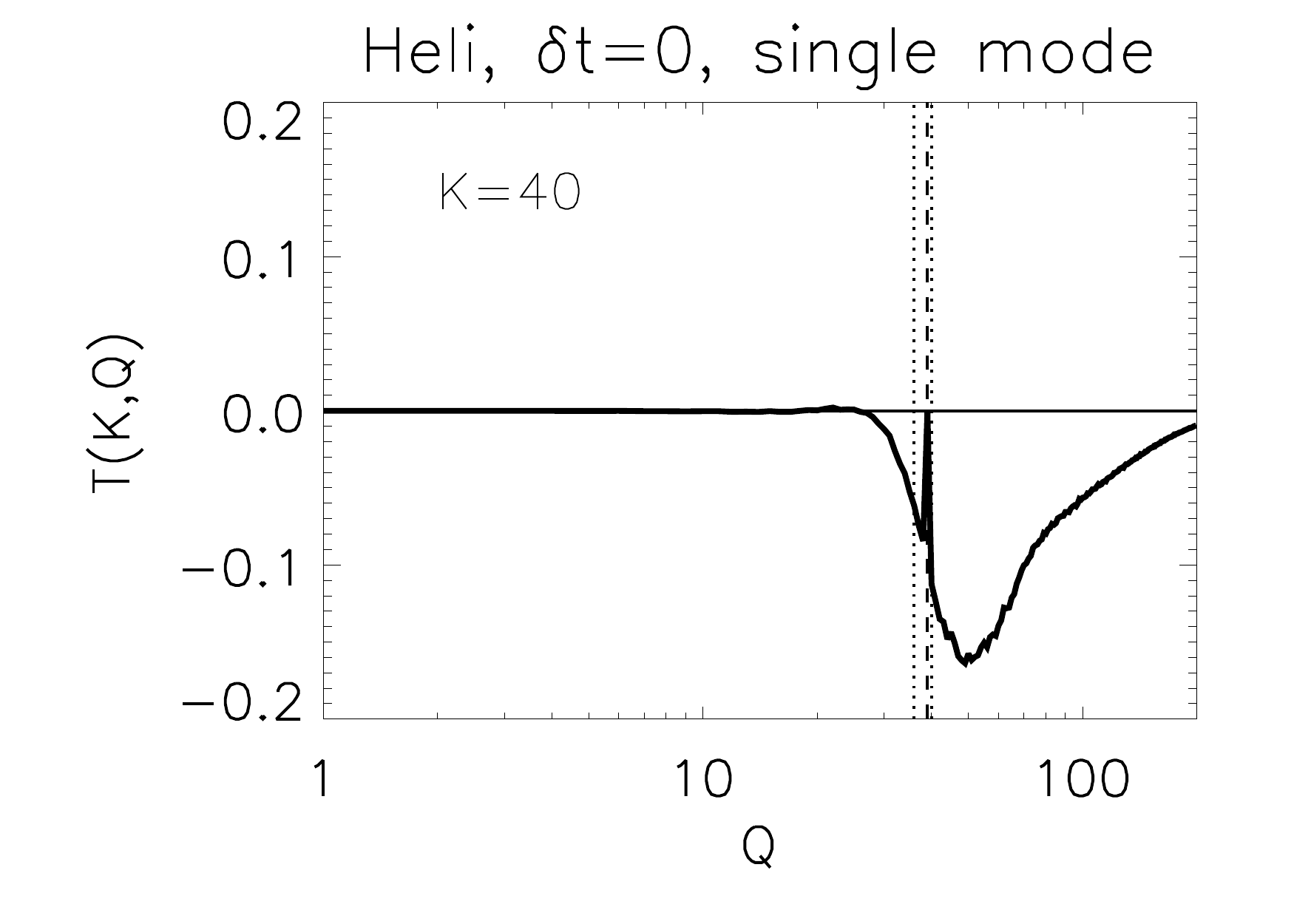}
  \includegraphics[width=0.48\textwidth]{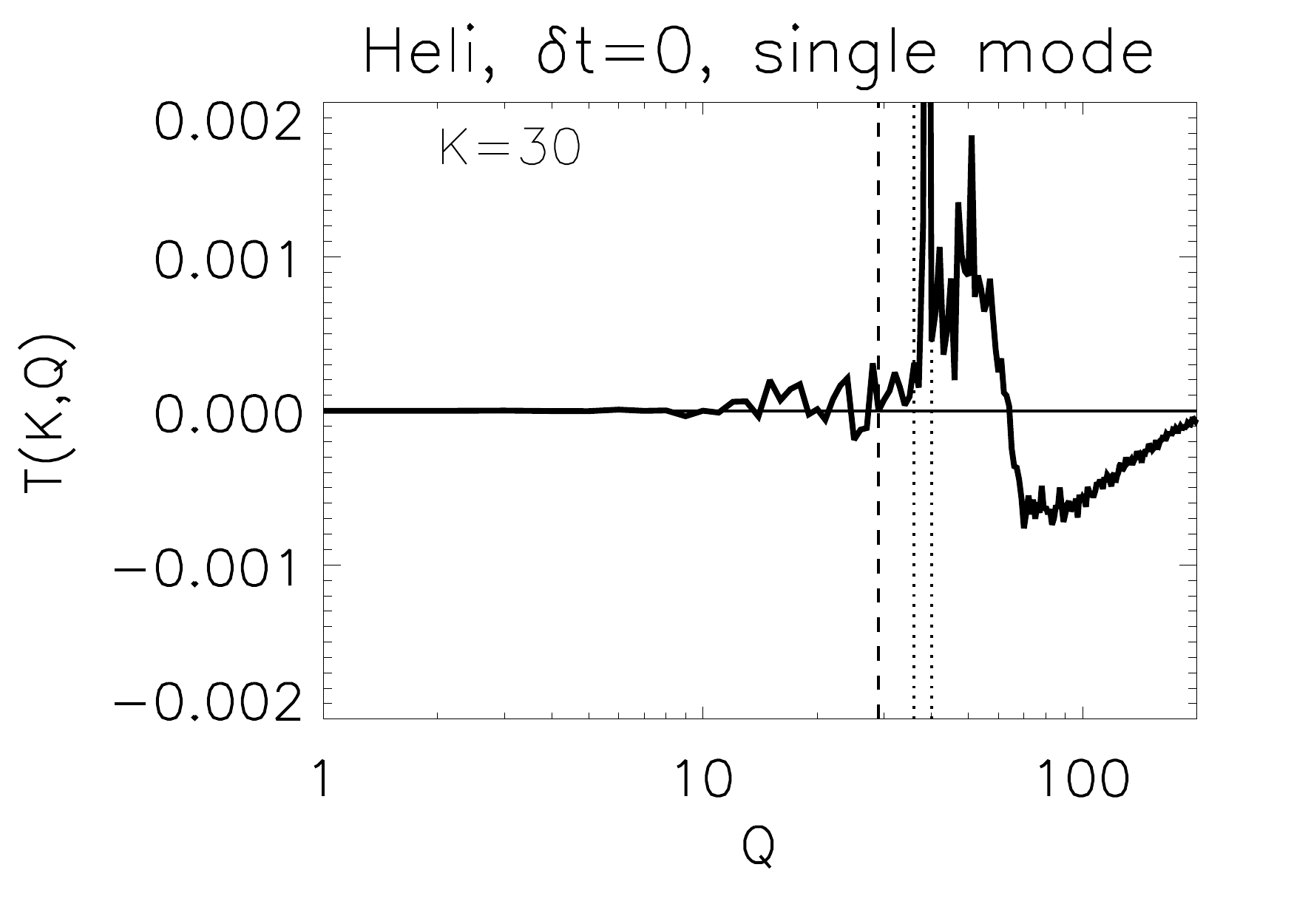}
  \includegraphics[width=0.48\textwidth]{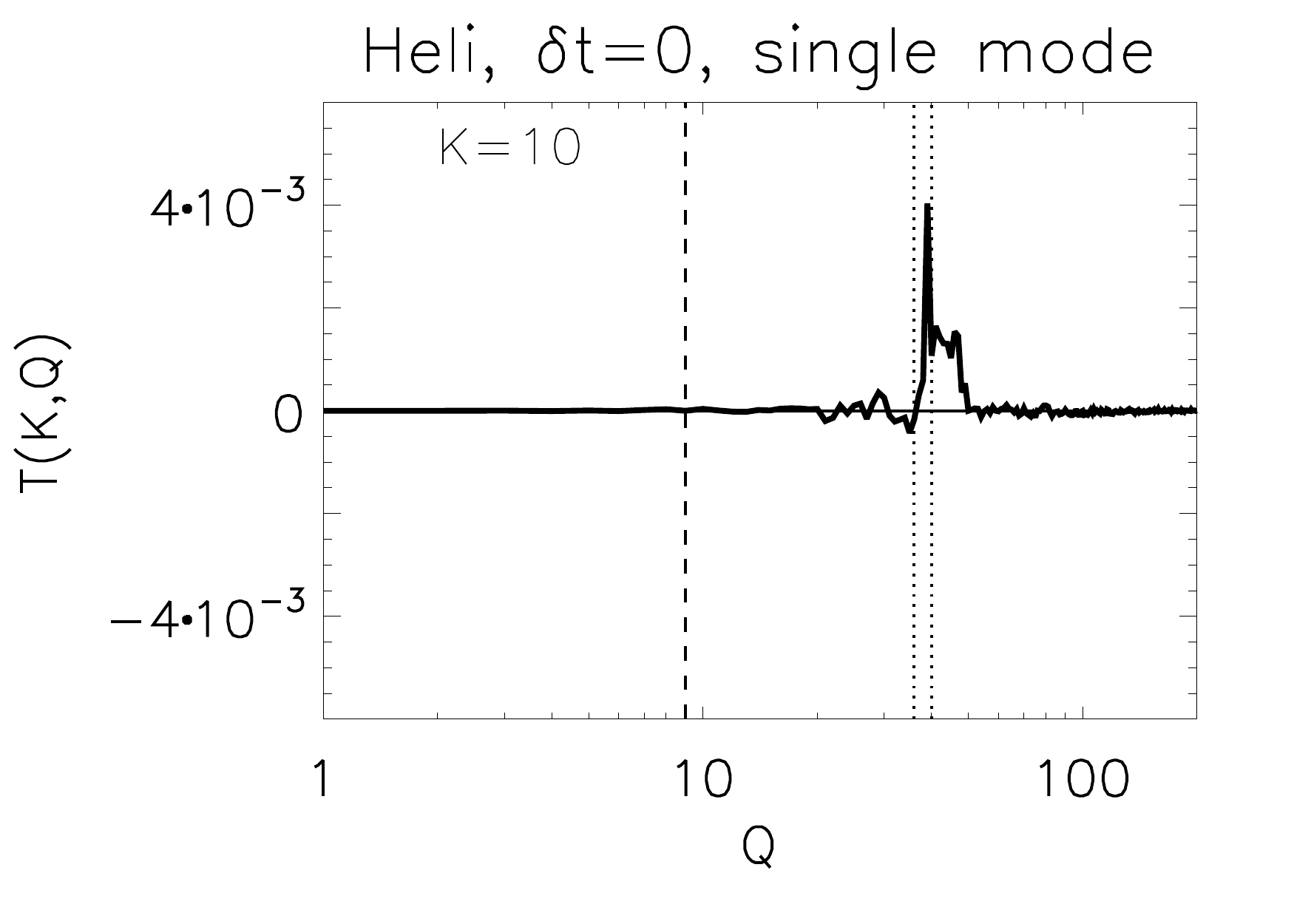}
  \caption{ Shell to Shell transfer function $\cT(K,Q)$ as a function of $Q$ for four different values of $K=80,40,30,10$ 
  obtained from the flow HS0.   }
  \label{fig:tranHb}
\end{figure}

The transfers are displayed more clearly if we examine particular values of $K$. 
In figure \ref{fig:tranHb} we plot  $\cT(K,Q)$ as a function of $Q$ for $K=80,40,30,10$. 
The value of $K$ is also indicated in the four panels by the vertical dashed line,
while the doted lines indicate the forcing scales. 
Positive values of  $\cT(K,Q)$ indicate the range of wavenumbers the examined $K$ receives energy 
while negative values indicate the range of wavenumbers it gives energy. 
The $K=80$ shell receives energy from all smaller wavenumbers $(Q<K)$ and gives energy to all larger wavenumbers $(Q>K)$.
The forcing scale $K=40$ gives energy to all wavenumbers small and large. 
The $K=30$ shell receives energy from the forcing scales while it looses energy to the largest wavenumbers $(Q>60)$,
while the largest scales $K=10$ only exchange energy with scales close to the forcing scales.

\begin{figure}
  \centering
  \includegraphics[width=0.68\textwidth]{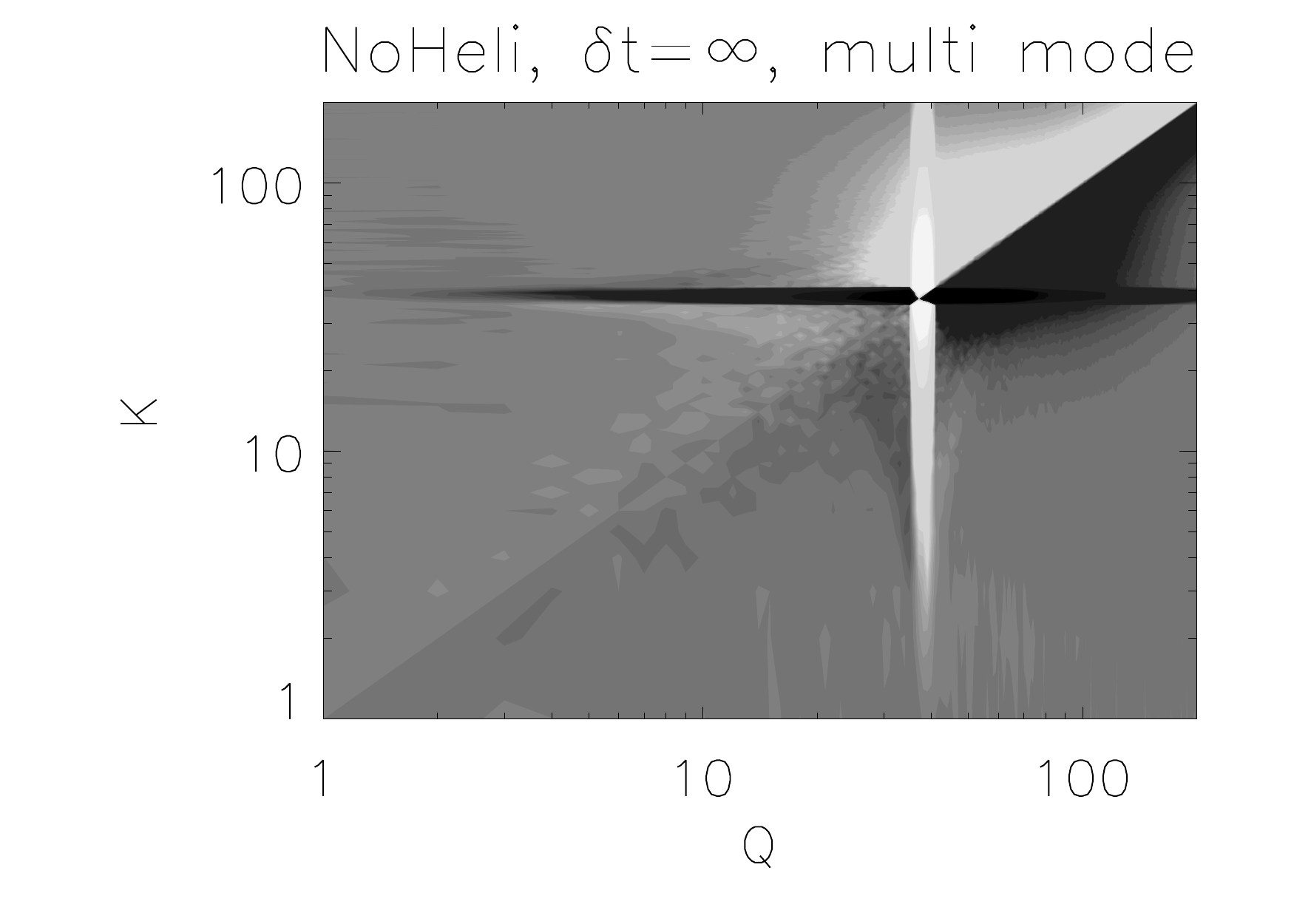}
  \caption{ A gray scale image of the shell to shell transfer function $\cT(K,Q)$ for the run NM8. }
  \label{fig:tranN8}
\end{figure}

In figure \ref{fig:tranN8} we plot the transfer function $\cT(K,Q)$ for the flow NM8 (non-helical, multi-mode, $\delta t=\infty$) that displayed 
strong deviations from the thermal energy spectrum in the large scales. 
The overal picture is similar to that of fig. \ref{fig:tranH0} although the interactions 
with the forcing scale are much more intence, and one can observe a local forward cascade
in the large scales indicated by the bright reagion above the diagonal and darck region bellow the diagonal for $K,Q<k_f$.

\begin{figure}
  \centering
  \includegraphics[width=0.48\textwidth]{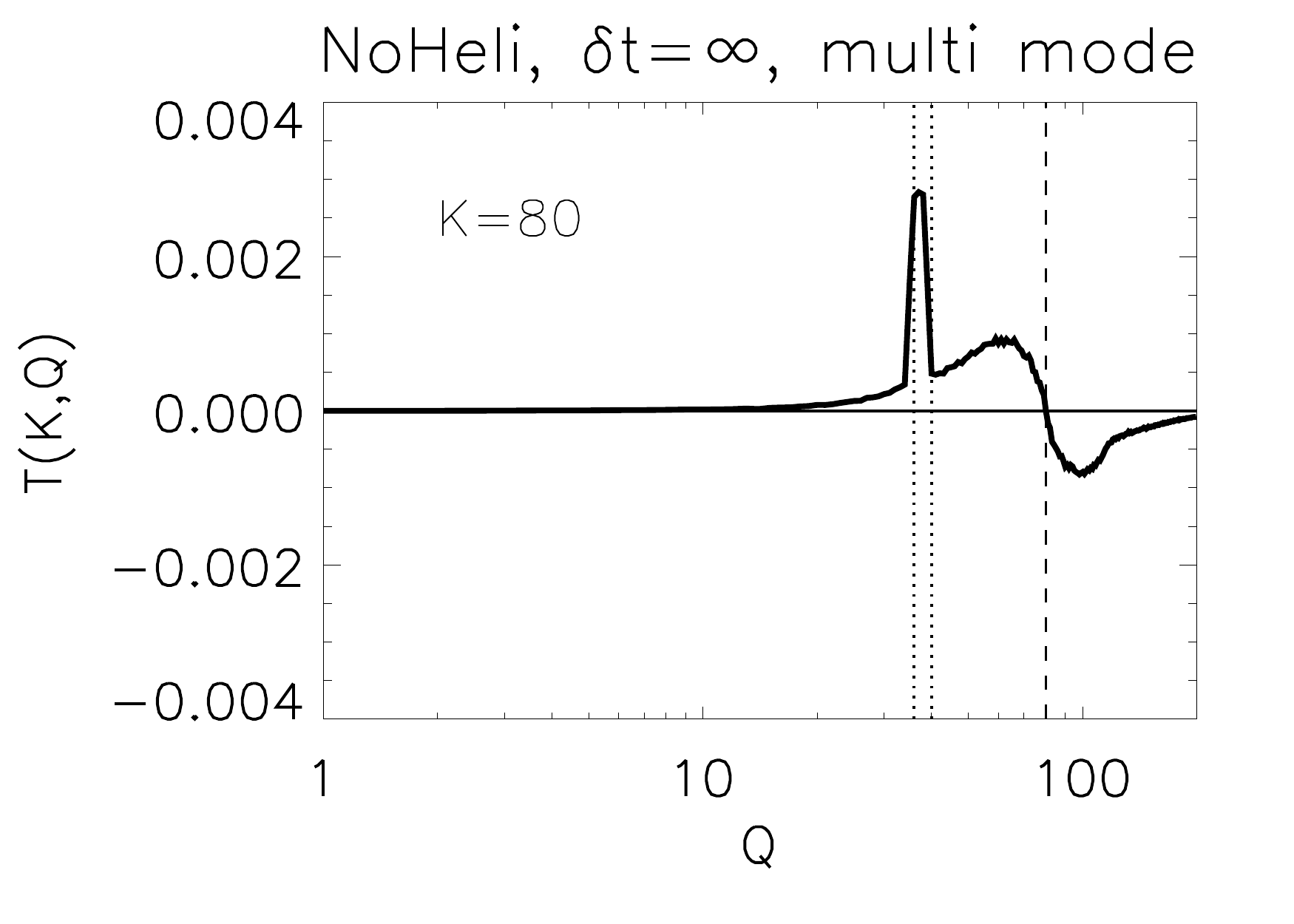}
  \includegraphics[width=0.48\textwidth]{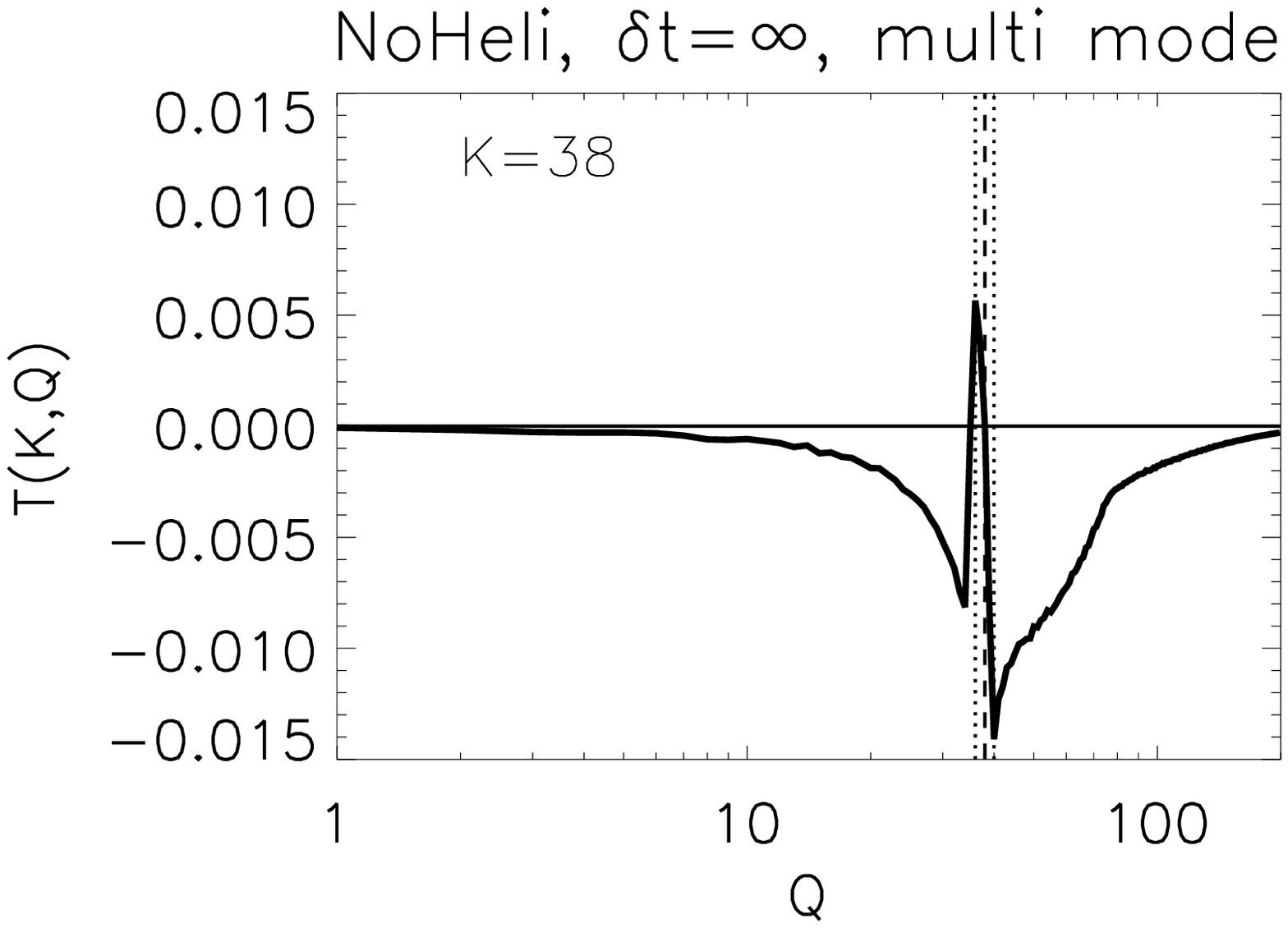}
  \includegraphics[width=0.48\textwidth]{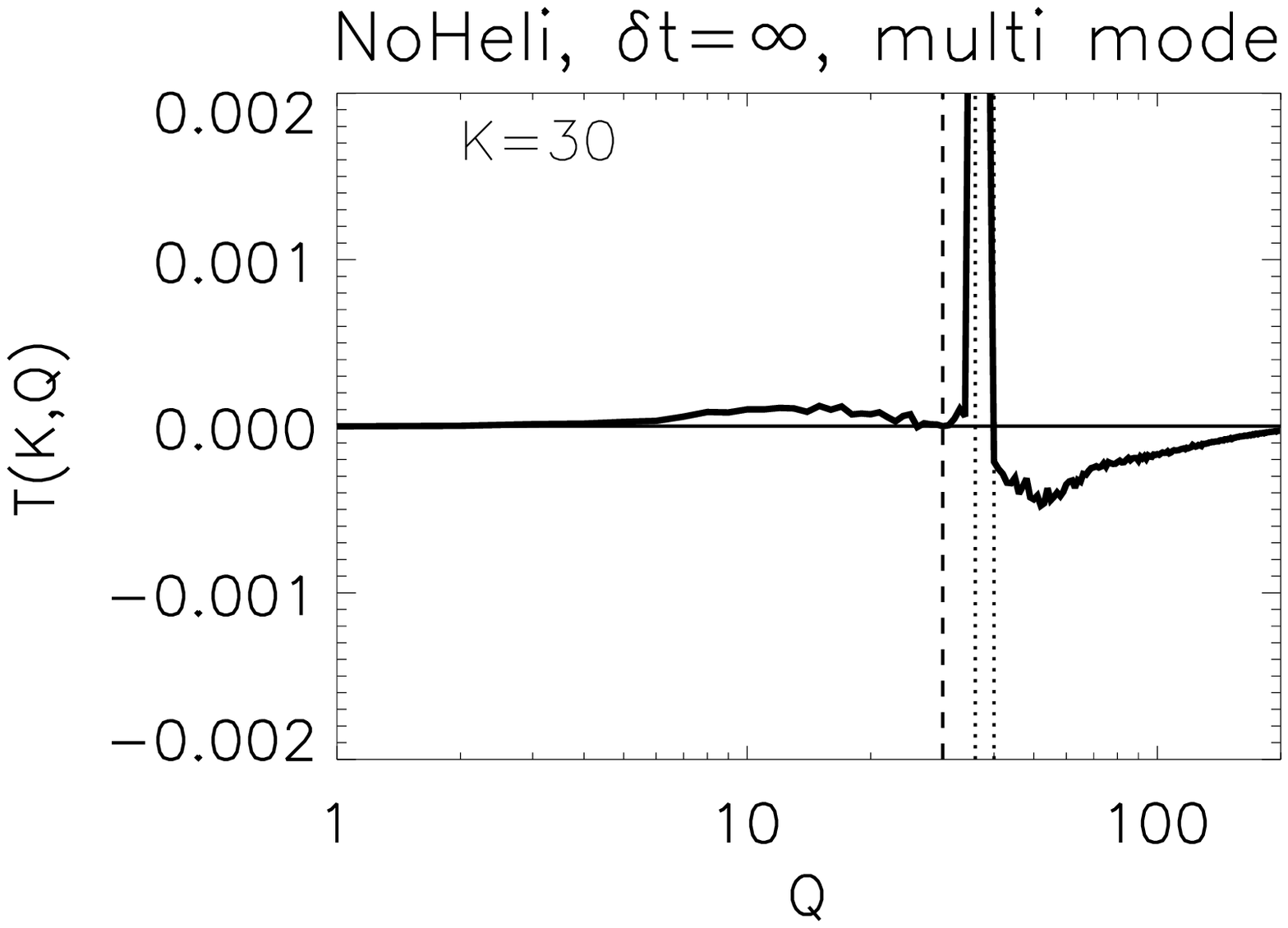}
  \includegraphics[width=0.48\textwidth]{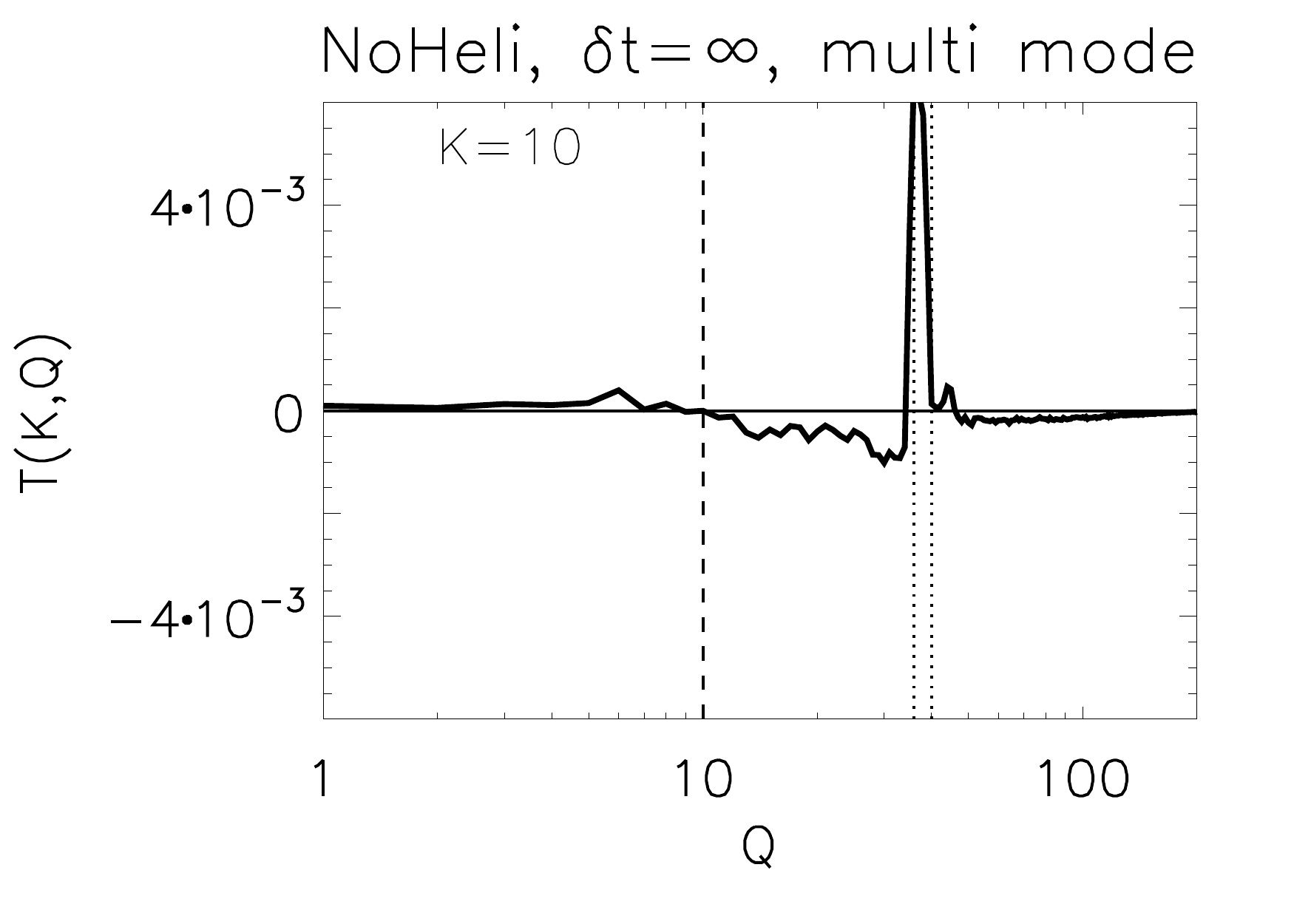}
  \caption{ Shell to Shell transfer function $\cT(K,Q)$ as a function of $Q$ for four different values of $K=80,40,30,10$ 
  obtained from the flow NM8. }
  \label{fig:tranNb}
\end{figure}

For more detail in figure \ref{fig:tranNb} we plot the transfer function $\cT(K,Q)$ for the same wavenumbers $K$ as in figure \ref{fig:tranHb}.
The overall picture for the turbulent and forced scales is the same as in \ref{fig:tranHb} with larger scales giving energy to smaller scales
and the forced scale giving energy to all. There are significant differences however if we look at the energy exchange at the large scales.
At these scales there is a sign of a local forward cascade at large scales: the shell $K=10$ receives energy from smaller wavenumbers and gives energy to nearby larger wavenumbers. At the same time the same shell receives energy non-locally from the forcing scale and loses energy non-locally to the turbulent scales. This strengthens the picture from previous sections that when large scales are away from the thermal equilibrium due to an excess of energy put by the forcing the large scales try to recover the thermal equilibrium by transporting (locally and non-locally) the energy to the smaller scales.

\section{Summary and Conclusions} 

The present work has examined 12 different simulated flows forced at small scales with scope to understand better the behavior of large scale flows and their relation to the absolute equilibrium solutions predicted by \cite{kraichnan1973helical}. The results were particularly interesting, revealing a variety of behaviors of the large scale components of turbulent flows. 
In particular it was shown in sec. \ref{sec:spectra} that the absolute equilibrium solutions are well reproduced by the large scales of turbulent flows when a few modes are are forced 
(spectrally sparse forcing), and when the forcing of these modes is sufficiently short time correlated. Small deviations from the absolute equilibrium solutions were observed when the forcing correlation time was increased. Helicity did not play a significant role  and even for fully helical forcing function very little helicity was injected in the large scales making the large scale flows to be almost non-helical.
Strong deviations from the  absolute equilibrium solutions were observed when the forcing was applied to all modes inside a spherical shell (spectrally dense forcing). 
In this case, and particularly for the infinitely time correlated forcing, the power law behavior of the energy spectrum was far from the  absolute equilibrium prediction $k^2$
and was closer to a $k$ independent behavior. 

The cause of this apparent lack of universality in the large scales was argued in sec.\ref{sec:analysis1} to be due to the number of triads that couple a large scale mode with two forced modes.
In the first case of sparse forcing these triads were absent while for the spectrally dense forcing they were shown to follow a power-law distribution with the large scale wavenumber $q$ that was $N_Q \propto q^1 $ for $\delta k_f \ll q \ll k_f $ and $N_Q \propto q^2 $ for $q \ll \delta k_f \ll k_f $. 

This difference altered the balance of the interactions in the large scales, something that was clearly reflected in the spectrum of the nonlinearity that was examined in sec.\ref{sec:analysis2}.
The spectrally sparse forcing lead to a $k^4$ spectrum for the nonlinearity in agreement with the the one calculated  for flows in absolute equilibrium,
while the spectrally dense forcing lead to nonlinearity spectrum closer to $k^3$ which was in agreement with our estimates obtained assuming that the forced 
modes play a dominant role in the large scales.
Decomposing the flow in section \ref{sec:analysis3} in different components verified that for the sparse forcing the interactions 
that coupled two forced modes with one large scale mode were absent, while for the spectrally dense forcing they were dominant
and lead to a $k^3$ spectrum for the nonlinearity. The interactions with the forced modes in the later case were shown to be balanced by 
the local  large scale interactions.

We further managed to identified the role played by the different scales by looking at the fluxes caused by different scales in sec. \ref{sec:analysis4} 
and the shell-to-shell transfers in sec. \ref{sec:analysis6}. 
This analysis revealed that interactions with the forced scales inject energy to the large scales while interactions with the turbulent scales and large scale self interactions 
tend to bring energy back to the small scales. Finally, we investigated the fluxes due to homo-chiral and hetero-chiral interactions in section \ref{sec:analysis5}. 
The first were shown to move energy to the large scales while the later moved energy away from the large scales in contrast with the absolute equilibrium flows for which these fluxes average to zero. The amplitude of these opposite directed fluxes decreased as $q\to 0$ with a high power-law.

Using a thermodynamics analogy the present analysis indicates that the large scales in a turbulent flow resemble a reservoir 
that is in a (non-local) contact with a second out-of equilibrium reservoir consisting of the smaller (forced, turbulent and dissipative) scales. 
When the energy injection to the large scales from the forced modes is relative weak (as is the case for the spectrally sparse forcing) 
then the large scale spectrum remains close to a thermal equilibrium and the role of long range interactions is to set the global energy (temperature) of the equilibrium state. 
If on the other hand the long-range interactions are dominant (as is the case for the spectrally dense forcing), the  large-scale self-interactions cannot respond fast enough  to bring the system in equilibrium and the large scales deviate from the equilibrium state and the energy spectrum can display different exponents.

\acknowledgments 
This work was granted access to the HPC resources of MesoPSL financed by the Region 
Ile de France and the project Equip@Meso (reference ANR-10-EQPX-29-01) of 
the programme Investissements d'Avenir supervised by the Agence Nationale pour 
la Recherche and the HPC resources of GENCI-TGCC-CURIE \& GENCI-CINES-JADE 
(Project No. A0010506421, A0030506421 \& A0050506421) where the present numerical simulations have been performed.
This work has also been supported by the Agence nationale de la recherche
(ANR DYSTURB project No. ANR-17-CE30-0004).

\appendix 

\section{Number of interacting triads}\label{app:nmodes} 

\begin{figure}                 
  \centering
  \includegraphics[width=0.32\textwidth]{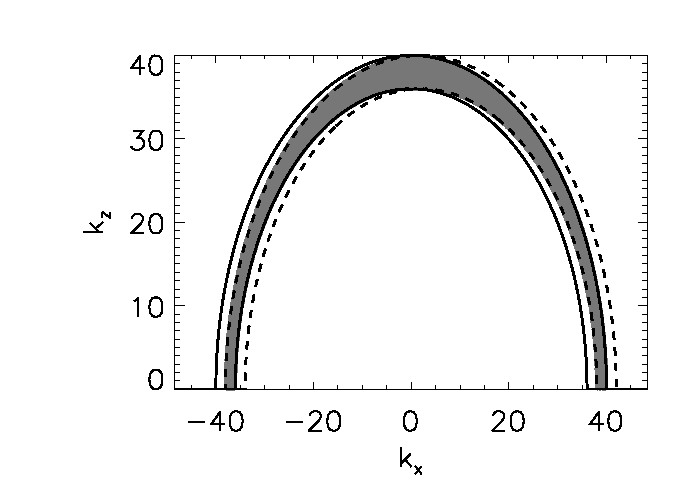}
  \includegraphics[width=0.32\textwidth]{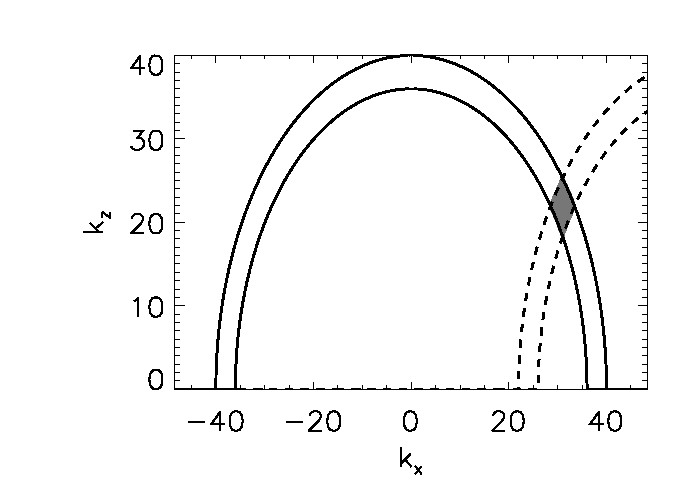}
  \includegraphics[width=0.32\textwidth]{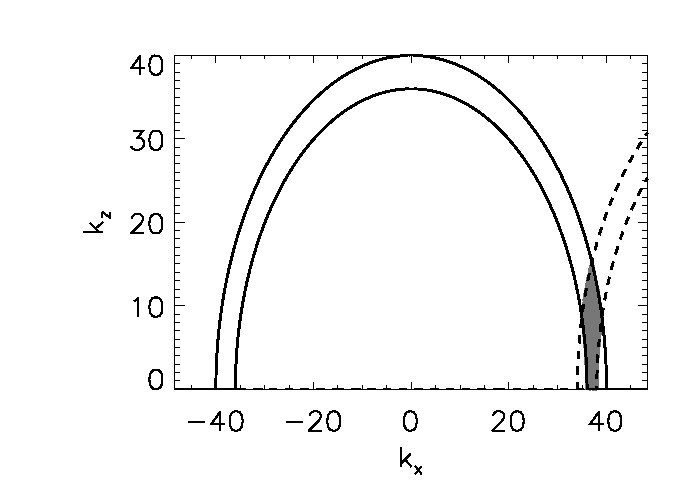}
  \caption{ The figure demonstrates how to calculate the number of triads that can be constructed with two modes inside the forcing shell 
  and a large scale mode $\bf q$, for three different values of $|\bf q|=2$ (left panel)) $|\bf q|=64$ (center panel) and   $|\bf q|=74$ (right panel). 
  The solid lines indicate the modes with $|{\bf k}|=k_f$ and $k_f-\delta k_f$ while the dashed lines indicate the circles $|{\bf k+q}|=k_f $ and $k_f-\delta k_f$. 
  The modes $\bf k$ for which both $\bf k$ and $\bf k+q$ are among the forced modes are given by the modes that lie in the intersection of the two spherical shells
  that is dipicted by the shaded area. Note that almost all forcing modes can form triads with small $\bf q$, while the allowed number of modes goes to zero when $|{\bf q}|=k_f$.  }
  \label{fig:modes1}
\end{figure}

For a given mode ${\bf q}$, the number of modes ${\bf k}_1,{\bf k}_2 \in K_F$ that satisfy ${\bf q}+{\bf k}_1+{\bf k}_2=0$ is given by the number of modes $N_{\bf q}$ that reside in the intersection of the two spherical shells $k_f-\delta k_f < |{\bf k}| \le k_f$  and $k_f-\delta k_f < |{\bf k+q}| \le k_f$. Since the density of wavenumbers in the Fourier space for a cubic domain of side $2\pi$ is uniform  and equal to unity $N_{\bf q}$ is approximately equal to the volume of the afore mentioned intersection (it becomes exactly equal to $N_{\bf q}$ when $k_f \to \infty$).  This intersection for three different values of ${\bf q} =(q,0,0)$ is demonstrated in figure \ref{fig:modes1}. Note that in fig. \ref{fig:modes1} only a plane cut is shown at $k_y=0$ but the intersection area is symmetric around the axis of $\bf q$ here taken to be the $x$-axis.

\begin{figure}                 
  \centering
  \includegraphics[width=0.32\textwidth]{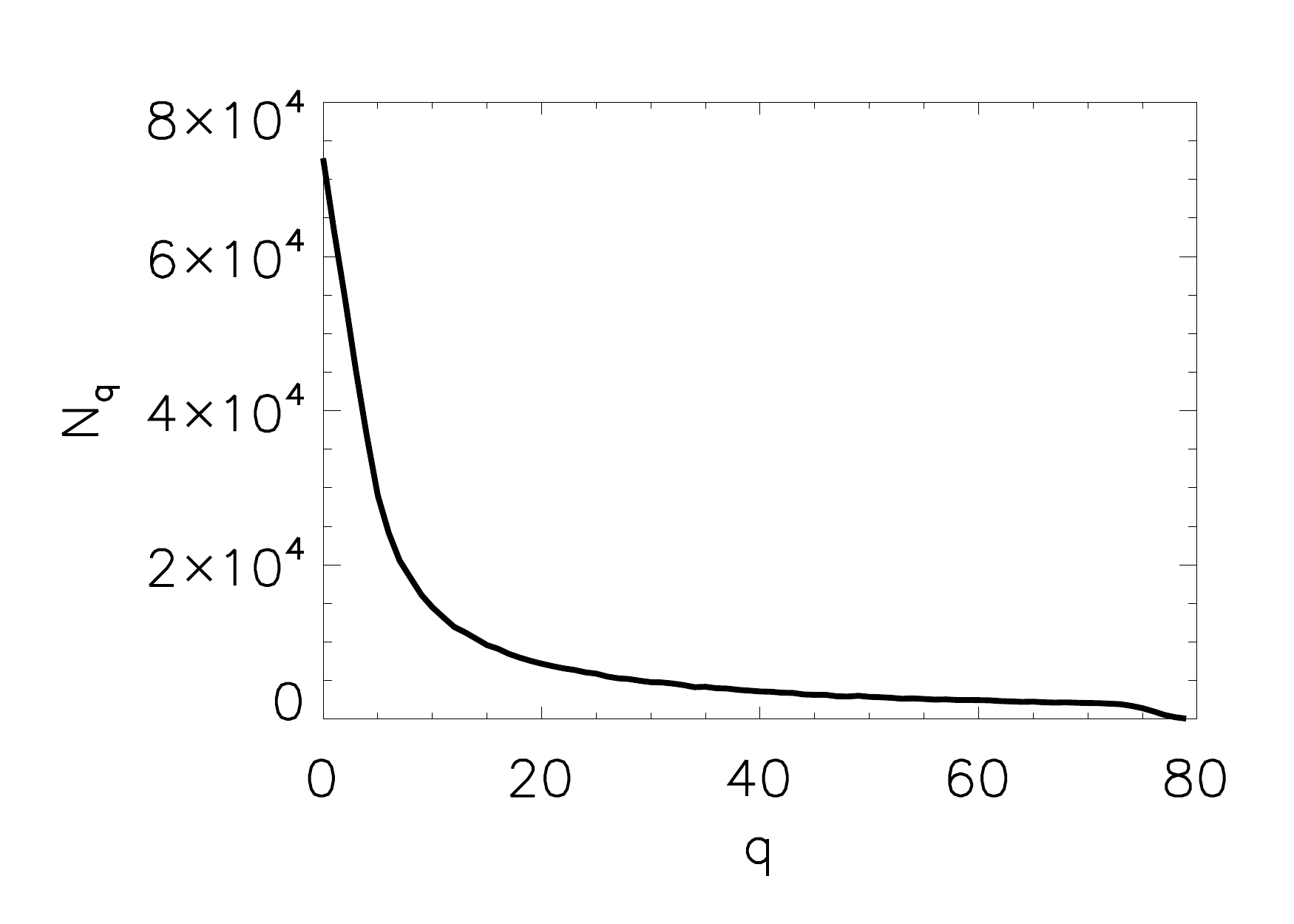}
  \includegraphics[width=0.32\textwidth]{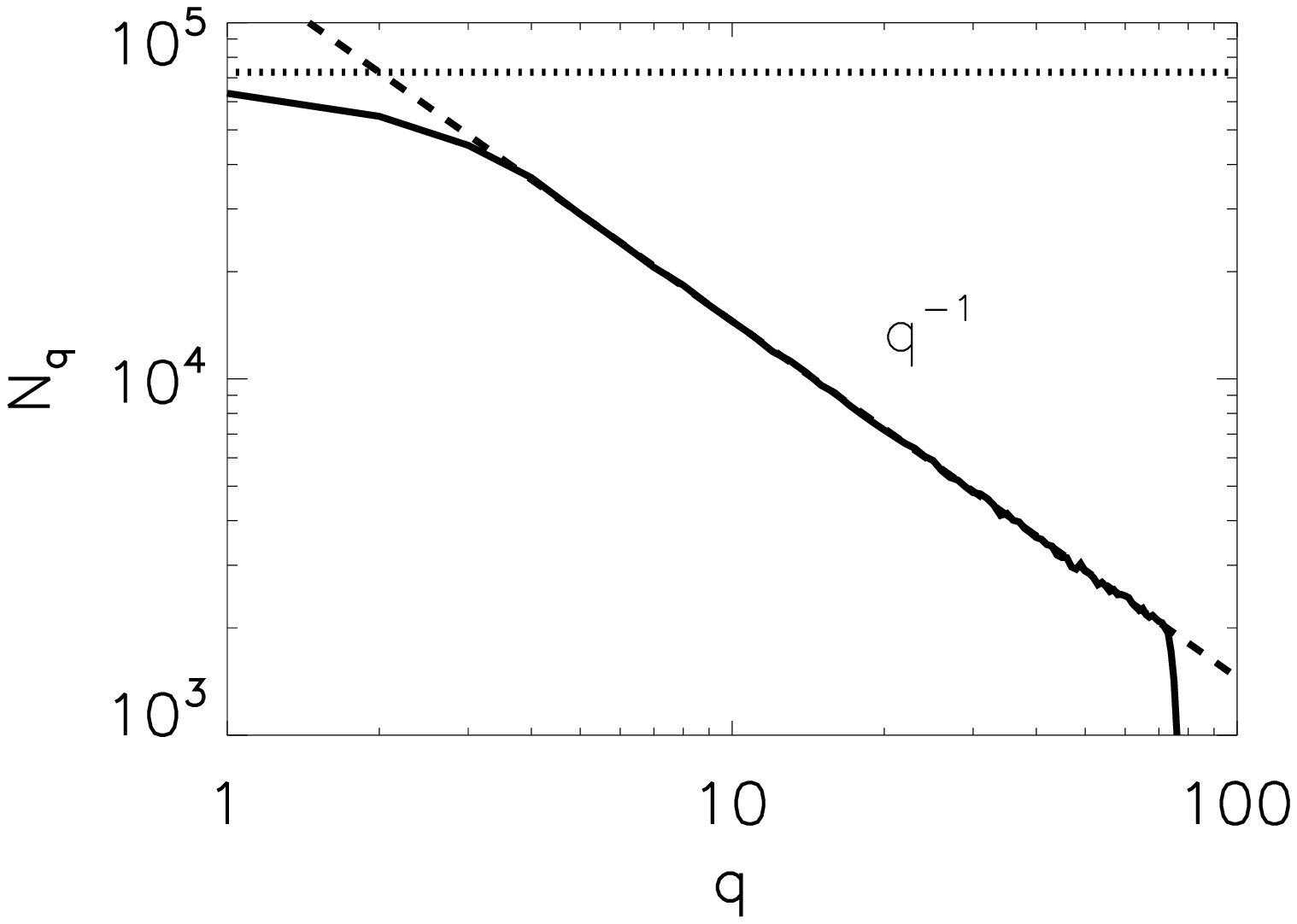}
  \includegraphics[width=0.32\textwidth]{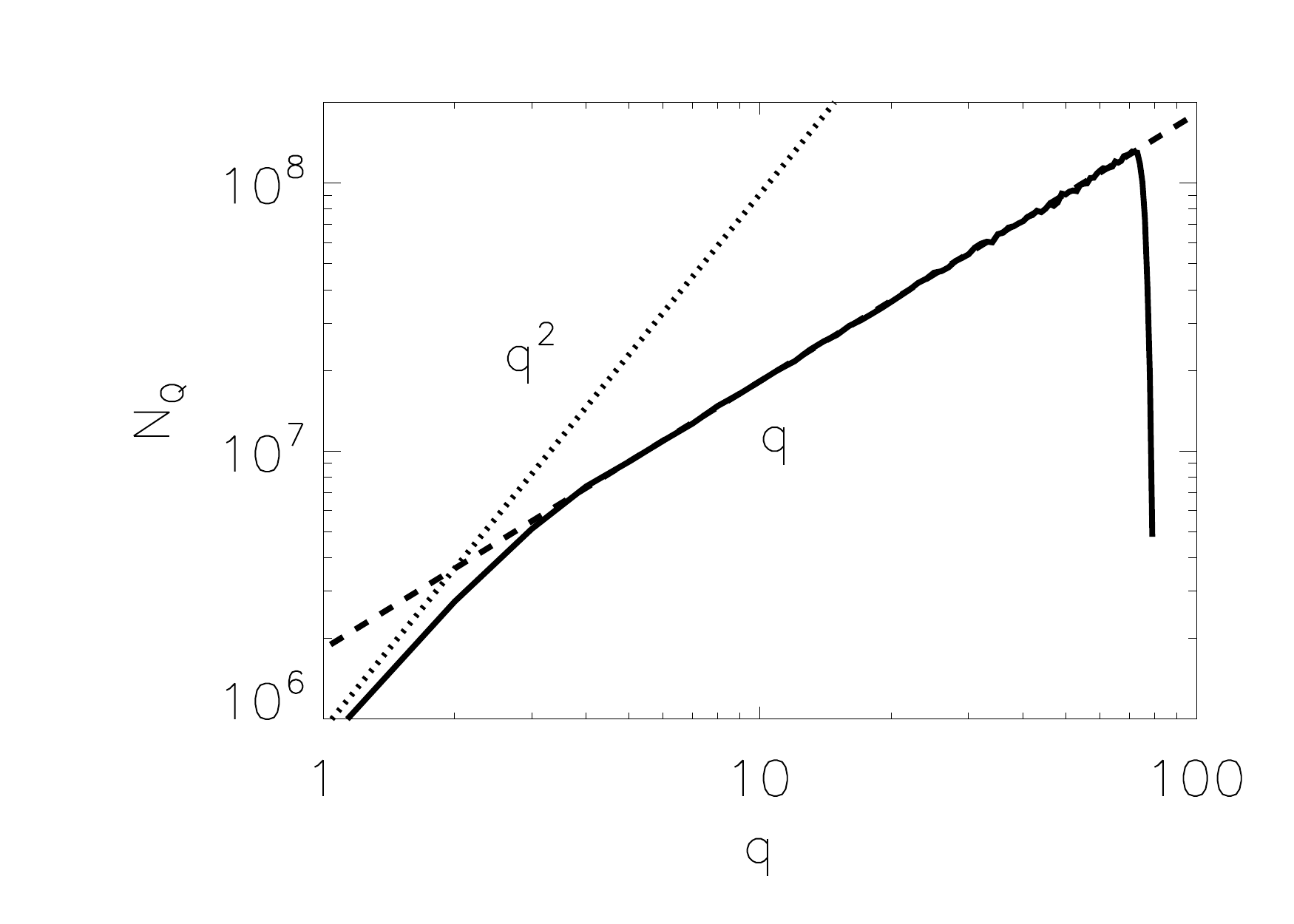}
  \caption{ Left panel: Number of interacting triads $N_{\bf q}$ between two forced modes ($|{\bf k}|\in[k_f-\delta k_f,k_f]$ 
  and a single large scale mode $\bf q$ as a function of it modulus  $q=|{\bf q}|$. Middle panel: Same plot in a log-log scale. 
  The dashed line indicates the power-law $q^{-1}$. Right panel: Number of interacting triads of the forcing modes with all 
  modes within a sphere of radius $q$. The dashed lines indicate the asymptotic predictions $N_{\bf q} = 2 \pi k_f^2 \delta k_f^2/q$
  and $N_Q=8 \pi^2 k_f^2 \delta k_f^2 q$ (see eq. \ref{NQ1}), while the doted line indicate the prediction for $q\to0$ (see eq. \ref{NQ2}).      }
  \label{fig:modes2}
\end{figure}

The volume of the intersection can be easily calculated (e.g. by a Monte Carlo method) the results of which for our case
($\kf=40$ and $\delta k_f=4$) are shown in figure \ref{fig:modes2} in linear scale (left panel) and a log-log scale in the right panel.
It results in a $q^{-1}$ power-law behavior  for $\delta k_f \ll q \ll k_f$. This power-law can easily be predicted by noting that in this range of $q$ the intersection volume is given by $N_{\bf q} = 2\pi A (k_f^2-q^2)^{1/2}$, where $A$ is the area of the small rectangle shown more clearly in the middle panel of fig. \ref{fig:modes1}, and rotational symmetry around $\bf q$ has been taken into account. 
This area $A$ is given by $A=\delta k_f^2/ \sin(2\theta)$ where $\theta = \arccos(q/k_f)$ the angle formed by $\bf k$ and $\bf q$. This leads to the prediction 
\beq N_{\bf q} \simeq 2 \pi k_f^2 \delta k_f^2/q, \qquad \mathrm{for} \qquad \delta k_f \ll q \ll k_f \label{NQ1} \eeq 
which is the dashed line shown in the middle panel of fig \ref{fig:modes2}. Then the total number of triads $N_Q$ having two modes in the forced shell and one  mode in a spherical shell $Q$ of unit width and of radius $q$, is given by $N_Q=4\pi q^2 N_q \simeq 8 \pi^2 k_f^2 \delta k_f^2 q$, where the last equality holds for $\delta k_f \ll q \ll k_f$. This number $N_Q$ is plotted in the right panel of fig \ref{fig:modes2} (solid line) along with the approximation (dashed line).


For $q \ll \delta k_f$ the number $N_q$ is approaching a finite value that corresponds to the case that the two spherical shells overlap.
Thus for $q\ll \delta k_f \ll k_f$ the number of interacting triads is equal to the volume  of the spherical shell 
\beq N_{\bf q}= 4\pi k_f^2 \delta k_f \qquad  \mathrm{for} \qquad q \ll \delta k_f \ll k_f \label{NQ2}\eeq 
shown by the horizontal doted line in midle panel fo figure \ref{fig:modes2}.
The total number of interacting triads with all modes in the spherical shell of radius between $q$ and $q+1$ is $N_Q= 16\pi^2 q^2 k_f^2\delta k_f$. 
Note that for $\delta k_f \ll q$ we have $N_Q \propto q$ while for   $q \ll \delta k_f $ we have $N_Q \propto q^2$.

\section{Nonlinearity spectrum and Pressure spectrum for flows in absolute equilibrium }\label{app:absSpec} 
Using the Fourier transform of the velocity field ${\bf v}({\bf x},t)=\sum {\bf \hat v}({\bf k},t) e^{i {\bf k}\cdot {\bf x}}$, 
the truncated Euler equations \eqref{TEE} can be expressed as
the finite system of ordinary differential equations for the complex variables ${\bf \hat v}({\bf k})$ 
\begin{equation}
{\partial_t { \hat v}_\alpha({\bf k},t)}  =  -\frac{i} {2} {\mathcal P}_{\alpha \beta \gamma}({\bf k}) \sum_{\bf p} {\hat v}_\beta({\bf p},t) {\hat v}_\gamma({\bf k-p},t)
\label{eq_discrt}
\end{equation}
where ${\mathcal P}_{\alpha \beta \gamma}=k_\beta P_{\alpha \gamma}+k_\gamma P_{\alpha \beta}$ with $P_{\alpha \beta}=\delta_{\alpha \beta}-k_\alpha k_\beta/k^2$ and the convolution in (\ref{eq_discrt}) is truncated to ${\bf k}^2 \leq k_{\rm max}^2$, ${\bf p}^2 \leq k_{\rm max}^2$ and ${\bf (k-p)}^2 \leq k_{\rm max}^2$.

Denoting by ${\bf f}({\bf k})$ the r.h.s. of \eqref{eq_discrt}, one has
\begin{equation}
\langle f_\alpha({\bf k})f_\delta({\bf k'})\rangle =  -\frac{1} {4} {\mathcal P}_{\alpha \beta \gamma}({\bf k}) {\mathcal P}_{\delta \mu \nu}({\bf k'}) \sum_{\bf p} \sum_{\bf p' }
\langle {\hat v}_\beta({\bf p}) {\hat v}_\gamma({\bf k-p}){\hat v}_\mu({\bf p'}) {\hat v}_\nu({\bf k'-p'}) \rangle
\label{eq_specfull}
\end{equation}
where here $\langle \cdot \rangle$ denotes the (ensemble) average is taken over the absolute equilibrium which is is a zero-mean gaussian field with second order moment given by
(see {\it e.g.} \cite{OrszagHouches}, (5-16))
\begin{equation}
\langle {\hat v}_\alpha ({\bf k}_1,t) {\hat v}_\beta ({\bf k}_2,t) \rangle =C P_{\alpha \beta}({\bf k}) \delta({\bf k}_1+{\bf k}_2)
\label{eq_orszag}
\end{equation}
for ${\bf k}^2\leq k_{\rm max}^2$.
Using the standard expression for fourth-order moment zero-mean jointly Gaussian random variables with covariance $\Gamma_{ij}$ (see {\it e.g.} \cite{frisch95}, Eq. (4-21)) 
\begin{equation}
\langle v_1 v_2 v_3 v_4 \rangle =\Gamma_{12}\Gamma_{34}+\Gamma_{13}\Gamma_{24}+\Gamma_{14}\Gamma_{23}. \label {eq:uriel}
\end{equation}
Two of the terms in \eqref{eq:uriel} are equal while the third is zero, yielding
\begin{equation}
\langle f_\alpha({\bf k})f_\delta({\bf -k}) \rangle=-\sum_{{\bf q}}  {\frac {C^2} 2}  {\mathcal P}_{\alpha \beta \gamma}({\bf k}) ~ {\mathcal P}_{\delta \mu \nu}(-{\bf k})~ P_{\beta \mu}({\bf k}/2-{\bf q}) 
 ~P_{\gamma \nu}({\bf k}/2+{\bf q}).
\label{eq:trace}
\end{equation}
Setting  ${\bf k}=(k,0,0)$ et ${\bf q}=(q_x,q_y,q_z)$, straightforward computation give
\begin{equation}
\langle f_\alpha({\bf k})f_\alpha({\bf -k}) \rangle=
\frac{4 C^2 k_x^2 \left(q_y^2+q_z^2\right) \left(k_x^2+4
   \left(3 q_x^2+q_y^2+q_z^2\right)\right)}{\left(k_x^2-4
   k_x q_x+4 \left(q_x^2+q_y^2+q_z^2\right)\right)
   \left(k_x^2+4 k_x q_x+4
   \left(q_x^2+q_y^2+q_z^2\right)\right)}.
    \nonumber
  \end{equation}
 For large $k_{\rm max}$, setting $(q_x,q_y,q_z)=(x k_{\rm max},y k_{\rm max},z k_{\rm max})$ and taking the dominant term in
 the limit $\epsilon = k/k_{\rm max} \rightarrow 0$, the integral over $(x,y,z)$ performed in polar coordinates, yields
\begin{equation}
\langle f_\alpha({\bf k})f_\alpha({\bf -k}) \rangle =\frac{56}{45}\pi C^2 k_{\rm max}^3 k^2. \nonumber 
 \end{equation}

The variable $C$ in \eqref{eq_orszag} can be related to the total thermalized energy $\cE$ by 
$\cE=\sum_{k \leq k_{\rm max}}E(k) = {\frac C 2} \sum_{|{\bf k}|\leq k_{\rm max}}  P_{\alpha \alpha}({\bf k}) 
=C \frac{4}{3} \pi k_{\rm max}^3$.
Thus, the thermal energy spectrum reads
 \begin{equation}
E(k)=3 \cE \frac {k^2}{k_{\rm max}^3}
\end{equation}
and one finally finds for the spectrum $E_{\cN} (k)= 4 \pi k^2 <f_\alpha({\bf k})f_\alpha({-\bf k})>$ of the nonlinear term 
 \begin{equation}
E_{\cN} (k)=\frac{14}{15} \cE k^2 E(k).
\end{equation}
A similar computation, starting with \eqref{eq_discrt} but keeping only the gradient terms in  ${\mathcal P}_{\alpha \beta \gamma}$ yields for the pressure gradient spectrum $E_{PG} (k)=\frac{16}{15} \cE k^2  E(k)$.

The pressure spectrum is thus given by
\begin{equation}
E_{P} (k)=\frac{16}{15} \cE E(k).  
 \end{equation}

\section{Spectrum of the nonlinearity for forced flows}\label{app:nonlin}  

We can estimate the spectrum of the nonlinearity for small wavenumbers $\bf q$ if some further simplification are made.
We begin by expressing the nonlinearity in terms of the helical mode decomposition.
In three dimensions, the three components of the Fourier modes $\bf \tilde{u}_k $
satisfy the incompressibility condition $\bf \tilde{u}_k \cdot k=0 $ leaving two independent complex
amplitudes. Therefore each Fourier mode can be further decomposed in two modes.
From all possible basis that a Fourier mode of an incompressible field can be decomposed
the most fruitful perhaps has been that of the decomposition to two helical modes (\cite{craya1958contributiona,Lesieur72,herring1974approach}).
It has been used in many classical papers \citep{Constantin1988,cambon1989spectral,waleffe1992nature}.
In this decomposition a Fourier mode $\bf \tilde{u}_k$ of the velocity field is written as
\beq
{\bf \tilde{u}_k} = \tilde{u}_{\bf k}^{+} {\bf h}_{\bf k}^{+}
                  + \tilde{u}_{\bf k}^{-} {\bf h}_{\bf k}^{-}.
\eeq
where the basis vectors ${\bf h}_{\bf k}^{+},{\bf h}_{\bf k}^{-}$ are
\beq
{\bf h}^s_{\bf k}= \frac{\bf k\times (e \times k) }{\sqrt{2}\bf|k \times (e \times k)|}
  +  i s\frac{\bf e \times k }{\sqrt{2}\bf |e \times k|} .
\eeq
Here $\bf e$ is an arbitrary unit vector. 
The sign index $s=\pm1$ indicates the sign of the helicity of ${\bf h}^s_{\bf k}$.
The basis vectors ${\bf h}^s_{\bf k}$ are eigenfunctions of the curl operator in Fourier space such that $i{\bf k \times h}^s_{\bf k} = s |{\bf k}| {\bf h}^s_{\bf k}$
and satisfy ${\bf h}^s_{\bf k} \cdot {\bf h}^s_{\bf k} =0$ and  ${\bf h}^s_{\bf k} \cdot {\bf h}^{-s}_{\bf k}={\bf h}^s_{\bf k} \cdot {\bf h}^{s}_{-\bf k} =1$ and form a complete base for incompressible vector fields.
The nonlinearity that acts on a mode $\tilde{u}^{s_{\bf q}}_{\bf q}$ is given by (see \cite{cambon1989spectral,waleffe1992nature}):
\beq
\tilde{\cN}^{s_{\bf q}}({\bf q})=
h_{\bf -q}^{s_q} \cdot \tilde{\cN}({\bf q}) = \sum_{\bf q=p+k} \sum_{s_{\bf k},s_{\bf p}} C_{\bq,\bk,\bp}^{s_\bq,s_\bk,s_\bp} u^{s_\bk}_\bk u^{s_\bp}_\bp
\eeq
where the pre-factor $C_{\bq,\bk,\bp}^{s_\bq,s_\bk,s_\bp}$ is given by
\beq C_{\bq,\bk,\bp}^{s_\bq,s_\bk,s_\bp} = \frac{1}{2} (s_\bk k - s_\bp p)[h_{\bf -q}^{s_q}\cdot(h_{\bk}^{s_\bk}\times h_{\bp}^{s_\bp})] \eeq 
For $q\ll k$ 
we have that  $p \simeq k (1 - {\bf q\cdot k}/k^2)$ and 
$h_{\bp}^{s_\bp} \simeq h_{-\bk}^{s_\bp} + \bq \cdot \partial_\bp  h_{\bp}^{s_\bp}|_{\bp=-\bk} +\mathcal{O}(q^2)$.
This implies that if $ s_{\bf k} = s_{\bf p}$ we would have $(s_\bk k - s_\bp p) \simeq  s_\bk {\bf q\cdot k}/k=\mathcal{O}(q)$.
On the other hand if $s_{\bf k} = - s_{\bf p}$ we have that 
$(h_{\bk}^{s_\bk}\times h_{\bp}^{s_\bp})=\mathcal{O}(q)$.
In both cases we thus have that for $q\ll k$ $$C_{\bq,\bk,\bp}^{s_\bq,s_\bk,s_\bp}=c\, q + \mathcal{O}(q^2)$$ where $c$ is an order one coefficient independent on the amplitude of $q$.
If we sum over all all triads assuming that the modes $u^{s_\bk}_\bk$ and $ u^{s_\bp}_\bp$ are independent and randomly distributed we then obtain the estimate
\beq \tilde{\cN}({\bf q}) \propto q u_{rms}^2 \sqrt{N_{\bf q}}. \label{estimate1} \eeq
Here $N_{\bf q}$ is the number of allowed triads and the square root is taken because have summed over $N_{\bf q}$ terms that take both negative and positive values.
With $u_{rms}$ we denote the root mean square amplitude of a mode inside the spherical shell $K$,
that is proportional the energy spectrum  $u_{rms}^2 \propto E(k)/k^2dk$. 
In writing eq. \ref{estimate1} we assumed that no further dependence on $q$ comes 
due to phase alignment between the modes $u^{s_\bk}_\bk$ and $ u^{s_\bp}_\bp$. 
This is a good assumption if the flow is in thermal equilibrium and this estimate can become more precise.
However it is not in general a good assumption for the turbulent scales and as we shall see differences can be present.
Squaring and summing over the  $4 \pi q^2 dq$ modes inside a spherical shell of radius $q$ and width $dq$ we obtain 
\beq                                      
E_{\cN} (q) \propto   q^4 u_{rms}^2 N_q   
\label{eq:SpecNonlin}                     
\eeq                                      

For a general flow however $E_\cN(q)$ will depend on the shape of the energy spectrum $E(k)$ and on the possible phase alignments of
all involved modes.
As was shown in the previous section if the shell of interacting wavenumbers is such that $q\ll dk$ then $N_q$  is independent of $q$ while if it 
is such that $q\ll dk$ then $N_q \propto 1/q$. 
This implies the following for an arbitrary energy spectrum $E(k)$ that varies with $k$ from 0 to $\infty$. 
If the energy spectrum is smooth (for variations $dk\gg q$) then the interactions can be considered 
as the sum of of interactions with different shells of width $dk \gg q$ that cover all $k$-space. This will result in $E_{N} (q) \propto   q^4$. 
If however there is a strong peak (e.g. at the forcing scale) of the energy spectrum that occurs over a variation of $k$ by $dk\ll q$ then the interactions with this peak should follow 
$E_{N} (q) \propto   q^3$ and could dominate the non-linearity spectrum if the peak is strong enough.

\bibliographystyle{jfm}
\bibliography{thermal}



\end{document}